\documentclass
[twocolumn,aps,prd,amsmath,amssymb,floatfix]
{revtex4-1}
\usepackage{CJK}
\usepackage[dvips]{graphicx}
\usepackage{bm}                    
\usepackage{mathptmx}               
\usepackage{dcolumn}                 
\usepackage{multirow}               
\usepackage{xcolor}
\usepackage[
breaklinks,pdfstartview=FitH,CJKbookmarks=true,
bookmarksnumbered=true,bookmarksopen=true,pdfborder={0 0 1},
colorlinks=true,linkcolor=blue,urlcolor=blue,anchorcolor=blue,citecolor=blue
            ]{hyperref}
\graphicspath{{eps/}}
\begin{document}

\def\bea{\begin{eqnarray}} \def\eea{\end{eqnarray}}
\def\be{\begin{equation}} \def\ee{\end{equation}}
\def\bal#1\eal{\begin{align}#1\end{align}}
\def\bse#1\ese{\begin{subequations}#1\end{subequations}}
\def\rra{\right\rangle} \def\lla{\left\langle}
\def\rv{\bm{r}} \def\tv{\bm{\tau}} \def\sv{\bm{\sigma}}
\def\tt{(\tv_1\cdot\tv_2)} \def\ss{(\sv_1\cdot\sv_2)}
\def\ra{\rightarrow}
\def\al{\alpha}
\def\la{\Lambda}
\def\eps{\epsilon}
\def\ms{M_\odot}
\def\esym{E_\text{sym}}
\def\mmax{M_\text{max}}
\def\fmax{f_\text{max}}
\def\l14{\Lambda_{1.4}}
\def\r14{R_{1.4}}
\def\r14{R^{(N)}_{1.4}}
\def\fm3{\;\text{fm}^{-3}}
\def\km{\;\text{km}}
\def\mev{\;\text{MeV}}
\def\hj#1{\textcolor{red}{#1}}
\def\OFF#1{}

\title{
Dark matter effects on the properties of neutron stars:
optical radii
}

\begin{CJK*}{UTF8}{gbsn}

\author{Hong-Ming Liu (刘宏铭)$^{1,4}$} 
\author{Jin-Biao Wei (魏金标)$^2$}
\author{\hbox{Zeng-Hua Li (李增花)$^{1,4}$}} \email[]{zhli09@fudan.edu.cn}
\author{G. F. Burgio$^3$}
\author{H.-J. Schulze$^3$}

\affiliation{
$^1$\hbox{
Institute of Modern Physics, Key Laboratory of Nuclear Physics
and Ion-Beam Application, MOE, Fudan University,}\\
{Shanghai 200433, People's Republic of China}\\
$^2$\hbox{
School of Mathematics and Physics, China University of Geosciences,
Lumo Road 388, 430074 Wuhan, China}\\
$^3$\hbox{
Istituto Nazionale di Fisica Nucleare, Sezione di Catania,
Dipartimento di Fisica, Universit\'a di Catania,}\\
{Via Santa Sofia 64, 95123 Catania, Italy}\\
$^4$\hbox{
Shanghai Research Center for Theoretical Nuclear Physics, NSFC and Fudan University, Shanghai 200438, China}\\
}

\date{\today}

\begin{abstract}
We study the effects of dark matter on the properties of neutron stars
by employing a DM-admixed model.
The Brueckner-Hartree-Fock theory with realistic three-body forces
and a generic bosonic self-interacting dark matter model
describe the equations of state for nuclear matter and DM, respectively.
We study the complete set of stable `dark' neutron stars
and in particular the observable radii of these objects.
A rich variety of stellar configurations is found and discussed in detail. 
\end{abstract}

\maketitle
\end{CJK*}

\section{Introduction}

The origin and nature of dark matter (DM) are among the most
challenging and fascinating issues
in modern particle physics and cosmology over the last few decades
\cite{Trimble87,Bergstrom00}.
There is now some consensus that DM is made up of yet unknown
elementary particles within or beyond the standard model \cite{Bertone05,Feng10}.
This new type of particle was first conjectured in 1933 by Zwicky \cite{Zwicky09}
and makes up more than 90\% of the matter in the Universe.
While its existence has been hinted at
from observations at galactic and super-galactic scales
\cite{Begeman91,Abdalla09,Abdalla10,Wittman00,Massey10},
most of its properties including its mass and
interactions with the rest of the standard-model particles
are presently unknown
\cite{Henriques90,Ciarcelluti11,Guever14,Raj18}.

Theoretically several kinds of bosonic and fermionic DM candidates
have been hypothesized,
such as a weakly-interacting massive particle (WIMP)
\cite{Goldman89,Andreas08,Kouvaris08,Kouvaris12,Bhat20},
especially neutralino
\cite{Martin98,Hooper04,Panotopoulos17,ADas19,
Das20,Das21b,Das21c,Das22,Kumar22,Lourenco22},
asymmetric dark matter (ADM)
\cite{Kouvaris11,McDermott12,Gresham19,Ivanytskyi20}
like mirror matter
\cite{Sandin09,Hippert22},
axion
\cite{Duffy09,Balatsky22}, and
strangelets \
\cite{Jacobs15,Ge19,Vandevender21,Zhitnitsky21}.
Experiments to detect DM particles are mainly based on three methods:
(i) the direct detection of the cross section between DM particles and nucleons
like DAMA/LIBRA \cite{Bernabei08,Bernabei10},
CRESST-I \cite{Bravin99}, and
XENON \cite{Aprile12,Aprile18};
(ii) indirect detection through scrutinizing the products of
DM candidate annihilation \cite{Abdalla09,Abdalla10,LeDelliou15};
and (iii) production by particle accelerator \cite{Sato16}.

Compact objects including white dwarfs, neutron stars (NSs), and strange stars
could efficiently capture DM particles by two mechanisms
because of their large baryonic density:
Capture might take place during their whole lifetime
and lead to a DM-admixed compact object,
if nongravitational interactions between ordinary nuclear matter (NM)
and DM exist
\cite{Lavallaz10,Lopes11,Bramante13,Bertoni13,Baryakhtar17,Bell21,Anzuini21}.
If annihilation reactions occur as well,
the compact objects may be heated and the kinematical properties
such as linear and angular momentum may be significantly affected
\cite{Gonzales10,Angeles12,Herrero19,Garani21,Bose22,Fujiwara22,Coffey22}.
On the other hand,
DM particles might be accumulated already during the stellar formation process
\cite{Ciarcelluti11,Ellis18,Gleason22}.

Among these compact objects,
NSs are the densest environments known in the Universe,
and thus provide unique astrophysical laboratories to probe the properties of DM,
following the continuous update of observational astronomical data.
Up to now, accurate observations of massive NSs such as
PSR J1614-2230 ($M = 1.908\pm0.016\ms$) \cite{Arzoumanian18},
PSR J0348+0432 ($M = 2.01\pm0.04\ms$) \cite{Antoniades13}, and
PSR J0740+6620 ($M = 2.08\pm0.07\ms$) \cite{Cromartie20},
imply that the maximum gravitational mass of NSs lies above $2\ms$.
The recent Neutron Star Interior Composition Explorer (NICER) data
also give the mass and radius of
PSR-J0030+0451 with
$R(1.44^{+0.15}_{-0.14}\ms) = 13.02^{+1.24}_{-1.06}\km$
\cite{Riley19} and
$R(1.34^{+0.15}_{-0.16}\ms) = 12.71^{+1.14}_{-1.19}\km$
\cite{Miller19},
and for PSR J0740+6620 with
$R(2.08\pm0.07\ms) = 13.7^{+2.6}_{-1.5}\km$
\cite{Riley21} and
$R(2.072^{+0.067}_{-0.066}\ms) = 12.39^{+1.30}_{-0.98}\km$
\cite{Miller21}.

It is thus of great interest to theoretically analyze
DM-admixed NS (DNS) models.
In general one employs a general-relativistic two-fluid approach
\cite{Kodama72,Comer99,Sandin09},
in which one fluid describes ordinary NM
and the other the different DM candidates.
Regarding the EOSs of ordinary NM,
many microscopic and phenomenological models have been used so far
for this purpose,
such as the
relativistic mean-field theory (RMF)
\cite{Panotopoulos17,Herrero19,ADas19,
Quddus20,Bhat20,Das20,Wang21,Das21b,Das22,Lourenco22,Miao22,Hippert22,Kumar22},
Brueckner-Hartree-Fock (BHF)
\cite{Li12},
variational
\cite{Leung11,Maselli17,Leung22},
chiral effective field theory
\cite{Tolos15,Dengler22},
and others
\cite{Ellis18,Ivanytskyi20,Kain21}.

Concerning the DM EOS,
considering the quasi total lack of knowledge regarding the nature of DM,
in the present work we employ
as a representative example
the frequently used simple one-parameter model
of bosons self interacting by a quartic term of the scalar field
in the Lagrangian density
\cite{Colpi86,Chavanis12,Maselli17,Leung22,Karkevandi22}.
It allows to study in a simple and transparent way
the qualitative features of DM in NSs.

The fundamental theoretical challenge of the eventual presence of DM in NSs
is the fact that the mass-radius relation is not anymore a unique function,
but depends on an additional degree of freedom, the DM content.
The absence of any knowledge regarding the nature of DM,
combined with the persisting uncertainty of the high-density NM EOS,
then renders any theoretical conclusion regarding either DM or high-density NM
doubtful.
The simplest example would be the observation of a very massive `NS',
which could simply be caused by the presence of a massive DM halo,
but also by a very stiff NM EOS.
Therefore theoretical methods have to be devised
to unequivocally identify the presence and quantity of DM in NSs.

We note at this point that estimates of the acquired DM content
by accretion during the NS lifetime yield extremely small results of
$\lesssim10^{-10}\ms$
\cite{Goldman89,Kouvaris08,Kouvaris10,Kouvaris11,McDermott12,Guever14,
Bramante15,Baryakhtar17,Deliyergiyev19,Ivanytskyi20},
which would be unobservable.
The existence of DNSs with large DM fractions
therefore requires exotic capture or formation mechanisms
\cite{Ciarcelluti11,Goldman13,Kouvaris15,Eby16,Maselli17,Ellis18,
Nelson19,Deliyergiyev19,DiGiovanni20},
which remain so far very speculative.
Keeping this in mind,
we study in this work in a qualitative manner
DNSs with arbitrary DM fraction up to 100\%,
corresponding to pure dark stars
\cite{Colpi86,Schunck03,Liebling12,Eby16}.

The effects of DM on the properties of NM and NSs
have been intensely investigated in recent years
\cite{Leung11,Li12,Tolos15,Maselli17,
Quddus20,Das20,Delpopolo20,Delpopolo20b,Das21,Yang21,Kain21,Leung22,Dengler22}.
A large number of works studied the possibility of DM capture by NSs
\cite{Lopes11,Bell21,Maity21,Anzuini21,Bose22}
and related phenomena like heating
\cite{Kouvaris08,Kouvaris10,Gonzales10,Bertoni13,Baryakhtar17,Raj18,
Garani21,Coffey22,Fujiwara22}
or internal black hole formation and collapse
\cite{Goldman89,Sandin09,Lavallaz10,Kouvaris12,McDermott12,
Bramante13,Bramante14,Bramante15,Ivanytskyi20}.
Recently, more quantitative studies have been performed.
For example, Ref.~\cite{Quddus20} indicated that the presence of DM in NSs
will soften the EOS and reduce the values of NSs observables
like mass, radius, tidal deformability, and moment of inertia.
Ref.~\cite{Das20} investigated the DM effects
on the derived NM parameters like incompressibility, symmetry energy,
and higher-order derivatives like
slope parameter $L$, isovector incompressibility $K_\text{sym}$,
and skewness parameter $Q_\text{sym}$.
Ref.~\cite{Maselli17} studied the equilibrium structure
of rotating bosonic and fermionic dark stars
and indicated the existence of universal relations between the
moment of inertia $I$,
tidal deformability (Love number),
and quadrupole moment $Q$ ($I$-Love-$Q$ relations),
similar to those of normal NSs
\cite{Yagi13,Yagi13b,Yagi17}.
Ref.~\cite{Zhang20} combined the bosonic DM EOS with several NM EOSs
and examined the possibility of the LIGO/Virgo events
GW170817 \cite{Abbott17} and GW190425 \cite{Abbott20}
being realized by this DNS scenario.
Many recent works
\cite{Maselli17,Ellis18,ADas19,Nelson19,Quddus20,Husain21,Das21b,Das21c,
Das22,Leung22,Lourenco22,Karkevandi22,Dengler22,Hippert22}
focused on DM effects on the NS tidal deformability
and related observables \cite{Das21d},
which are directly accessible by recent and future GW detectors.
The impact of DM on the pulsar x-ray profile \cite{Miao22}
and on NS cooling processes \cite{Bhat20,Kumar22}
were also examined recently.

While thus many sophisticated and difficult-to-measure features of DNSs
have recently been examined,
we would like in this work to take a closer look at one of the most
fundamental and direct observables of a NS,
its optical radius.
Many other properties, like moment of inertia and tidal deformability,
are closely correlated with this simple quantity.
It turns out that there is a rich scenario of possible variations
of the optical radii of DNSs due to added DM,
mainly because the DM EOS is nearly unknown presently.
We will thus carefully examine this quantity
over the full parameter space of our theoretical framework.

This article is organized as follows.
In the next section, the EOSs
of ordinary NM and DM used in this work are briefly described.
The detailed calculations and discussion are presented in Sec.~\ref{Sec.3}.
A summary is given in Sec.~\ref{Sec.4}.

\section{Formalism}

\subsection{Equation of state for nuclear matter}

The nuclear EOS of our calculations is derived in the framework of the
Brueckner-Bethe-Goldstone theory,
which is based on a linked-cluster expansion of the energy per nucleon of
nuclear matter \cite{Jeukenne76,Baldo99,Baldo12}.
The basic ingredient in this many-body approach is the in-medium Brueckner
reaction matrix $K$,
which is the solution of the Bethe-Goldstone equation
\be
 K(\rho,x;E) = V + \text{Re}\sum_{1,2} V \frac
 {\left|1 2\right\rangle Q \left\langle 12\right|}
 { E - e_1 - e_2 } K(\rho,x;E) \:,
\label{e:g}
\ee
where $V$ is the bare nucleon-nucleon ($NN$) interaction,
$E$ is the starting energy,
and the multi-indices $1,2$ denote in general momentum, isospin, and spin.
$x=\rho_p/\rho$ is the proton fraction, and
$\rho_p$ and $\rho$ are the proton and the total nucleon density, respectively.
The propagation of intermediate nucleon pairs is determined by the
Pauli operator $Q$ and the single-particle (s.p.) energy
\be
 e_1 = e(1;\rho,x) =
 \frac{k_1^2}{2m_1} + U_1 \:.
\label{e:e}
\ee
The BHF approximation for the s.p.~potential
using the continuous choice is
\be
 U_1(\rho,x) = \text{Re} \sum_{2<k_F^{(2)}}
 \langle 1 2| G(\rho,x;e_1+e_2) | 1 2 \rangle_a \:,
\label{e:u}
\ee
where the matrix element is antisymmetrized.
Due to the occurrence of $U(k)$ in Eq.~(\ref{e:e}),
the coupled system of equations (\ref{e:g}) to (\ref{e:u})
must be solved in a self-consistent manner.
The corresponding BHF energy density is
\be
 \eps_N = \sum_{i=n,p} 2\sum_{k<k_F^{(i)}}
 \left( {k^2\over 2m_i} + {1\over 2}U_i(k) \right) \:,
\ee
and all other thermodynamic quantities can be derived in a consistent way,
in particular the chemical potentials and the pressure are
\be
 \mu_i = \frac{\partial\eps_N}{\partial\rho_i} \:,
\ee
\be
 p_N = \rho^2\frac{\partial(\eps_N/\rho)}{\partial\rho}
 = \sum_i \mu_i \rho_i - \eps_N \:.
\ee
It has been shown that the nuclear EOS can be calculated with good accuracy
in this BHF two-hole-line approximation with the continuous choice for the
s.p.~potential,
since the results in this scheme are quite close to the calculations
which include also the three-hole-line contributions
\cite{Song98,Song00,Lu17,Lu18}.

In this formalism,
the only input quantity needed is the bare interaction $V$
in the Bethe-Goldstone equation (\ref{e:g}).
In the present work, we use the Argonne $V_{18}$ potential \cite{Wiringa95}
as the two-nucleon interaction,
supplemented by a consistent meson-exchange three-nucleon force,
which allows to reproduce correctly the nuclear matter saturation point
\cite{Grange89,Zuo02,Li08a,Li08b}
and other properties of nuclear matter around saturation
\cite{Wei20,Burgio21b}.
In practice we use the analytical parametrizations given in \cite{Liu22}
to compute the EOS for homogeneous nuclear matter,
$\rho > \rho_t \approx 0.08\fm3$,
and employ for the low-density clustered part
the well-known Negele-Vautherin EOS \cite{Negele71}
for the inner crust in the medium-density regime
($0.001\fm3 < \rho < \rho_t$),
and the ones by Baym-Pethick-Sutherland \cite{Baym71}
and Feynman-Metropolis-Teller \cite{Feynman49} for the outer crust
($\rho < 0.001\fm3$).

After adding the lepton contributions to the nucleonic energy density,
the structure of pure NSs can be computed by solving the standard TOV equations
(\ref{e:tovn},\ref{e:tovm},\ref{e:tovnu})
for beta-stable and charge-neutral matter.
We remark here that the value of the maximum NS mass $\mmax=2.36\ms$
of the V18 EOS
is larger than the current observational lower limits
\cite{Demorest10,Antoniades13,Fonseca16,Cromartie20}.
Some theoretical analyses of the GW170817 event
indicate also an upper limit on the maximum mass
of $\sim(2.2$--$2.4)\ms$
\cite{Shibata17,Margalit17,Rezzolla18,Shibata19},
with which the V18 EOS would be compatible as well.
However, those are very model dependent,
in particular on the still to-be-determined temperature dependence of the EOS
\cite{Khadkikar21,Bauswein21,Figura21,Liu22}.

Regarding the radius, we found in \cite{Burgio18,Wei19} that
for the V18 EOS the predicted value of a 1.4-solar-mass NS, $R_{1.4}=12.3\km$,
fulfills the constraints derived from
the tidal deformability in the GW170817 merger event.
It is also compatible with
recent mass-radius results of the NICER mission
for the pulsars J0030+0451
\cite{Riley19,Miller19}
and J0740+6620
\cite{Riley21,Miller21,Pang21,Raaijmakers21}.
The combined (strongly model-dependent) analysis
of both pulsars together with GW170817 event observations
\cite{Abbott17,Abbott18}
yields improved limits on $R_{2.08}=12.35\pm0.75\km$ \cite{Miller21},
but in particular on the radius $R_{1.4}$, namely
$12.45\pm0.65\km$ \cite{Miller21}, 
$11.94^{+0.76}_{-0.87}\km$ \cite{Pang21}, and
$12.33^{+0.76}_{-0.81}\km$ or
$12.18^{+0.56}_{-0.79}\km$ \cite{Raaijmakers21}.
In the following theoretical analysis we will for simplicity assume
$R_{1.4}=11$--$13\km$ as an estimate of the pure NS radius.

Therefore the V18 BHF EOS is a realistic nucleonic model
compatible with all current (astro)physical constraints,
which we will use now in combination with a schematic DM EOS.

\begin{figure}[t]
\vskip-12mm
\centerline{\hskip4mm\includegraphics[scale=0.55,angle=0,clip]{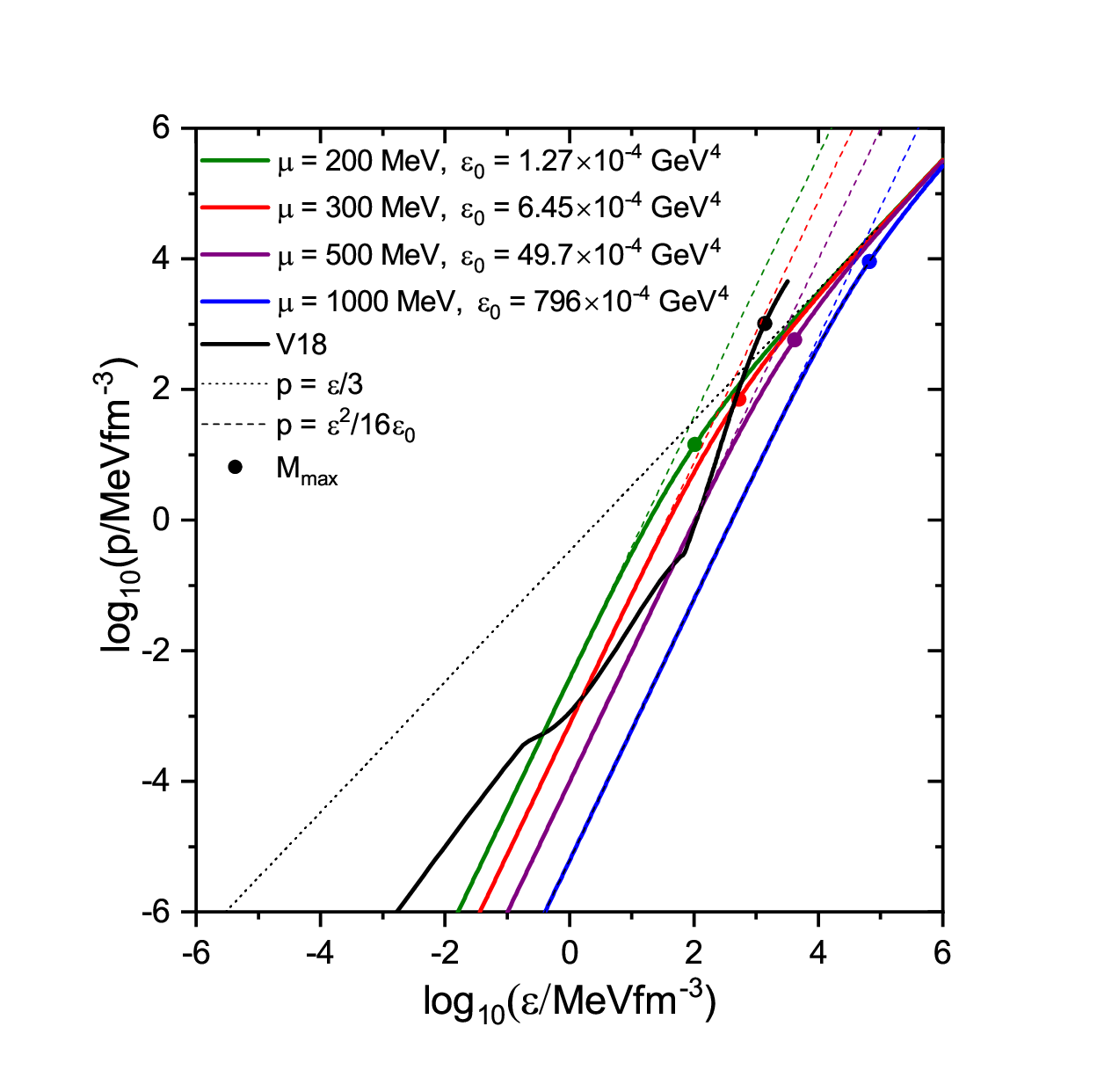}}
\vskip-9mm
\caption{
EOSs of pure DM for four DM models
and the nuclear V18 EOS.
The markers indicate $\mmax$ configurations.
The low-density and high-density asymptotics are also shown.
\hj{}
}
\label{f:eos}
\end{figure}

\begin{figure*}[t]
\vskip-8mm
\centerline{\hskip12mm\includegraphics[scale=0.34]{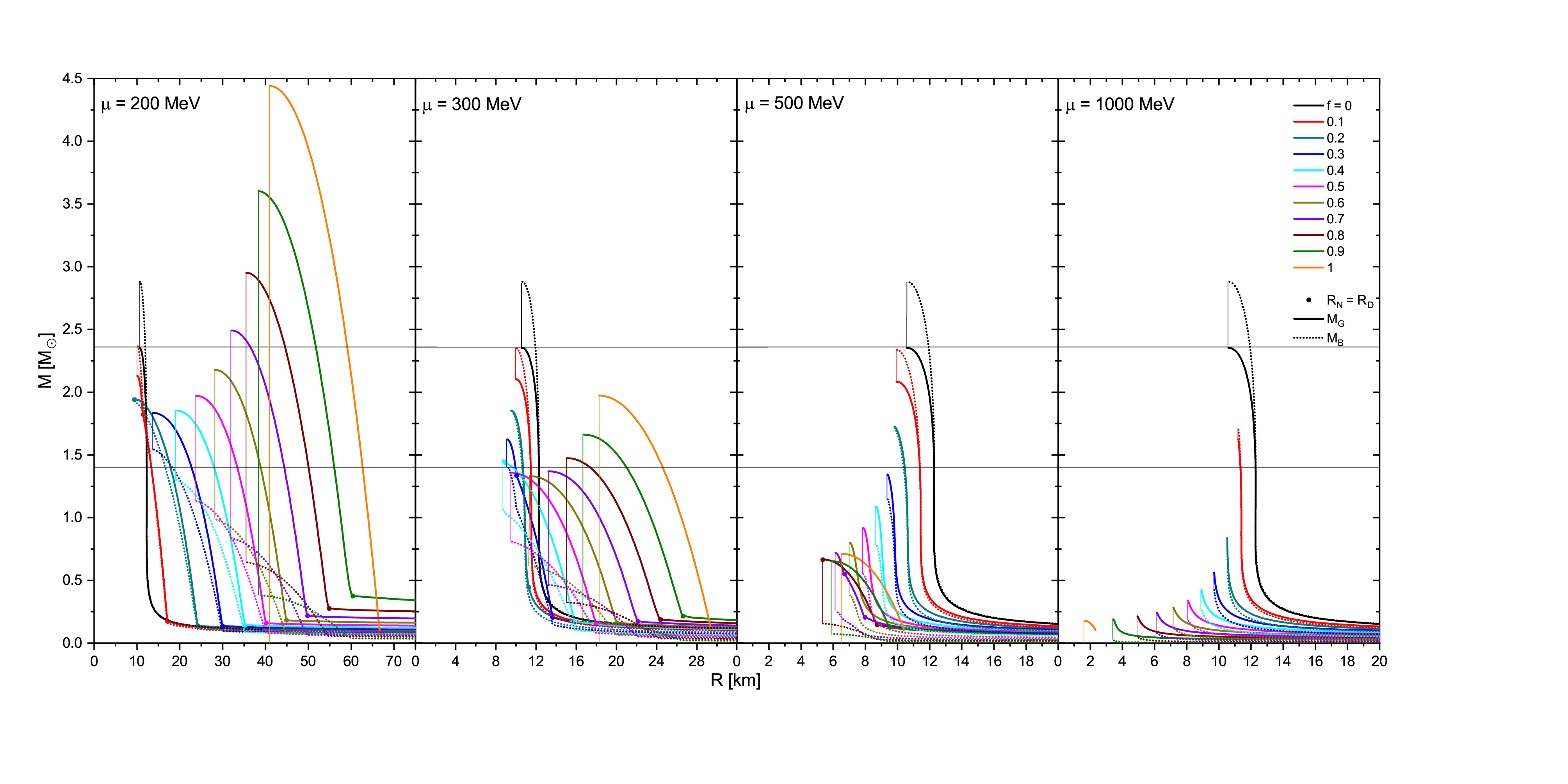}}
\vskip-16mm
\caption{
Mass-radius relations
as a function of DM fraction $f=M_D/M$
for four values of the DM particle mass $\mu=200,300,500,1000\mev$.
Total gravitational mass $M$ (solid curves)
and baryonic mass $M_B$ (dotted curves)
are shown as functions of the outer radius $R=\max(R_N,R_D)$.
Vertical lines are to guide the eye.
Horizontal lines indicate $M=1.4\ms$ and $M=\mmax=2.36\ms$ for the pure NS.
Markers indicate the $R_D=R_N$ stars.
Note the different $R$ scales.
See text for further details.
\hj{}
}
\label{f:mr}
\end{figure*}

\begin{figure}[t]
\vskip-12mm
\centerline{\hskip3mm\includegraphics[scale=0.46,angle=0,clip]{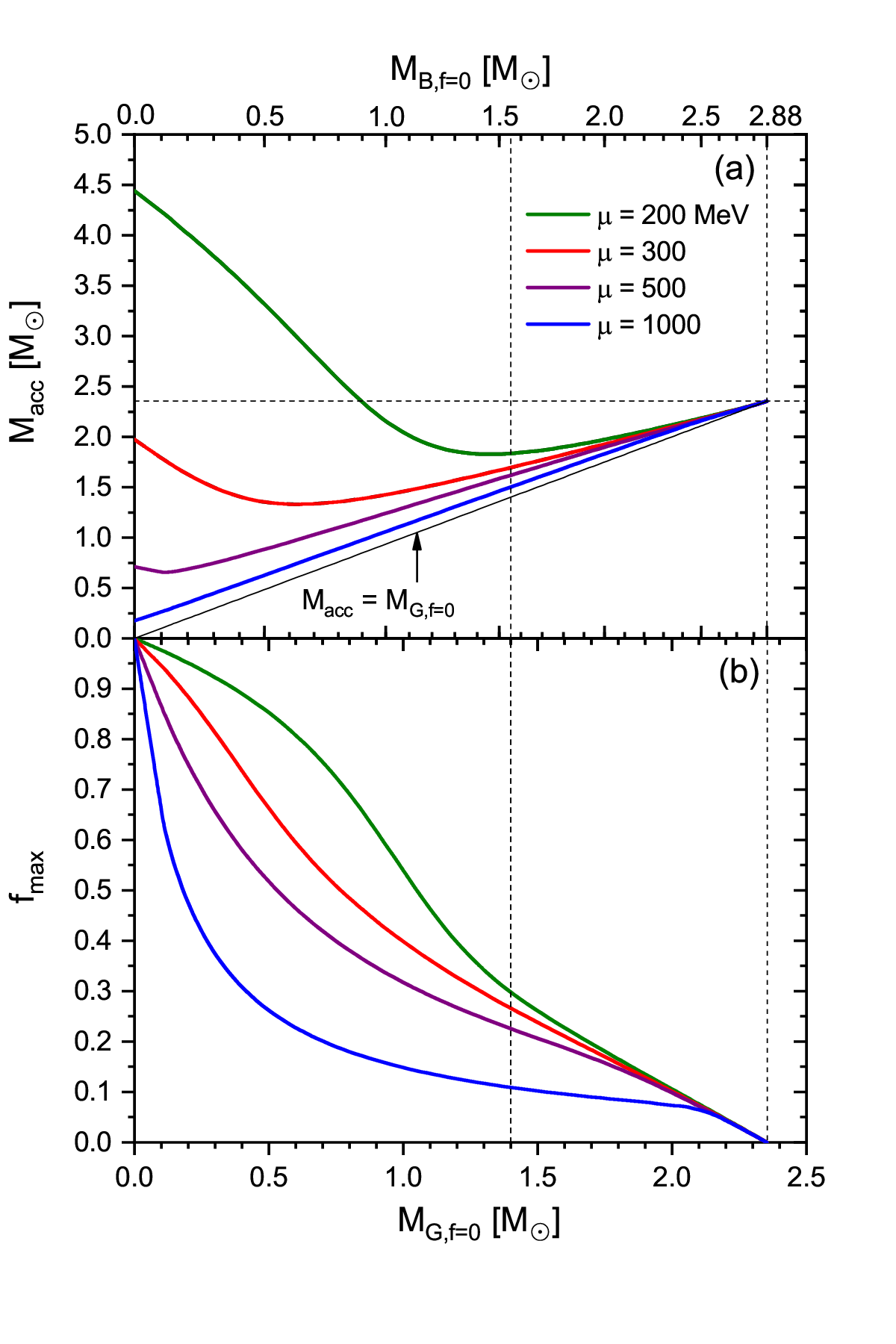}}
\vskip-12mm
\caption{
The maximum DM fraction $\fmax$ (b)
and the corresponding gravitational mass $M_\text{acc}$ (a)
reachable by accretion onto a pure NS ($f=0$)
of gravitational mass $M_G$ 
or baryonic mass $M_B$ 
for four DM models.
}
\label{f:fmax}
\end{figure}

\begin{figure}[t]
\vskip-13mm
\centerline{\hskip1mm\includegraphics[scale=0.51,angle=0,clip]{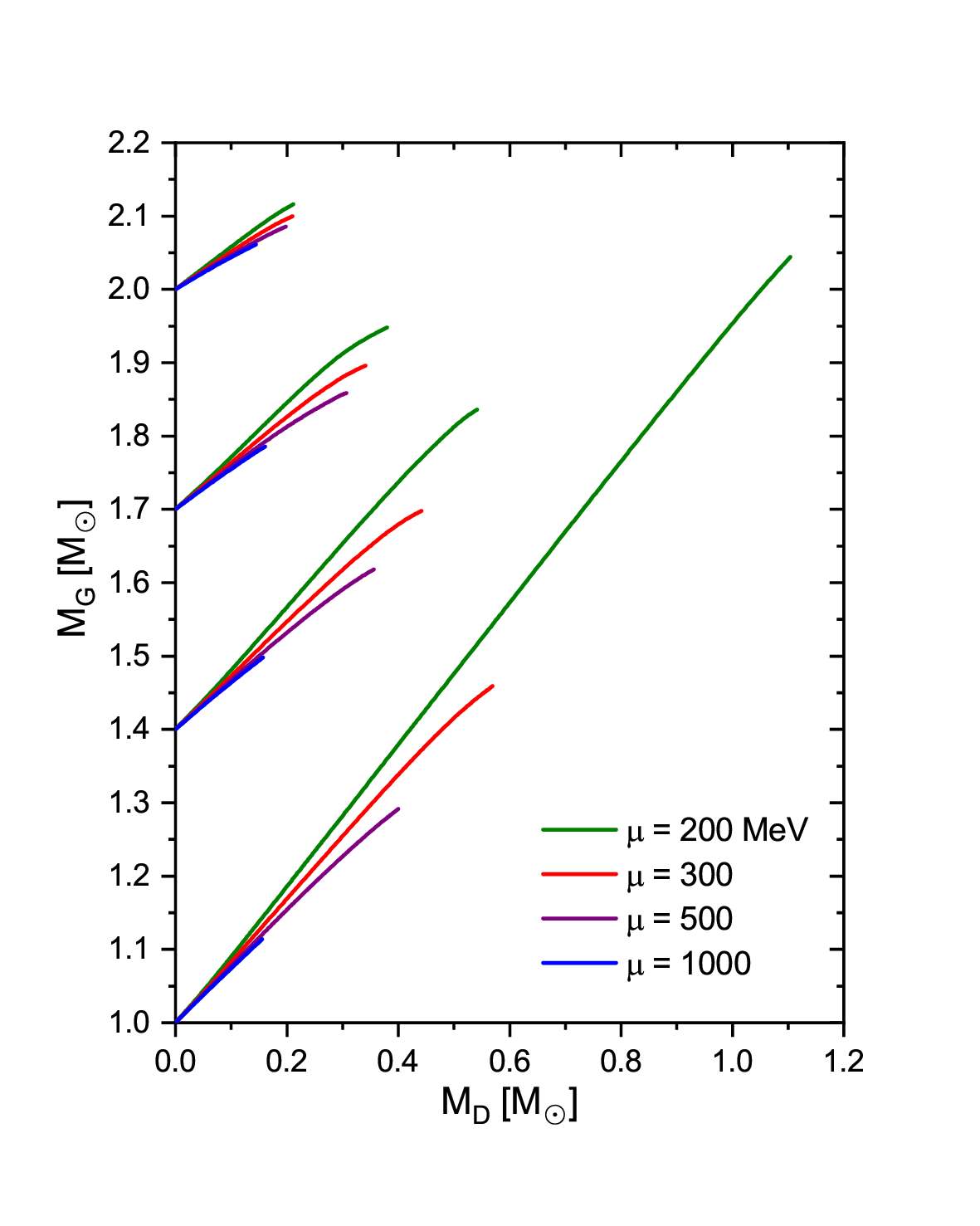}}
\vskip-12mm
\caption{
Total gravitational mass $M_G$
as function of DM mass $M_D$,
reachable by accretion
for four fixed baryonic masses  
and different DM models.
\hj{}
}
\label{f:mgmb}
\end{figure}

\begin{figure*}[t]
\vskip-12mm
\centerline{\hskip-1mm\includegraphics[scale=0.46,angle=0,clip]{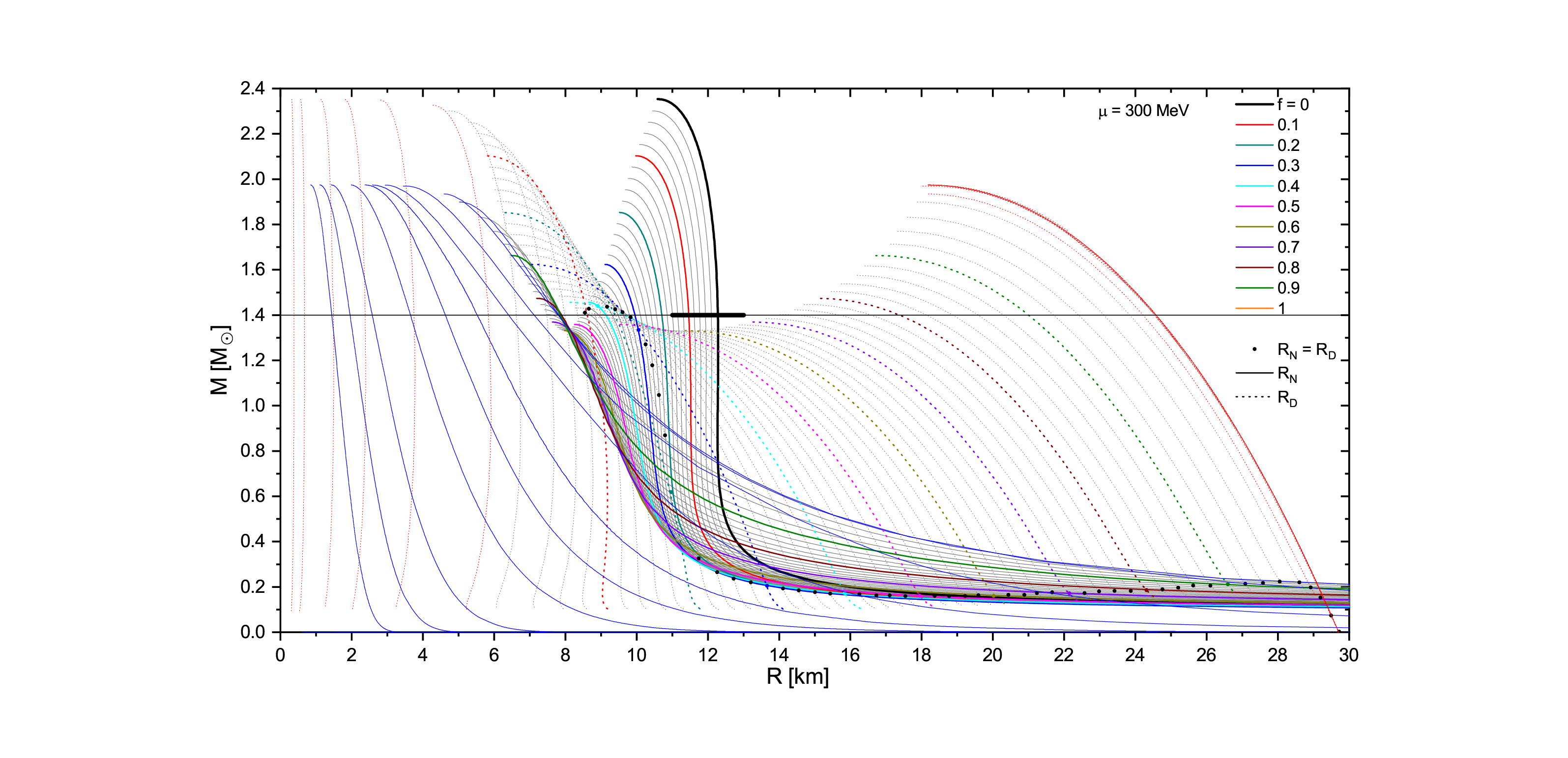}}
\vskip-14mm
\caption{
Total gravitational mass $M$ as function of nuclear (solid curves)
or dark (dotted curves) radius
for different DM fractions $f=M_D/M$
and the $\mu=300\mev$ model.
$R_N=R_D$ configurations are indicated by markers.
The range of NS radii $R_{1.4}=11$--$13\km$ is represented by a horizontal bar.
The values of $f$ are in intervals of 0.02,
apart from those close to the boundaries:
$f=10^{-2,-3,\ldots,-6}$
(thin dotted red curves)
and
$f=1-10^{-2,\ldots,-10}$
(thin solid blue curves).
\hj{}
}
\label{f:mrnd}
\end{figure*}

\begin{figure}[b]
\vskip-31mm
\centerline{\includegraphics[scale=0.48,angle=0,clip]{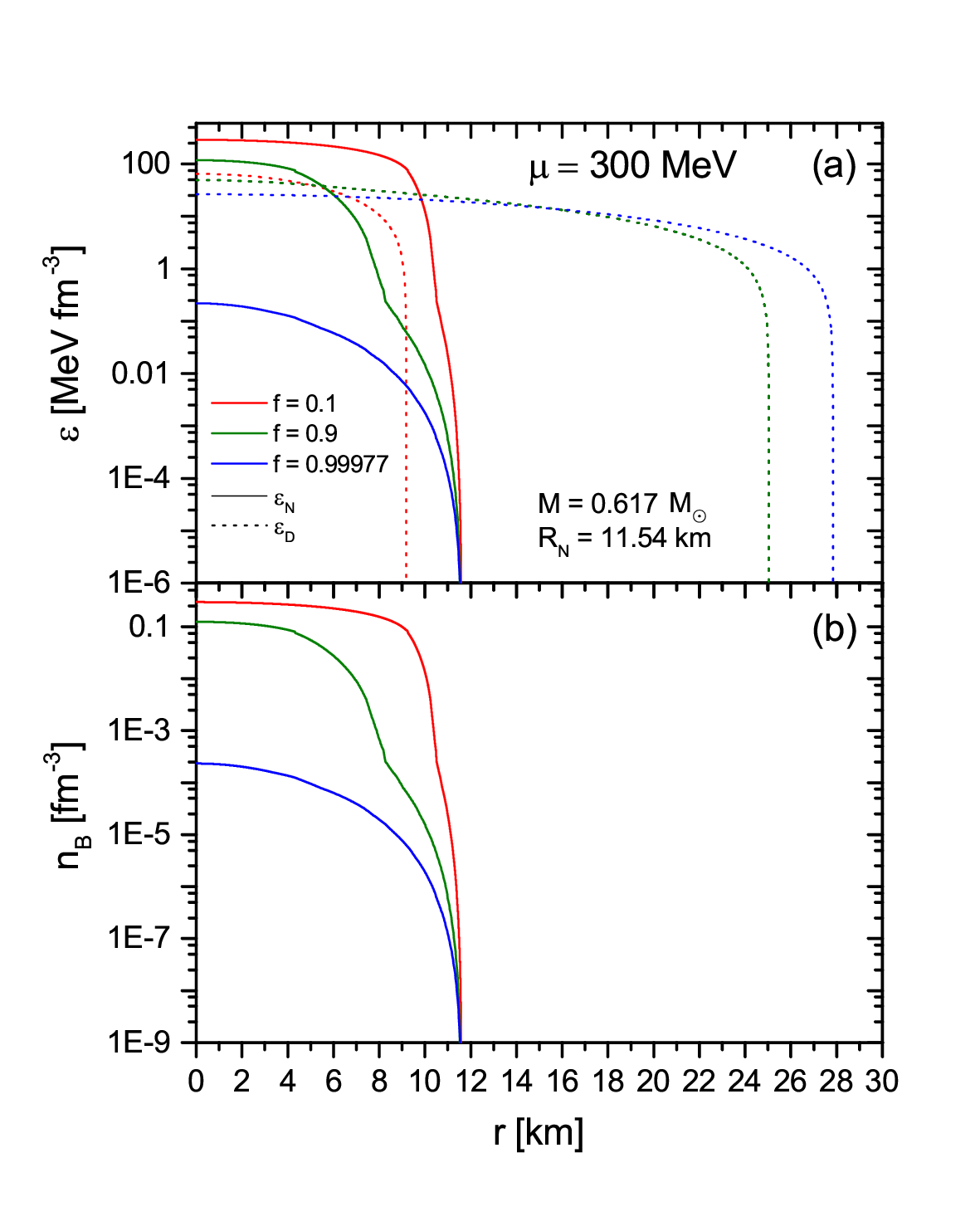}}
\vskip-10mm
\caption{
Profiles of
(a) nuclear and dark energy density, and
(b) baryon density
of the $f=0.1,0.9,0.99977$ DNSs
with the same $(M,R_N)=(0.617\ms,11.54\km)$
in Fig.~\ref{f:mrnd}.
}
\label{f:nb}
\end{figure}

\begin{figure*}[t]
\vskip-7mm
\centerline{\hskip-1mm\includegraphics[scale=0.3,angle=0,clip]{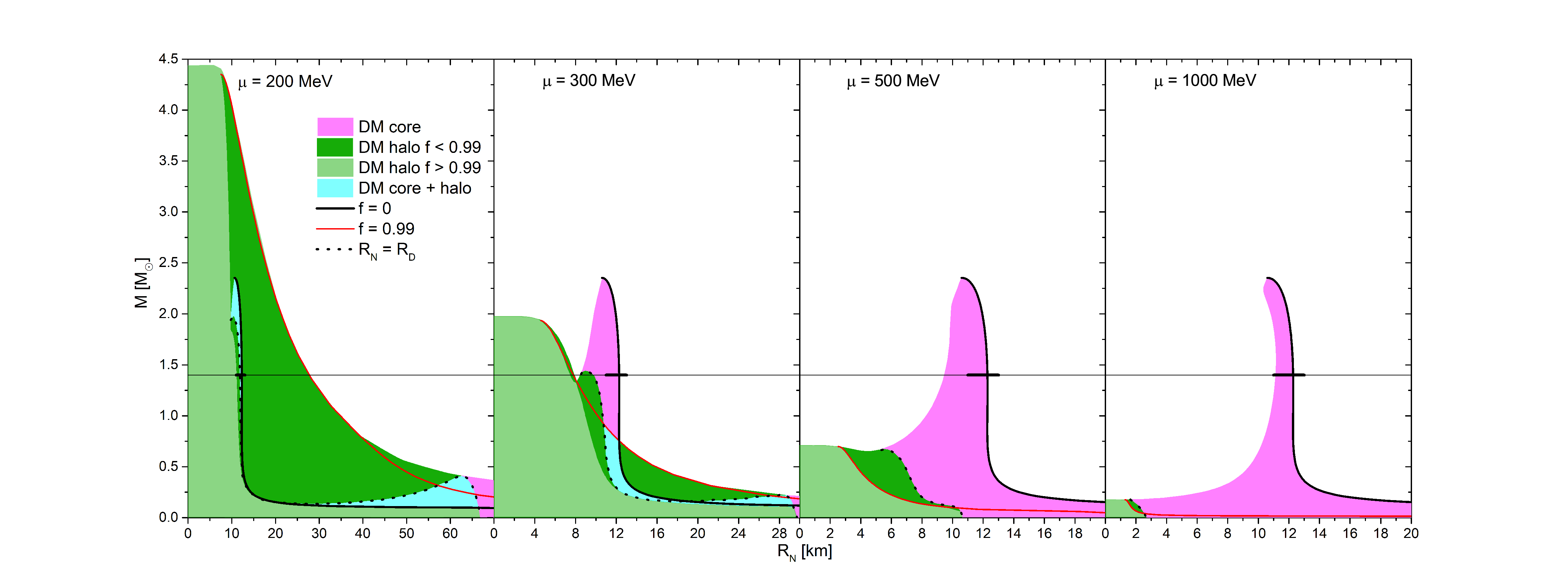}}
\vskip-6mm
\caption{
The domains of stable DM-core (red shading),
DM-halo (green),
and both (blue)
DNSs in the $(M,R_N)$ plane
for different DM models.
Note the different $R_N$ axes.
See extended discussion in the text.
}
\label{f:dns}
\end{figure*}

\begin{figure*}[t]
\vskip-10mm
\centerline{
\hskip-40mm\includegraphics[scale=0.52,angle=0,clip]{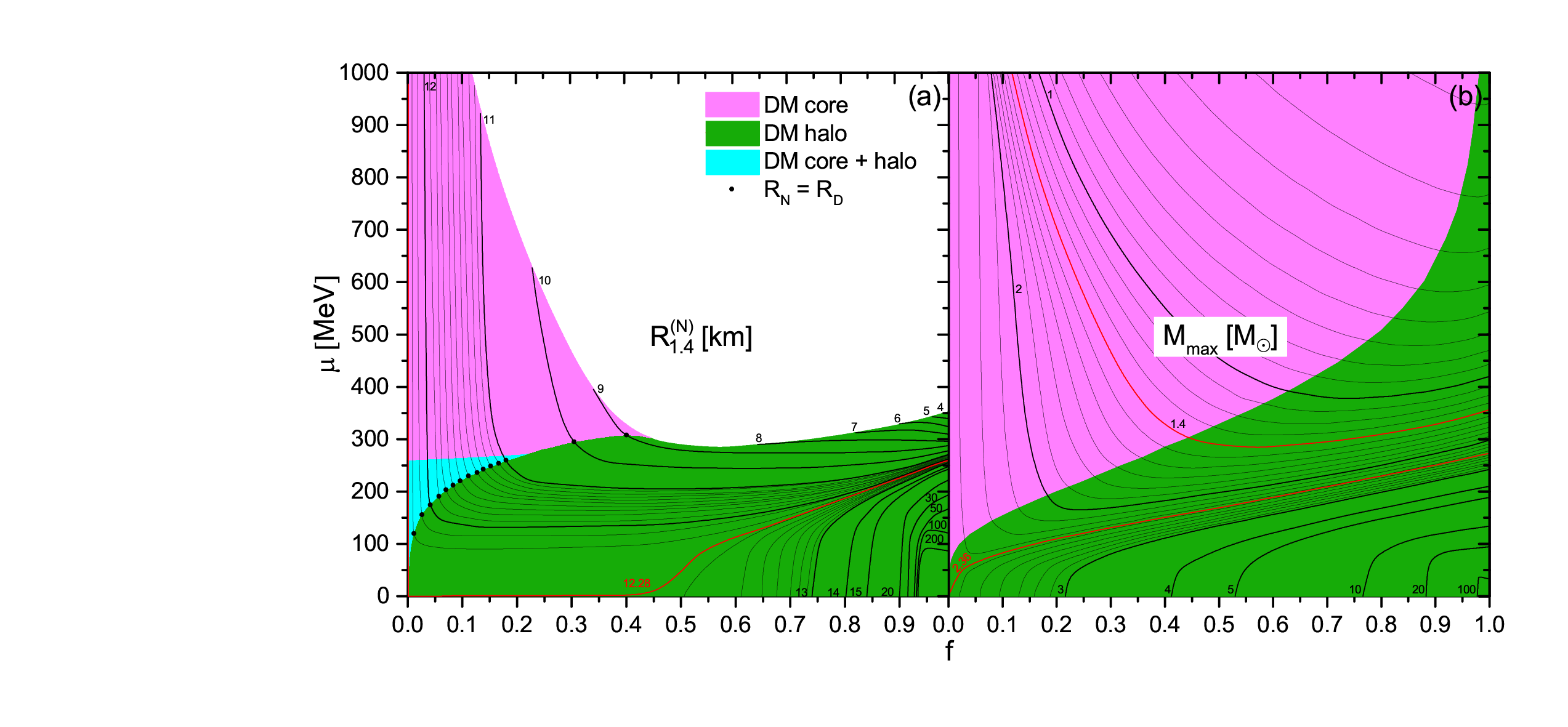}}
\vskip-9mm
\caption{
Contour plots of optical $\r14$
(indicated by numbers in km;
thin contours are in intervals of 0.1 km;
12.28 is the value for the pure NS),
and $\mmax$
(thin contours are in intervals of $0.1\ms$;
2.36 is the value for the pure NS),
as functions of $(\mu,f)$.
The color scheme is as in Fig.~\ref{f:dns}.
\\
}
\label{f:r14}
\label{f:mmx}
\end{figure*}

\begin{figure*}[t]
\vskip-10mm
\centerline{
\hskip-40mm\includegraphics[scale=0.52,angle=0,clip]{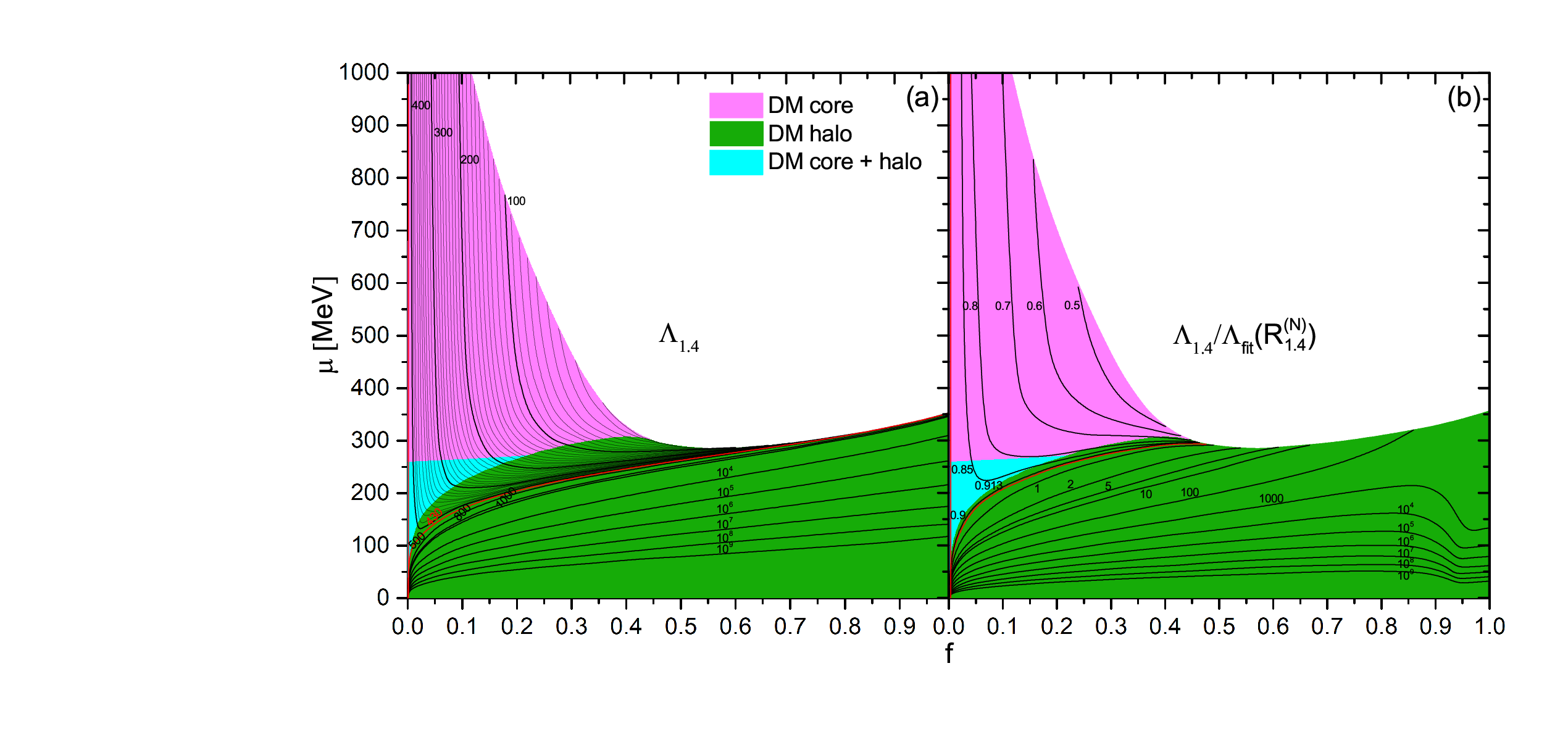}
}
\vskip-9mm
\caption{
Contour plots of
(a) tidal deformability $\Lambda_{1.4}$
(thin contours are in intervals of 10;
430 is the value for the pure NS),
and
(b) the ratio
$ \l14 / \la_\text{fit}(\r14)$,
Eq.~(\ref{e:univ}),
as functions of $(\mu,f)$.
The color scheme is as in Fig.~\ref{f:dns}.
The observation of GW170817 was reported as $\l14=70$--580 \cite{Abbott18}.
}
\label{f:l14}
\end{figure*}

\subsection{Equation of state for dark matter}

In this work we employ the frequently used generic DM model of
massive self-interacting bosons
\cite{Colpi86,Chavanis12,Maselli17,Leung22,Karkevandi22}
that features only one parameter, $\eps_0$.
The corresponding EOS relating pressure and energy density of pure DM is
\be
 p_D = \frac49 \eps_0
 \left( \sqrt{\frac34 \frac{\eps_D}{\eps_0} + 1} - 1 \right)^2 \:,
\ee
where $\eps_0\equiv\mu^4\!/4\lambda$,
$\mu$ is the DM particle mass, and
$\lambda$ is a dimensionless coupling constant.

Currently, there are basically no constraints on the DM particle mass
\cite{Kouvaris11,Bramante13}. 
However, it turns out that only masses of the order of the nucleon mass
yield sizeable observable effects on typical NS observables by admixed DM,
as will be demonstrated later.
We therefore study in the following in more detail
the four cases $\mu=200,300,500,1000\mev$
(together with the canonical choice $\lambda=\pi$;
the precise value is of minor importance due to $\mu$ entering with fourth
power in the definition of the single scale parameter $\eps_0$).
In the Newtonian limit,
the asymptotic expressions for this EOS are
$p_D = \eps_D^2/16\eps_0$ for low density and
$p_D = \eps_D/3$ for high density.

\subsection{Hydrostatic configuration}

The stable configurations of the DNSs are obtained from a
two-fluid version of the TOV equations
\cite{Kodama72,Comer99,Sandin09}:
\bal
 \frac{dp_D}{dr} &= -[p_D + \eps_D]\frac{d\nu}{dr} \:,
\\
 \frac{dp_N}{dr} &= -[p_N + \eps_N]\frac{d\nu}{dr} \:,
\label{e:tovn}
\\
 \frac{dm}{dr}   &= 4\pi r^2 \eps \:,
\label{e:tovm}
\\
 \frac{d\nu}{dr} &= \frac{m + 4\pi r^3p}{r(r - 2m)} \:,
\label{e:tovnu}
\eal
where $r$ is the radial coordinate from the center of the star, and
$p=p_N+p_D$,
$\eps=\eps_N+\eps_D$,
$m=m_N+m_D$
are the total pressure, energy density, and enclosed mass, respectively.

The total gravitational mass of the DNS is
\be
 M = m_N(R_N) + m_D(R_D) \:,
\ee
where the stellar radii $R_N$ and $R_D$
are defined by the vanishing of the respective pressures.
There are thus in general two scenarios:
DM core ($R_D<R_N$) or DM halo ($R_D>R_N$) stars.


\section{Results}
\label{Sec.3}

\subsection{Equations of state}

We begin in Fig.~\ref{f:eos} with a plot of the DM EOS
for the four choices of particle mass $\mu$
in comparison with the nucleonic V18 EOS.
The maximum-mass configurations of each case are indicated
and limit the relevant range of the EOS.
Larger values of $\mu$ correspond to larger maximum energy densities
and pressures,
but smaller maximum masses of pure DM stars.
The relevant range of the EOSs lies in between
the low-density and high-density regimes
and the universal high-density limit $p_D=\eps_D/3$
is never reached in practice.
For the nuclear EOS one notes in particular the different domains
corresponding to high-density BHF bulk matter
and low-density crustal matter.

\subsection{DNS structure and formation}

The mass-radius relations pertaining to DNSs with these EOSs
are shown in Fig.~\ref{f:mr},
namely curves with constant DM fraction $f\equiv M_D/M$
showing the gradual transition from pure NSs ($f=0$) to pure DM stars ($f=1$).
$R=\max(R_N,R_D)$ is always the outer radius in these plots.
Note that $f$ is not necessarily a physically relevant quantity,
but just a degree of freedom that allows to cover the full range of
stellar configurations with variable NM and DM contents.
In particular, it has been shown in stability analyses
\cite{Leung11,Kouvaris15,Kain21,Leung22,Gleason22}
that the maxima of the fixed-$f$ curves correspond
to the last stable configurations,
just as in the pure NS limit.
Thus only the stable stellar configurations are shown in the figure.

In this regard, we display also the baryonic stellar mass
as a function of radius in the figure (dotted curves).
This is relevant for the formation of DNSs
by DM capture or accretion onto an initial NS with fixed baryon number.
It can be seen that this process can only reach limited values of $f$
before meeting instability
\cite{McDermott12,Bramante13,Bramante14,Bramante15,Delpopolo20}.
Thus most high-$f$ configurations
are unattainable by accretion onto a fixed-$M_B$ NS
(if baryon number is conserved \cite{Davoudiasl11,Baym18} and
if such strong accretion mechanisms would exist),
but collapse would occur before at a certain smaller DM fraction.
For example,
for the $\mu=200\mev$ model,
an initial $(M_G,M_B)=(1.4,1.56)\ms$ pure NS
would allow a maximum DM fraction $\fmax\approx0.30$
of a $M_G\approx1.84\ms$ DNS    
before collapse.

The results are summarized in Fig.~\ref{f:fmax},
which shows the maximal DM fraction $\fmax$
together with the corresponding gravitational mass $M_\text{acc}$
reachable by accretion
onto a pure NS with given gravitational or baryonic mass.
Stable \hbox{high-$f$} stars could only be produced by
DM accretion onto very light NSs (if they exist)
or by exotic mechanisms
such as accretion of normal matter onto a dark star \cite{Ciarcelluti11,Ellis18,Gleason22}.
Those high-$f$ stars may have larger masses than pure NSs
only for DM models with $\mu\lesssim275\mev$,
as seen in Fig.~\ref{f:mr}.

As a further illustration,
we display in Fig.~\ref{f:mgmb} the total gravitational mass
as a function of accreted DM mass
for different fixed-$M_B$ configurations.
Again it is clearly seen that low-mass NSs allow more accreted DM
and that the effect grows with decreasing $\mu$.

\subsection{Observable radii}

After this discussion of some aspects of DNS formation dynamics,
we now focus on the properties of the observable nuclear radii $R_N$
of those objects.
For this purpose, we plot in Fig.~\ref{f:mrnd}
for the $\mu=300\mev$ model
both $M(R_N)$ and $M(R_D)$ relations
for varying values of $f=0,...,1$.
The domain covered by the solid curves represents the complete set
of stable DNS $(M,R_N)$ configurations for this model.
Some of these are DM-core ($R_D<R_N$) stars,
and the others DM-halo ($R_D>R_N$) stars.
The $R_D=R_N$ boundary between these two domains
is indicated by markers in the figure.

Note that some $(M,R_N)$ configurations correspond to two possible stars:
a low-$f$ DM-core star and a high-$f$ DM-halo star \cite{Leung11},
see, e.g., the intersection of the $f=0.1$ and $f=0.9$ curves
at $(M,R_N)=(0.617\ms,11.54\km)$.
Such `twin' stars would be indiscernable
by only mass and optical radius measurements,
but obviously their internal structure and related properties
are completely different,
which we illustrate in Fig.~\ref{f:nb},
where the respective DM core and halo structure of both configurations
with identical mass and nuclear radius are clearly evident.

Typical NSs ($M\gtrsim1\ms$) always experience a reduction of their optical radii
by the successive addition of DM
(see the $R_{1.4}$ radii in Fig.~\ref{f:mrnd}).
There are also low-mass stars with larger nuclear radii,
but those are high-$f$ configurations of low nuclear density embedded in DM,
very different from normal NSs.
A special case is the behavior in the limit \hbox{$f\ra 1$},
for which we plot the $f=1-10^{-2,-3,-4,\ldots}$ curves.
The masses and/or nuclear radii of these stars may become arbitrarily small,
corresponding to always smaller nuclear density profiles
trapped inside or held together by the DM `container' star with a radius of
$R_D\approx30\km$ in this case.
Note that including these configurations,
the $(M,R_N)=(0.617\ms,11.54\km)$ point mentioned before
becomes triple-occupied with an additional $f=0.99977$ DM-halo star
(also shown in Fig.~\ref{f:nb}).
In the limit of very small masses/nuclear densities and small radii,
so-called ``dark compact planets" are encountered \cite{Tolos15,Dengler22}.
All these are very exotic and speculative objects,
which might perhaps be created by trapping or accretion of normal matter
on a pre-existing dark star \cite{Ciarcelluti11,Ellis18,Gleason22}.
In the absence of clear ideas about the nature of the trapped matter,
in this work the provisory relevant EOS is the one of the NS crust.

\subsection{DNS configurations}

There is thus a very rich scenario of DNS configurations depending on the
parameters $\mu$ and $f$.
Following Ref.~\cite{Kain21},
the results in Figs.~\ref{f:mr} and \ref{f:mrnd}
allow to plot the domains of stable DM-core and DM-halo DNSs
in the $(M,R_N)$ plane,
which is shown in Fig.~\ref{f:dns}
for the four DM models.

In the case of heavy DM particles, $\mu\gtrsim1\,$GeV,
the configuration diagram is relatively simple:
DM accumulates in the core of the star up to very large fractions $f$,
and thus nearly all DNSs are DM-core objects,
apart from very light and small stars.
For regular NS masses there is a possible limited variation (decrease)
of the radius,
corresponding to `small' values of $f$.
With increasing $\mu$,
this band of DM-core stars becomes increasingly narrow,
corresponding to a very small
possible reduction of the radius.
Therefore, visible effects of added DM on NSs
can only be expected for `small' DM particle masses
$\mu\lesssim{\cal O}(10^3\mev)$.

In this case, the configuration plot becomes more complex:
as just discussed for $\mu=300\mev$,
there are double- or even triple-occupied $(M,R_N)$ points.
DM-halo configurations with `large' values of $f$ become increasingly important
and occupy always larger domains of the plot,
including those with very small nuclear radii and $f\ra 1$.
Those stars are characterized by
(a) masses (much) larger than $\mmax$ of pure NSs,
and/or
(b) large possible variations of the radius at any value of $M$.
In particular, large increases of $R_N$ are now also possible.

As for large $\mu$, also with decreasing $\mu$,
the band of DM-core stars attached to the $f=0$ pure NS curve,
becomes increasingly narrow
(e.g., blue zone for $\mu=200\mev$),
and this domain is triple-occupied by two other DM-halo configurations,
as in Fig.~\ref{f:nb}, for example.
Therefore, notable effects by a limited DM fraction in NSs
can only be expected for a restricted range of DM particle mass,
$\mu\approx10^2-10^3\mev$,
of the same order as the nucleon mass.
For lower values, there are possible huge variations of both $M$ and $R_N$,
but corresponding to very speculative high-$f$ DM-halo stars.
We will analyze these effects in more detail in the following.

\subsection{DNS maximum mass and radii}

Fig.~\ref{f:r14}(a) summarizes the results for possible optical $\r14$
configurations as a contour plot in the $(f,\mu)$ plane.
In accordance with Fig.~\ref{f:dns},
for `large' values of $\mu>355\mev$,
there are only DM-core configurations with reduced radii down to
$\r14\approx8.5\km$
(red domain).
The radii decrease with increasing DM fraction $f$.
For lower $\mu$, also DM-halo configurations with arbitrarily small
nuclear radii become available at very large $f$
(green domain).
For $\mu<263\mev$, $\r14$ can also increase at large $f$,
beyond the pure NS value $12.28\km$ (red contour),
up to increasingly large values with $\mu\ra 0$.
Note that,
following the previous discussion of
Figs.~\ref{f:mrnd},\ref{f:nb},\ref{f:dns},
in this case there are three possible configurations
for given radii slightly smaller than $\r14=12.28\km$ of the pure NS:
one DM-core star at low $f$ and two DM-halo stars
(the latter one with $f\approx1$ is not visible in Fig.~\ref{f:r14}).
The possible radius reduction (and associated $f$) of the DM-core star
become vanishingly small with $\mu\ra0$
(blue domain).

The same color scheme as in Fig.~\ref{f:dns}
illustrates clearly the predominance of DM-core stars
with `small' possible reduction of $\r14$
for $\mu\gtrsim300\mev$,
while for smaller $\mu$ nearly all stars are DM-halo configurations
with an increasingly wide range of $\r14$.
The discovery of compact stars with abnormaly large radii could
thus hint to a realization of DM with `low' particle mass $\mu$
below ${\cal O}(100\mev)$.
This would then also imply the existence of DNSs with abnormaly large masses.

For completeness we show the contour plot of the
maximum stable gravitational DNS mass as a function of $(\mu,f)$
in Fig.~\ref{f:mmx}(b).
While of theoretical interest,
such a plot is of limited practical use,
because in order to draw any conclusion by confrontation with masses
of observed objects,
the DM fractions of these objects should be well known.
In the absence of such information,
only few qualitative conclusions could be drawn,
for example, the ascertained observation of a very heavy DNS would
put certain upper limits on the DM particle mass $\mu$.
For the present bosonic model,
DNSs heavier than normal NSs
($2.36\ms$, red contour)
could only exist for $\mu<275\mev$.
The $1.4\ms$ contour is also emphasized in the plot and corresponds
to the upper color-white boundary in Fig.~\ref{f:r14}(a).

\subsection{DNS tidal deformability}

As a final application and demonstration
of the great importance of stellar radii
we discuss the frequently studied
\cite{Maselli17,Ellis18,ADas19,Nelson19,Quddus20,Husain21,Das21b,Das21c,
Das22,Leung22,Lourenco22,Karkevandi22,Dengler22,Hippert22}
tidal deformability (quadrupole polarizability) $\l14$ in Fig.~\ref{f:l14}(a).
This quantity is related to the tidal Love number $k_2$
\cite{Hartle67,Hinderer10,Binnington09,Damour10,Postnikov10},
\be
 \la = \frac23 \bigg(\frac{R}{M}\bigg)^5 k_2 \:,
\ee
where $R$ is the gravitational-mass radius, i.e.,
the outer DNS radius in our formalism.
The relevant equations to compute $k_2$
in the two-fluid formalism can be found in
\cite{Leung22,Karkevandi22,Dengler22},
for example.
The dependence of $\l14$
on the fifth power of the radius $R_{1.4}$ then
implies also a very strong dependence on $(\mu,f)$
that can clearly be seen in the figure.
For DM-core configurations $\l14$ is reduced
compared to the pure NS value 430
(red contour),
together with the radius $R=R_N$ reported in Fig.~\ref{f:r14}(a),
whereas for DM-halo configurations with large $f$ and $R=R_D$,
$\l14$ becomes enormous.
Thus any admixture of DM causes a substantial effect
that we analyze as follows.

A universal relation between the tidal deformabilities
and the compactness of pure NSs
was introduced in Ref.~\cite{Yagi13},
and in Ref.~\cite{Yagi17} the following fit was proposed
\be
 \frac{M}{R} = 0.36 - 0.0355 \ln\la + 0.000705 (\ln\la)^2 \:,
\label{e:univ}
\ee
or equivalently
\be
 \ln\la_\text{fit}(R) = 25.2 - 37.7 \sqrt{\frac{M}{R} + 0.087} \:,
\label{e:univ}
\ee
which holds to within 7\% for a large set of NS nucleonic EOSs \cite{Yagi17}.
Substantial deviations from this fit formula therefore indicate
`non-nucleonic' compact stars,
and we illustrate this by displaying in Fig.~\ref{f:l14}(b)
the ratio
$ \l14 / \la_\text{fit}(\r14)$
between the DNS value
and the expected value of a pure NS with the same optical radius $\r14$
as the DNS,
as reported in Fig.~\ref{f:r14}(a).
The results are in line with those in panel (a),
namely for DNS-core stars the ratio is reduced,
whereas for DM-halo stars there might be an enormous enhancement.
Independent measurements of $\r14$ and $\l14$
could thus potentially reveal the presence of DM
and even provide information on $(\mu,f)$
by just comparing $\l14$ and $\la_\text{fit}(\r14)$.
For the sake of demonstration a mass $M=1.4\ms$ was chosen here,
but the kind of analysis remains valid for other mass values.



\section{Summary}
\label{Sec.4}

We have analyzed the properties of DM-admixed NSs in a theoretical approach
combining a realistic and well-constrained BHF EOS of pure nuclear matter
with a generic DM EOS of interacting bosons involving only one parameter.
This allows to scan the qualitative properties of these compact objects
in a general way.
We focussed in particular on the observable optical (nuclear) radii
of these stars and their modification by added DM.

A rich variety of stellar configurations exists,
depending on the DM EOS,
expressed by the DM particle mass in our case.
Only if this mass is of similar order as the nucleon mass,
visible effects on the NS optical radii,
caused by small amounts of added DM,
can be observed.
For these DM-core stars, radii (and maximum masses) are always reduced.
For smaller DM particle mass ($\mu\lesssim300\mev$),
the radius reduction of DM-core stars becomes very small,
but for DM-halo stars
very large modifications of both maximum masses and radii are possible;
in particular, both can increase to very large values with $\mu\ra0$.
However, creation of these DM-rich objects requires
very exotic and speculative mechanisms.
Thus in both limits of large and small DM particle mass
(relative to $\mu\approx10^2-10^3\mev$),
the observable effects on the radii of DM-core stars become very small.
The same comments apply to observables related to stellar radii,
such as the tidal deformability analyzed here and in several recent publications.

These results are specific to the particular
(albeit somewhat generic)
bosonic DM EOS employed here,
and it will be very interesting to carry out the same kind of analysis
for different models, in particular fermionic ones.
In fact, whether any of these DNS objects exist,
is still a completely open question,
linked to the precise nature and realization of DM,
which might only be answered by future refined optical and gravitational-wave
observations addressing the issues discussed in this and other recent articles.
As it stands, the nature of dark matter and any details
like the related particle mass are unknown.
Until the situation improves,
detailed investigations of specific and intricate physical effects
are of limited use,
and we have preferred a general but rather complete analysis
regarding the impact of DM on the most `visible' and characteristic NS feature,
its optical radius.

Furthermore, our schematic investigation neglected
other important physical effects,
first of all finite temperature and rotation,
which are essential for many astronomical objects relevant here.
In particular,
those might impose additional lower limits on the mass of dark NSs.


\section*{Acknowledgments}

This work is sponsored by the National Key R\&D Program of China No.~2022YFA1602303 and
the National Natural Science Foundation of China under Grant
Nos.~12205260, 11975077,11475045,12147101.

\def\aap{A\&A}
\def\apjl{Astrophys. J. Lett.}
\def\araa{Annu. Rev. Astron. Astrophys.}
\def\epja{EPJA}
\def\epjc{EPJC}
\def\jcap{Journal of Cosmology and Astroparticle Physics}
\def\jcap{JCAP}
\def\mnras{MNRAS}
\def\npb{Nucl. Phys. B}
\def\physrep{Phys. Rep.}
\def\plb{Phys. Lett. B}
\def\rpp{Rep. Prog. Phys.}
\bibliographystyle{apsrev4-1}
\bibliography{dmb}

\begin{thebibliography}{149}%
\makeatletter
\providecommand \@ifxundefined [1]{%
 \@ifx{#1\undefined}
}%
\providecommand \@ifnum [1]{%
 \ifnum #1\expandafter \@firstoftwo
 \else \expandafter \@secondoftwo
 \fi
}%
\providecommand \@ifx [1]{%
 \ifx #1\expandafter \@firstoftwo
 \else \expandafter \@secondoftwo
 \fi
}%
\providecommand \natexlab [1]{#1}%
\providecommand \enquote  [1]{``#1''}%
\providecommand \bibnamefont  [1]{#1}%
\providecommand \bibfnamefont [1]{#1}%
\providecommand \citenamefont [1]{#1}%
\providecommand \href@noop [0]{\@secondoftwo}%
\providecommand \href [0]{\begingroup \@sanitize@url \@href}%
\providecommand \@href[1]{\@@startlink{#1}\@@href}%
\providecommand \@@href[1]{\endgroup#1\@@endlink}%
\providecommand \@sanitize@url [0]{\catcode `\\12\catcode `\$12\catcode
  `\&12\catcode `\#12\catcode `\^12\catcode `\_12\catcode `\%12\relax}%
\providecommand \@@startlink[1]{}%
\providecommand \@@endlink[0]{}%
\providecommand \url  [0]{\begingroup\@sanitize@url \@url }%
\providecommand \@url [1]{\endgroup\@href {#1}{\urlprefix }}%
\providecommand \urlprefix  [0]{URL }%
\providecommand \Eprint [0]{\href }%
\providecommand \doibase [0]{http://dx.doi.org/}%
\providecommand \selectlanguage [0]{\@gobble}%
\providecommand \bibinfo  [0]{\@secondoftwo}%
\providecommand \bibfield  [0]{\@secondoftwo}%
\providecommand \translation [1]{[#1]}%
\providecommand \BibitemOpen [0]{}%
\providecommand \bibitemStop [0]{}%
\providecommand \bibitemNoStop [0]{.\EOS\space}%
\providecommand \EOS [0]{\spacefactor3000\relax}%
\providecommand \BibitemShut  [1]{\csname bibitem#1\endcsname}%
\let\auto@bib@innerbib\@empty
\bibitem [{\citenamefont {Trimble}(1987)}]{Trimble87}%
  \BibitemOpen
  \bibfield  {author} {\bibinfo {author} {\bibfnamefont {V.}~\bibnamefont
  {Trimble}},\ }\href {\doibase 10.1146/annurev.aa.25.090187.002233} {\bibfield
   {journal} {\bibinfo  {journal} {\araa}\ }\textbf {\bibinfo {volume} {25}},\
  \bibinfo {pages} {425} (\bibinfo {year} {1987})}\BibitemShut {NoStop}%
\bibitem [{\citenamefont {{Bergstr{\"o}m}}(2000)}]{Bergstrom00}%
  \BibitemOpen
  \bibfield  {author} {\bibinfo {author} {\bibfnamefont {L.}~\bibnamefont
  {{Bergstr{\"o}m}}},\ }\href {\doibase 10.1088/0034-4885/63/5/2r3} {\bibfield
  {journal} {\bibinfo  {journal} {Rep. Prog. Phys.}\ }\textbf {\bibinfo
  {volume} {63}},\ \bibinfo {pages} {793} (\bibinfo {year} {2000})}\BibitemShut
  {NoStop}%
\bibitem [{\citenamefont {{Bertone}}\ \emph {et~al.}(2005)\citenamefont
  {{Bertone}}, \citenamefont {{Hooper}},\ and\ \citenamefont
  {{Silk}}}]{Bertone05}%
  \BibitemOpen
  \bibfield  {author} {\bibinfo {author} {\bibfnamefont {G.}~\bibnamefont
  {{Bertone}}}, \bibinfo {author} {\bibfnamefont {D.}~\bibnamefont {{Hooper}}},
  \ and\ \bibinfo {author} {\bibfnamefont {J.}~\bibnamefont {{Silk}}},\ }\href
  {\doibase 10.1016/j.physrep.2004.08.031} {\bibfield  {journal} {\bibinfo
  {journal} {\physrep}\ }\textbf {\bibinfo {volume} {405}},\ \bibinfo {pages}
  {279} (\bibinfo {year} {2005})}\BibitemShut {NoStop}%
\bibitem [{\citenamefont {Feng}(2010)}]{Feng10}%
  \BibitemOpen
  \bibfield  {author} {\bibinfo {author} {\bibfnamefont {J.~L.}\ \bibnamefont
  {Feng}},\ }\href {\doibase 10.1146/annurev-astro-082708-101659} {\bibfield
  {journal} {\bibinfo  {journal} {\araa}\ }\textbf {\bibinfo {volume} {48}},\
  \bibinfo {pages} {495} (\bibinfo {year} {2010})}\BibitemShut {NoStop}%
\bibitem [{\citenamefont {Zwicky}(2009)}]{Zwicky09}%
  \BibitemOpen
  \bibfield  {author} {\bibinfo {author} {\bibfnamefont {F.}~\bibnamefont
  {Zwicky}},\ }\href {\doibase 10.1007/s10714-008-0707-4} {\bibfield  {journal}
  {\bibinfo  {journal} {Gen. Relativ. Gravit.}\ }\textbf {\bibinfo {volume}
  {41}},\ \bibinfo {pages} {207} (\bibinfo {year} {2009})}\BibitemShut
  {NoStop}%
\bibitem [{\citenamefont {Begeman}\ \emph {et~al.}(1991)\citenamefont
  {Begeman}, \citenamefont {Broeils},\ and\ \citenamefont
  {Sanders}}]{Begeman91}%
  \BibitemOpen
  \bibfield  {author} {\bibinfo {author} {\bibfnamefont {K.~G.}\ \bibnamefont
  {Begeman}}, \bibinfo {author} {\bibfnamefont {A.~H.}\ \bibnamefont
  {Broeils}}, \ and\ \bibinfo {author} {\bibfnamefont {R.~H.}\ \bibnamefont
  {Sanders}},\ }\href {\doibase 10.1093/mnras/249.3.523} {\bibfield  {journal}
  {\bibinfo  {journal} {Mon. Not. R. Astron. Soc.}\ }\textbf {\bibinfo {volume}
  {249}},\ \bibinfo {pages} {523} (\bibinfo {year} {1991})}\BibitemShut
  {NoStop}%
\bibitem [{\citenamefont {Abdalla}\ \emph {et~al.}(2009)\citenamefont
  {Abdalla}, \citenamefont {Abramo}, \citenamefont {Sodr$\acute{\rm{e}}$},\
  and\ \citenamefont {Wang}}]{Abdalla09}%
  \BibitemOpen
  \bibfield  {author} {\bibinfo {author} {\bibfnamefont {E.}~\bibnamefont
  {Abdalla}}, \bibinfo {author} {\bibfnamefont {L.~R.}\ \bibnamefont {Abramo}},
  \bibinfo {author} {\bibfnamefont {L.}~\bibnamefont {Sodr$\acute{\rm{e}}$}}, \
  and\ \bibinfo {author} {\bibfnamefont {B.}~\bibnamefont {Wang}},\ }\href
  {\doibase https://doi.org/10.1016/j.physletb.2009.02.008} {\bibfield
  {journal} {\bibinfo  {journal} {Phys. Lett. B}\ }\textbf {\bibinfo {volume}
  {673}},\ \bibinfo {pages} {107} (\bibinfo {year} {2009})}\BibitemShut
  {NoStop}%
\bibitem [{\citenamefont {Abdalla}\ \emph {et~al.}(2010)\citenamefont
  {Abdalla}, \citenamefont {Abramo},\ and\ \citenamefont
  {de~Souza}}]{Abdalla10}%
  \BibitemOpen
  \bibfield  {author} {\bibinfo {author} {\bibfnamefont {E.}~\bibnamefont
  {Abdalla}}, \bibinfo {author} {\bibfnamefont {L.~R.}\ \bibnamefont {Abramo}},
  \ and\ \bibinfo {author} {\bibfnamefont {J.~C.~C.}\ \bibnamefont
  {de~Souza}},\ }\href {\doibase 10.1103/PhysRevD.82.023508} {\bibfield
  {journal} {\bibinfo  {journal} {Phys. Rev. D}\ }\textbf {\bibinfo {volume}
  {82}},\ \bibinfo {pages} {023508} (\bibinfo {year} {2010})}\BibitemShut
  {NoStop}%
\bibitem [{\citenamefont {Wittman}\ \emph {et~al.}(2000)\citenamefont
  {Wittman}, \citenamefont {Tyson}, \citenamefont {Kirkman}, \citenamefont
  {Dell'Antonio},\ and\ \citenamefont {Bernstein}}]{Wittman00}%
  \BibitemOpen
  \bibfield  {author} {\bibinfo {author} {\bibfnamefont {D.~M.}\ \bibnamefont
  {Wittman}}, \bibinfo {author} {\bibfnamefont {J.~A.}\ \bibnamefont {Tyson}},
  \bibinfo {author} {\bibfnamefont {D.}~\bibnamefont {Kirkman}}, \bibinfo
  {author} {\bibfnamefont {I.}~\bibnamefont {Dell'Antonio}}, \ and\ \bibinfo
  {author} {\bibfnamefont {G.}~\bibnamefont {Bernstein}},\ }\href {\doibase
  10.1038/35012001} {\bibfield  {journal} {\bibinfo  {journal} {Nature
  (London)}\ }\textbf {\bibinfo {volume} {405}},\ \bibinfo {pages} {143}
  (\bibinfo {year} {2000})}\BibitemShut {NoStop}%
\bibitem [{\citenamefont {Massey}\ \emph {et~al.}(2010)\citenamefont {Massey},
  \citenamefont {Kitching},\ and\ \citenamefont {Richard}}]{Massey10}%
  \BibitemOpen
  \bibfield  {author} {\bibinfo {author} {\bibfnamefont {R.}~\bibnamefont
  {Massey}}, \bibinfo {author} {\bibfnamefont {T.}~\bibnamefont {Kitching}}, \
  and\ \bibinfo {author} {\bibfnamefont {J.}~\bibnamefont {Richard}},\ }\href
  {\doibase 10.1088/0034-4885/73/8/086901} {\bibfield  {journal} {\bibinfo
  {journal} {Rep. Prog. Phys.}\ }\textbf {\bibinfo {volume} {73}},\ \bibinfo
  {pages} {086901} (\bibinfo {year} {2010})}\BibitemShut {NoStop}%
\bibitem [{\citenamefont {{Henriques}}\ \emph {et~al.}(1990)\citenamefont
  {{Henriques}}, \citenamefont {{Liddle}},\ and\ \citenamefont
  {{Moorhouse}}}]{Henriques90}%
  \BibitemOpen
  \bibfield  {author} {\bibinfo {author} {\bibfnamefont {A.~B.}\ \bibnamefont
  {{Henriques}}}, \bibinfo {author} {\bibfnamefont {A.~R.}\ \bibnamefont
  {{Liddle}}}, \ and\ \bibinfo {author} {\bibfnamefont {R.~G.}\ \bibnamefont
  {{Moorhouse}}},\ }\href {\doibase 10.1016/0550-3213(90)90514-E} {\bibfield
  {journal} {\bibinfo  {journal} {\npb}\ }\textbf {\bibinfo {volume} {337}},\
  \bibinfo {pages} {737} (\bibinfo {year} {1990})}\BibitemShut {NoStop}%
\bibitem [{\citenamefont {Ciarcelluti}\ and\ \citenamefont
  {Sandin}(2011)}]{Ciarcelluti11}%
  \BibitemOpen
  \bibfield  {author} {\bibinfo {author} {\bibfnamefont {P.}~\bibnamefont
  {Ciarcelluti}}\ and\ \bibinfo {author} {\bibfnamefont {F.}~\bibnamefont
  {Sandin}},\ }\href {\doibase https://doi.org/10.1016/j.physletb.2010.11.021}
  {\bibfield  {journal} {\bibinfo  {journal} {Phys. Lett. B}\ }\textbf
  {\bibinfo {volume} {695}},\ \bibinfo {pages} {19} (\bibinfo {year}
  {2011})}\BibitemShut {NoStop}%
\bibitem [{\citenamefont {G\"uver}\ \emph {et~al.}(2014)\citenamefont
  {G\"uver}, \citenamefont {Erkoca}, \citenamefont {Reno},\ and\ \citenamefont
  {Sarcevic}}]{Guever14}%
  \BibitemOpen
  \bibfield  {author} {\bibinfo {author} {\bibfnamefont {T.}~\bibnamefont
  {G\"uver}}, \bibinfo {author} {\bibfnamefont {A.~E.}\ \bibnamefont {Erkoca}},
  \bibinfo {author} {\bibfnamefont {M.~H.}\ \bibnamefont {Reno}}, \ and\
  \bibinfo {author} {\bibfnamefont {I.}~\bibnamefont {Sarcevic}},\ }\href
  {\doibase 10.1088/1475-7516/2014/05/013} {\bibfield  {journal} {\bibinfo
  {journal} {J. Cosmol. Astropart. Phys.}\ }\textbf {\bibinfo {volume}
  {2014}},\ \bibinfo {pages} {013} (\bibinfo {year} {2014})}\BibitemShut
  {NoStop}%
\bibitem [{\citenamefont {Raj}\ \emph {et~al.}(2018)\citenamefont {Raj},
  \citenamefont {Tanedo},\ and\ \citenamefont {Yu}}]{Raj18}%
  \BibitemOpen
  \bibfield  {author} {\bibinfo {author} {\bibfnamefont {N.}~\bibnamefont
  {Raj}}, \bibinfo {author} {\bibfnamefont {P.}~\bibnamefont {Tanedo}}, \ and\
  \bibinfo {author} {\bibfnamefont {H.-B.}\ \bibnamefont {Yu}},\ }\href
  {\doibase 10.1103/PhysRevD.97.043006} {\bibfield  {journal} {\bibinfo
  {journal} {Phys. Rev. D}\ }\textbf {\bibinfo {volume} {97}},\ \bibinfo
  {pages} {043006} (\bibinfo {year} {2018})}\BibitemShut {NoStop}%
\bibitem [{\citenamefont {{Goldman}}\ and\ \citenamefont
  {{Nussinov}}(1989)}]{Goldman89}%
  \BibitemOpen
  \bibfield  {author} {\bibinfo {author} {\bibfnamefont {I.}~\bibnamefont
  {{Goldman}}}\ and\ \bibinfo {author} {\bibfnamefont {S.}~\bibnamefont
  {{Nussinov}}},\ }\href {\doibase 10.1103/PhysRevD.40.3221} {\bibfield
  {journal} {\bibinfo  {journal} {\prd}\ }\textbf {\bibinfo {volume} {40}},\
  \bibinfo {pages} {3221} (\bibinfo {year} {1989})}\BibitemShut {NoStop}%
\bibitem [{\citenamefont {Andreas}\ \emph {et~al.}(2008)\citenamefont
  {Andreas}, \citenamefont {Hambye},\ and\ \citenamefont {Tytgat}}]{Andreas08}%
  \BibitemOpen
  \bibfield  {author} {\bibinfo {author} {\bibfnamefont {S.}~\bibnamefont
  {Andreas}}, \bibinfo {author} {\bibfnamefont {T.}~\bibnamefont {Hambye}}, \
  and\ \bibinfo {author} {\bibfnamefont {M.~H.~G.}\ \bibnamefont {Tytgat}},\
  }\href {\doibase 10.1088/1475-7516/2008/10/034} {\bibfield  {journal}
  {\bibinfo  {journal} {J. Cosmol. Astropart. Phys.}\ }\textbf {\bibinfo
  {volume} {2008}},\ \bibinfo {pages} {034} (\bibinfo {year}
  {2008})}\BibitemShut {NoStop}%
\bibitem [{\citenamefont {{Kouvaris}}(2008)}]{Kouvaris08}%
  \BibitemOpen
  \bibfield  {author} {\bibinfo {author} {\bibfnamefont {C.}~\bibnamefont
  {{Kouvaris}}},\ }\href {\doibase 10.1103/PhysRevD.77.023006} {\bibfield
  {journal} {\bibinfo  {journal} {\prd}\ }\textbf {\bibinfo {volume} {77}},\
  \bibinfo {eid} {023006} (\bibinfo {year} {2008})}\BibitemShut {NoStop}%
\bibitem [{\citenamefont {{Kouvaris}}(2012)}]{Kouvaris12}%
  \BibitemOpen
  \bibfield  {author} {\bibinfo {author} {\bibfnamefont {C.}~\bibnamefont
  {{Kouvaris}}},\ }\href {\doibase 10.1103/PhysRevLett.108.191301} {\bibfield
  {journal} {\bibinfo  {journal} {\prl}\ }\textbf {\bibinfo {volume} {108}},\
  \bibinfo {eid} {191301} (\bibinfo {year} {2012})}\BibitemShut {NoStop}%
\bibitem [{\citenamefont {{Bhat}}\ and\ \citenamefont {{Paul}}(2020)}]{Bhat20}%
  \BibitemOpen
  \bibfield  {author} {\bibinfo {author} {\bibfnamefont {S.~A.}\ \bibnamefont
  {{Bhat}}}\ and\ \bibinfo {author} {\bibfnamefont {A.}~\bibnamefont
  {{Paul}}},\ }\href {\doibase 10.1140/epjc/s10052-020-8072-x} {\bibfield
  {journal} {\bibinfo  {journal} {Eur. Phys. J. C}\ }\textbf {\bibinfo {volume}
  {80}},\ \bibinfo {eid} {544} (\bibinfo {year} {2020})}\BibitemShut {NoStop}%
\bibitem [{\citenamefont {Martin}(1998)}]{Martin98}%
  \BibitemOpen
  \bibfield  {author} {\bibinfo {author} {\bibfnamefont {S.~P.}\ \bibnamefont
  {Martin}},\ }\href {\doibase 10.1142/9789812839657_0001} {\bibfield
  {journal} {\bibinfo  {journal} {Adv. Ser. Direct. High Energy Phys.}\
  }\textbf {\bibinfo {volume} {18}},\ \bibinfo {pages} {1} (\bibinfo {year}
  {1998})}\BibitemShut {NoStop}%
\bibitem [{\citenamefont {Hooper}\ and\ \citenamefont {Wang}(2004)}]{Hooper04}%
  \BibitemOpen
  \bibfield  {author} {\bibinfo {author} {\bibfnamefont {D.}~\bibnamefont
  {Hooper}}\ and\ \bibinfo {author} {\bibfnamefont {L.-T.}\ \bibnamefont
  {Wang}},\ }\href {\doibase 10.1103/PhysRevD.69.035001} {\bibfield  {journal}
  {\bibinfo  {journal} {Phys. Rev. D}\ }\textbf {\bibinfo {volume} {69}},\
  \bibinfo {pages} {035001} (\bibinfo {year} {2004})}\BibitemShut {NoStop}%
\bibitem [{\citenamefont {Panotopoulos}\ and\ \citenamefont
  {Lopes}(2017)}]{Panotopoulos17}%
  \BibitemOpen
  \bibfield  {author} {\bibinfo {author} {\bibfnamefont {G.}~\bibnamefont
  {Panotopoulos}}\ and\ \bibinfo {author} {\bibfnamefont {I.}~\bibnamefont
  {Lopes}},\ }\href {\doibase 10.1103/PhysRevD.96.083004} {\bibfield  {journal}
  {\bibinfo  {journal} {Phys. Rev. D}\ }\textbf {\bibinfo {volume} {96}},\
  \bibinfo {pages} {083004} (\bibinfo {year} {2017})}\BibitemShut {NoStop}%
\bibitem [{\citenamefont {Das}\ \emph {et~al.}(2019)\citenamefont {Das},
  \citenamefont {Malik},\ and\ \citenamefont {Nayak}}]{ADas19}%
  \BibitemOpen
  \bibfield  {author} {\bibinfo {author} {\bibfnamefont {A.}~\bibnamefont
  {Das}}, \bibinfo {author} {\bibfnamefont {T.}~\bibnamefont {Malik}}, \ and\
  \bibinfo {author} {\bibfnamefont {A.~C.}\ \bibnamefont {Nayak}},\ }\href
  {\doibase 10.1103/PhysRevD.99.043016} {\bibfield  {journal} {\bibinfo
  {journal} {Phys. Rev. D}\ }\textbf {\bibinfo {volume} {99}},\ \bibinfo
  {pages} {043016} (\bibinfo {year} {2019})}\BibitemShut {NoStop}%
\bibitem [{\citenamefont {Das}\ \emph {et~al.}(2020)\citenamefont {Das},
  \citenamefont {Kumar}, \citenamefont {Kumar}, \citenamefont {Biswal},
  \citenamefont {Nakatsukasa}, \citenamefont {Li},\ and\ \citenamefont
  {Patra}}]{Das20}%
  \BibitemOpen
  \bibfield  {author} {\bibinfo {author} {\bibfnamefont {H.~C.}\ \bibnamefont
  {Das}}, \bibinfo {author} {\bibfnamefont {A.}~\bibnamefont {Kumar}}, \bibinfo
  {author} {\bibfnamefont {B.}~\bibnamefont {Kumar}}, \bibinfo {author}
  {\bibfnamefont {S.~K.}\ \bibnamefont {Biswal}}, \bibinfo {author}
  {\bibfnamefont {T.}~\bibnamefont {Nakatsukasa}}, \bibinfo {author}
  {\bibfnamefont {A.}~\bibnamefont {Li}}, \ and\ \bibinfo {author}
  {\bibfnamefont {S.~K.}\ \bibnamefont {Patra}},\ }\href {\doibase
  10.1093/mnras/staa1435} {\bibfield  {journal} {\bibinfo  {journal} {Mon. Not.
  R. Astron. Soc.}\ }\textbf {\bibinfo {volume} {495}},\ \bibinfo {pages}
  {4893} (\bibinfo {year} {2020})}\BibitemShut {NoStop}%
\bibitem [{\citenamefont {Das}\ \emph {et~al.}(2021)\citenamefont {Das},
  \citenamefont {Kumar},\ and\ \citenamefont {Patra}}]{Das21b}%
  \BibitemOpen
  \bibfield  {author} {\bibinfo {author} {\bibfnamefont {H.~C.}\ \bibnamefont
  {Das}}, \bibinfo {author} {\bibfnamefont {A.}~\bibnamefont {Kumar}}, \ and\
  \bibinfo {author} {\bibfnamefont {S.~K.}\ \bibnamefont {Patra}},\ }\href
  {\doibase 10.1103/PhysRevD.104.063028} {\bibfield  {journal} {\bibinfo
  {journal} {Phys. Rev. D}\ }\textbf {\bibinfo {volume} {104}},\ \bibinfo
  {pages} {063028} (\bibinfo {year} {2021})}\BibitemShut {NoStop}%
\bibitem [{\citenamefont {{Das}}\ \emph {et~al.}(2021)\citenamefont {{Das}},
  \citenamefont {{Kumar}},\ and\ \citenamefont {{Patra}}}]{Das21c}%
  \BibitemOpen
  \bibfield  {author} {\bibinfo {author} {\bibfnamefont {H.~C.}\ \bibnamefont
  {{Das}}}, \bibinfo {author} {\bibfnamefont {A.}~\bibnamefont {{Kumar}}}, \
  and\ \bibinfo {author} {\bibfnamefont {S.~K.}\ \bibnamefont {{Patra}}},\
  }\href {\doibase 10.1093/mnras/stab2387} {\bibfield  {journal} {\bibinfo
  {journal} {Mon. Not. R. Astron. Soc.}\ }\textbf {\bibinfo {volume} {507}},\
  \bibinfo {pages} {4053} (\bibinfo {year} {2021})}\BibitemShut {NoStop}%
\bibitem [{\citenamefont {{Das}}\ \emph {et~al.}(2022)\citenamefont {{Das}},
  \citenamefont {{Kumar}}, \citenamefont {{Kumar}},\ and\ \citenamefont
  {{Patra}}}]{Das22}%
  \BibitemOpen
  \bibfield  {author} {\bibinfo {author} {\bibfnamefont {H.~C.}\ \bibnamefont
  {{Das}}}, \bibinfo {author} {\bibfnamefont {A.}~\bibnamefont {{Kumar}}},
  \bibinfo {author} {\bibfnamefont {B.}~\bibnamefont {{Kumar}}}, \ and\
  \bibinfo {author} {\bibfnamefont {S.~K.}\ \bibnamefont {{Patra}}},\ }\href
  {\doibase 10.3390/galaxies10010014} {\bibfield  {journal} {\bibinfo
  {journal} {Galaxies}\ }\textbf {\bibinfo {volume} {10}},\ \bibinfo {pages}
  {14} (\bibinfo {year} {2022})}\BibitemShut {NoStop}%
\bibitem [{\citenamefont {{Kumar}}\ \emph {et~al.}(2022)\citenamefont
  {{Kumar}}, \citenamefont {{Das}},\ and\ \citenamefont {{Patra}}}]{Kumar22}%
  \BibitemOpen
  \bibfield  {author} {\bibinfo {author} {\bibfnamefont {A.}~\bibnamefont
  {{Kumar}}}, \bibinfo {author} {\bibfnamefont {H.~C.}\ \bibnamefont {{Das}}},
  \ and\ \bibinfo {author} {\bibfnamefont {S.~K.}\ \bibnamefont {{Patra}}},\
  }\href {\doibase 10.1093/mnras/stac1013} {\bibfield  {journal} {\bibinfo
  {journal} {Mon. Not. R. Astron. Soc.}\ }\textbf {\bibinfo {volume} {513}},\
  \bibinfo {pages} {1820} (\bibinfo {year} {2022})}\BibitemShut {NoStop}%
\bibitem [{\citenamefont {{Louren{\c{c}}o}}\ \emph {et~al.}(2022)\citenamefont
  {{Louren{\c{c}}o}}, \citenamefont {{Lenzi}}, \citenamefont {{Frederico}},\
  and\ \citenamefont {{Dutra}}}]{Lourenco22}%
  \BibitemOpen
  \bibfield  {author} {\bibinfo {author} {\bibfnamefont {O.}~\bibnamefont
  {{Louren{\c{c}}o}}}, \bibinfo {author} {\bibfnamefont {C.~H.}\ \bibnamefont
  {{Lenzi}}}, \bibinfo {author} {\bibfnamefont {T.}~\bibnamefont
  {{Frederico}}}, \ and\ \bibinfo {author} {\bibfnamefont {M.}~\bibnamefont
  {{Dutra}}},\ }\href {\doibase 10.1103/PhysRevD.106.043010} {\bibfield
  {journal} {\bibinfo  {journal} {\prd}\ }\textbf {\bibinfo {volume} {106}},\
  \bibinfo {eid} {043010} (\bibinfo {year} {2022})}\BibitemShut {NoStop}%
\bibitem [{\citenamefont {Kouvaris}\ and\ \citenamefont
  {Tinyakov}(2011)}]{Kouvaris11}%
  \BibitemOpen
  \bibfield  {author} {\bibinfo {author} {\bibfnamefont {C.}~\bibnamefont
  {Kouvaris}}\ and\ \bibinfo {author} {\bibfnamefont {P.}~\bibnamefont
  {Tinyakov}},\ }\href {\doibase 10.1103/PhysRevD.83.083512} {\bibfield
  {journal} {\bibinfo  {journal} {Phys. Rev. D}\ }\textbf {\bibinfo {volume}
  {83}},\ \bibinfo {pages} {083512} (\bibinfo {year} {2011})}\BibitemShut
  {NoStop}%
\bibitem [{\citenamefont {{McDermott}}\ \emph {et~al.}(2012)\citenamefont
  {{McDermott}}, \citenamefont {{Yu}},\ and\ \citenamefont
  {{Zurek}}}]{McDermott12}%
  \BibitemOpen
  \bibfield  {author} {\bibinfo {author} {\bibfnamefont {S.~D.}\ \bibnamefont
  {{McDermott}}}, \bibinfo {author} {\bibfnamefont {H.-B.}\ \bibnamefont
  {{Yu}}}, \ and\ \bibinfo {author} {\bibfnamefont {K.~M.}\ \bibnamefont
  {{Zurek}}},\ }\href {\doibase 10.1103/PhysRevD.85.023519} {\bibfield
  {journal} {\bibinfo  {journal} {\prd}\ }\textbf {\bibinfo {volume} {85}},\
  \bibinfo {eid} {023519} (\bibinfo {year} {2012})}\BibitemShut {NoStop}%
\bibitem [{\citenamefont {Gresham}\ and\ \citenamefont
  {Zurek}(2019)}]{Gresham19}%
  \BibitemOpen
  \bibfield  {author} {\bibinfo {author} {\bibfnamefont {M.~I.}\ \bibnamefont
  {Gresham}}\ and\ \bibinfo {author} {\bibfnamefont {K.~M.}\ \bibnamefont
  {Zurek}},\ }\href {\doibase 10.1103/PhysRevD.99.083008} {\bibfield  {journal}
  {\bibinfo  {journal} {Phys. Rev. D}\ }\textbf {\bibinfo {volume} {99}},\
  \bibinfo {pages} {083008} (\bibinfo {year} {2019})}\BibitemShut {NoStop}%
\bibitem [{\citenamefont {Ivanytskyi}\ \emph {et~al.}(2020)\citenamefont
  {Ivanytskyi}, \citenamefont {Sagun},\ and\ \citenamefont
  {Lopes}}]{Ivanytskyi20}%
  \BibitemOpen
  \bibfield  {author} {\bibinfo {author} {\bibfnamefont {O.}~\bibnamefont
  {Ivanytskyi}}, \bibinfo {author} {\bibfnamefont {V.}~\bibnamefont {Sagun}}, \
  and\ \bibinfo {author} {\bibfnamefont {I.}~\bibnamefont {Lopes}},\ }\href
  {\doibase 10.1103/PhysRevD.102.063028} {\bibfield  {journal} {\bibinfo
  {journal} {Phys. Rev. D}\ }\textbf {\bibinfo {volume} {102}},\ \bibinfo
  {pages} {063028} (\bibinfo {year} {2020})}\BibitemShut {NoStop}%
\bibitem [{\citenamefont {{Sandin}}\ and\ \citenamefont
  {{Ciarcelluti}}(2009)}]{Sandin09}%
  \BibitemOpen
  \bibfield  {author} {\bibinfo {author} {\bibfnamefont {F.}~\bibnamefont
  {{Sandin}}}\ and\ \bibinfo {author} {\bibfnamefont {P.}~\bibnamefont
  {{Ciarcelluti}}},\ }\href {\doibase 10.1016/j.astropartphys.2009.09.005}
  {\bibfield  {journal} {\bibinfo  {journal} {Astropart. Phys.}\ }\textbf
  {\bibinfo {volume} {32}},\ \bibinfo {pages} {278} (\bibinfo {year}
  {2009})}\BibitemShut {NoStop}%
\bibitem [{\citenamefont {{Hippert}}\ \emph {et~al.}(2022)\citenamefont
  {{Hippert}}, \citenamefont {{Setford}}, \citenamefont {{Tan}}, \citenamefont
  {{Curtin}}, \citenamefont {{Noronha-Hostler}},\ and\ \citenamefont
  {{Yunes}}}]{Hippert22}%
  \BibitemOpen
  \bibfield  {author} {\bibinfo {author} {\bibfnamefont {M.}~\bibnamefont
  {{Hippert}}}, \bibinfo {author} {\bibfnamefont {J.}~\bibnamefont
  {{Setford}}}, \bibinfo {author} {\bibfnamefont {H.}~\bibnamefont {{Tan}}},
  \bibinfo {author} {\bibfnamefont {D.}~\bibnamefont {{Curtin}}}, \bibinfo
  {author} {\bibfnamefont {J.}~\bibnamefont {{Noronha-Hostler}}}, \ and\
  \bibinfo {author} {\bibfnamefont {N.}~\bibnamefont {{Yunes}}},\ }\href
  {\doibase 10.1103/PhysRevD.106.035025} {\bibfield  {journal} {\bibinfo
  {journal} {\prd}\ }\textbf {\bibinfo {volume} {106}},\ \bibinfo {eid}
  {035025} (\bibinfo {year} {2022})}\BibitemShut {NoStop}%
\bibitem [{\citenamefont {Duffy}\ and\ \citenamefont {van
  Bibber}(2009)}]{Duffy09}%
  \BibitemOpen
  \bibfield  {author} {\bibinfo {author} {\bibfnamefont {L.~D.}\ \bibnamefont
  {Duffy}}\ and\ \bibinfo {author} {\bibfnamefont {K.}~\bibnamefont {van
  Bibber}},\ }\href {\doibase 10.1088/1367-2630/11/10/105008} {\bibfield
  {journal} {\bibinfo  {journal} {New. J. Phys}\ }\textbf {\bibinfo {volume}
  {11}},\ \bibinfo {pages} {105008} (\bibinfo {year} {2009})}\BibitemShut
  {NoStop}%
\bibitem [{\citenamefont {{Balatsky}}\ \emph {et~al.}(2022)\citenamefont
  {{Balatsky}}, \citenamefont {{Fraser}},\ and\ \citenamefont
  {{R{\o}ising}}}]{Balatsky22}%
  \BibitemOpen
  \bibfield  {author} {\bibinfo {author} {\bibfnamefont {A.~V.}\ \bibnamefont
  {{Balatsky}}}, \bibinfo {author} {\bibfnamefont {B.}~\bibnamefont
  {{Fraser}}}, \ and\ \bibinfo {author} {\bibfnamefont {H.~S.}\ \bibnamefont
  {{R{\o}ising}}},\ }\href {\doibase 10.1103/PhysRevD.105.023504} {\bibfield
  {journal} {\bibinfo  {journal} {\prd}\ }\textbf {\bibinfo {volume} {105}},\
  \bibinfo {eid} {023504} (\bibinfo {year} {2022})}\BibitemShut {NoStop}%
\bibitem [{\citenamefont {{Jacobs}}\ \emph {et~al.}(2015)\citenamefont
  {{Jacobs}}, \citenamefont {{Starkman}},\ and\ \citenamefont
  {{Lynn}}}]{Jacobs15}%
  \BibitemOpen
  \bibfield  {author} {\bibinfo {author} {\bibfnamefont {D.~M.}\ \bibnamefont
  {{Jacobs}}}, \bibinfo {author} {\bibfnamefont {G.~D.}\ \bibnamefont
  {{Starkman}}}, \ and\ \bibinfo {author} {\bibfnamefont {B.~W.}\ \bibnamefont
  {{Lynn}}},\ }\href {\doibase 10.1093/mnras/stv774} {\bibfield  {journal}
  {\bibinfo  {journal} {Mon. Not. R. Astron. Soc.}\ }\textbf {\bibinfo {volume}
  {450}},\ \bibinfo {pages} {3418} (\bibinfo {year} {2015})}\BibitemShut
  {NoStop}%
\bibitem [{\citenamefont {{Ge}}\ \emph {et~al.}(2019)\citenamefont {{Ge}},
  \citenamefont {{Lawson}},\ and\ \citenamefont {{Zhitnitsky}}}]{Ge19}%
  \BibitemOpen
  \bibfield  {author} {\bibinfo {author} {\bibfnamefont {S.}~\bibnamefont
  {{Ge}}}, \bibinfo {author} {\bibfnamefont {K.}~\bibnamefont {{Lawson}}}, \
  and\ \bibinfo {author} {\bibfnamefont {A.}~\bibnamefont {{Zhitnitsky}}},\
  }\href {\doibase 10.1103/PhysRevD.99.116017} {\bibfield  {journal} {\bibinfo
  {journal} {\prd}\ }\textbf {\bibinfo {volume} {99}},\ \bibinfo {eid} {116017}
  (\bibinfo {year} {2019})}\BibitemShut {NoStop}%
\bibitem [{\citenamefont {{VanDevender}}\ \emph {et~al.}(2021)\citenamefont
  {{VanDevender}}, \citenamefont {{Schmitt}}, \citenamefont {{McGinley}},
  \citenamefont {{Duggan}}, \citenamefont {{McGinty}}, \citenamefont
  {{VanDevender}}, \citenamefont {{Wilson}}, \citenamefont {{Dixon}},
  \citenamefont {{Girard}},\ and\ \citenamefont {{McRae}}}]{Vandevender21}%
  \BibitemOpen
  \bibfield  {author} {\bibinfo {author} {\bibfnamefont {J.~P.}\ \bibnamefont
  {{VanDevender}}}, \bibinfo {author} {\bibfnamefont {R.~G.}\ \bibnamefont
  {{Schmitt}}}, \bibinfo {author} {\bibfnamefont {N.}~\bibnamefont
  {{McGinley}}}, \bibinfo {author} {\bibfnamefont {D.~G.}\ \bibnamefont
  {{Duggan}}}, \bibinfo {author} {\bibfnamefont {S.}~\bibnamefont {{McGinty}}},
  \bibinfo {author} {\bibfnamefont {A.~P.}\ \bibnamefont {{VanDevender}}},
  \bibinfo {author} {\bibfnamefont {P.}~\bibnamefont {{Wilson}}}, \bibinfo
  {author} {\bibfnamefont {D.}~\bibnamefont {{Dixon}}}, \bibinfo {author}
  {\bibfnamefont {H.}~\bibnamefont {{Girard}}}, \ and\ \bibinfo {author}
  {\bibfnamefont {J.}~\bibnamefont {{McRae}}},\ }\href {\doibase
  10.3390/universe7050116} {\bibfield  {journal} {\bibinfo  {journal}
  {Universe}\ }\textbf {\bibinfo {volume} {7}},\ \bibinfo {pages} {116}
  (\bibinfo {year} {2021})}\BibitemShut {NoStop}%
\bibitem [{\citenamefont {{Zhitnitsky}}(2021)}]{Zhitnitsky21}%
  \BibitemOpen
  \bibfield  {author} {\bibinfo {author} {\bibfnamefont {A.}~\bibnamefont
  {{Zhitnitsky}}},\ }\href {\doibase 10.1142/S0217732321300172} {\bibfield
  {journal} {\bibinfo  {journal} {Mod. Phys. Lett. A}\ }\textbf {\bibinfo
  {volume} {36}},\ \bibinfo {eid} {2130017} (\bibinfo {year}
  {2021})}\BibitemShut {NoStop}%
\bibitem [{\citenamefont {Bernabei}\ \emph {et~al.}(2008)\citenamefont
  {Bernabei} \emph {et~al.}}]{Bernabei08}%
  \BibitemOpen
  \bibfield  {author} {\bibinfo {author} {\bibfnamefont {R.}~\bibnamefont
  {Bernabei}} \emph {et~al.},\ }\href {\doibase 10.1140/epjc/s10052-008-0662-y}
  {\bibfield  {journal} {\bibinfo  {journal} {Eur. Phys. J. C}\ }\textbf
  {\bibinfo {volume} {56}},\ \bibinfo {pages} {333} (\bibinfo {year}
  {2008})}\BibitemShut {NoStop}%
\bibitem [{\citenamefont {Bernabei}\ \emph {et~al.}(2010)\citenamefont
  {Bernabei} \emph {et~al.}}]{Bernabei10}%
  \BibitemOpen
  \bibfield  {author} {\bibinfo {author} {\bibfnamefont {R.}~\bibnamefont
  {Bernabei}} \emph {et~al.},\ }\href {\doibase 10.1140/epjc/s10052-010-1303-9}
  {\bibfield  {journal} {\bibinfo  {journal} {Eur. Phys. J. C}\ }\textbf
  {\bibinfo {volume} {67}},\ \bibinfo {pages} {30} (\bibinfo {year}
  {2010})}\BibitemShut {NoStop}%
\bibitem [{\citenamefont {Bravin}\ \emph {et~al.}(1999)\citenamefont {Bravin}
  \emph {et~al.}}]{Bravin99}%
  \BibitemOpen
  \bibfield  {author} {\bibinfo {author} {\bibfnamefont {M.}~\bibnamefont
  {Bravin}} \emph {et~al.},\ }\href {\doibase
  https://doi.org/10.1016/S0927-6505(99)00073-0} {\bibfield  {journal}
  {\bibinfo  {journal} {Astropart. Phys.}\ }\textbf {\bibinfo {volume} {12}},\
  \bibinfo {pages} {107} (\bibinfo {year} {1999})}\BibitemShut {NoStop}%
\bibitem [{\citenamefont {Aprile}\ \emph {et~al.}(2012)\citenamefont {Aprile}
  \emph {et~al.}}]{Aprile12}%
  \BibitemOpen
  \bibfield  {author} {\bibinfo {author} {\bibfnamefont {E.}~\bibnamefont
  {Aprile}} \emph {et~al.} (\bibinfo {collaboration} {XENON100
  Collaboration}),\ }\href {\doibase 10.1103/PhysRevLett.109.181301} {\bibfield
   {journal} {\bibinfo  {journal} {Phys. Rev. Lett.}\ }\textbf {\bibinfo
  {volume} {109}},\ \bibinfo {pages} {181301} (\bibinfo {year}
  {2012})}\BibitemShut {NoStop}%
\bibitem [{\citenamefont {Aprile}\ \emph {et~al.}(2018)\citenamefont {Aprile}
  \emph {et~al.}}]{Aprile18}%
  \BibitemOpen
  \bibfield  {author} {\bibinfo {author} {\bibfnamefont {E.}~\bibnamefont
  {Aprile}} \emph {et~al.} (\bibinfo {collaboration} {XENON Collaboration 7}),\
  }\href {\doibase 10.1103/PhysRevLett.121.111302} {\bibfield  {journal}
  {\bibinfo  {journal} {Phys. Rev. Lett.}\ }\textbf {\bibinfo {volume} {121}},\
  \bibinfo {pages} {111302} (\bibinfo {year} {2018})}\BibitemShut {NoStop}%
\bibitem [{\citenamefont {Le~Delliou}\ \emph {et~al.}(2015)\citenamefont
  {Le~Delliou}, \citenamefont {Marcondes}, \citenamefont {Lima~Neto},\ and\
  \citenamefont {Abdalla}}]{LeDelliou15}%
  \BibitemOpen
  \bibfield  {author} {\bibinfo {author} {\bibfnamefont {M.}~\bibnamefont
  {Le~Delliou}}, \bibinfo {author} {\bibfnamefont {R.~J.~F.}\ \bibnamefont
  {Marcondes}}, \bibinfo {author} {\bibfnamefont {G.~B.}\ \bibnamefont
  {Lima~Neto}}, \ and\ \bibinfo {author} {\bibfnamefont {E.}~\bibnamefont
  {Abdalla}},\ }\href {\doibase 10.1093/mnras/stv1561} {\bibfield  {journal}
  {\bibinfo  {journal} {Mon. Not. R. Astron. Soc.}\ }\textbf {\bibinfo {volume}
  {453}},\ \bibinfo {pages} {2} (\bibinfo {year} {2015})}\BibitemShut {NoStop}%
\bibitem [{\citenamefont {Sato}\ and\ \citenamefont {Tobioka}(2016)}]{Sato16}%
  \BibitemOpen
  \bibfield  {author} {\bibinfo {author} {\bibfnamefont {R.}~\bibnamefont
  {Sato}}\ and\ \bibinfo {author} {\bibfnamefont {K.}~\bibnamefont {Tobioka}},\
  }\href {\doibase 10.1016/j.physletb.2016.07.051} {\bibfield  {journal}
  {\bibinfo  {journal} {Phys. Lett. B}\ }\textbf {\bibinfo {volume} {760}},\
  \bibinfo {pages} {590} (\bibinfo {year} {2016})}\BibitemShut {NoStop}%
\bibitem [{\citenamefont {de~Lavallaz}\ and\ \citenamefont
  {Fairbairn}(2010)}]{Lavallaz10}%
  \BibitemOpen
  \bibfield  {author} {\bibinfo {author} {\bibfnamefont {A.}~\bibnamefont
  {de~Lavallaz}}\ and\ \bibinfo {author} {\bibfnamefont {M.}~\bibnamefont
  {Fairbairn}},\ }\href {\doibase 10.1103/PhysRevD.81.123521} {\bibfield
  {journal} {\bibinfo  {journal} {Phys. Rev. D}\ }\textbf {\bibinfo {volume}
  {81}},\ \bibinfo {pages} {123521} (\bibinfo {year} {2010})}\BibitemShut
  {NoStop}%
\bibitem [{\citenamefont {{Lopes}}\ \emph {et~al.}(2011)\citenamefont
  {{Lopes}}, \citenamefont {{Casanellas}},\ and\ \citenamefont
  {{Eug{\'e}nio}}}]{Lopes11}%
  \BibitemOpen
  \bibfield  {author} {\bibinfo {author} {\bibfnamefont {I.}~\bibnamefont
  {{Lopes}}}, \bibinfo {author} {\bibfnamefont {J.}~\bibnamefont
  {{Casanellas}}}, \ and\ \bibinfo {author} {\bibfnamefont {D.}~\bibnamefont
  {{Eug{\'e}nio}}},\ }\href {\doibase 10.1103/PhysRevD.83.063521} {\bibfield
  {journal} {\bibinfo  {journal} {\prd}\ }\textbf {\bibinfo {volume} {83}},\
  \bibinfo {eid} {063521} (\bibinfo {year} {2011})}\BibitemShut {NoStop}%
\bibitem [{\citenamefont {{Bramante}}\ \emph {et~al.}(2013)\citenamefont
  {{Bramante}}, \citenamefont {{Fukushima}},\ and\ \citenamefont
  {{Kumar}}}]{Bramante13}%
  \BibitemOpen
  \bibfield  {author} {\bibinfo {author} {\bibfnamefont {J.}~\bibnamefont
  {{Bramante}}}, \bibinfo {author} {\bibfnamefont {K.}~\bibnamefont
  {{Fukushima}}}, \ and\ \bibinfo {author} {\bibfnamefont {J.}~\bibnamefont
  {{Kumar}}},\ }\href {\doibase 10.1103/PhysRevD.87.055012} {\bibfield
  {journal} {\bibinfo  {journal} {\prd}\ }\textbf {\bibinfo {volume} {87}},\
  \bibinfo {eid} {055012} (\bibinfo {year} {2013})}\BibitemShut {NoStop}%
\bibitem [{\citenamefont {{Bertoni}}\ \emph {et~al.}(2013)\citenamefont
  {{Bertoni}}, \citenamefont {{Nelson}},\ and\ \citenamefont
  {{Reddy}}}]{Bertoni13}%
  \BibitemOpen
  \bibfield  {author} {\bibinfo {author} {\bibfnamefont {B.}~\bibnamefont
  {{Bertoni}}}, \bibinfo {author} {\bibfnamefont {A.~E.}\ \bibnamefont
  {{Nelson}}}, \ and\ \bibinfo {author} {\bibfnamefont {S.}~\bibnamefont
  {{Reddy}}},\ }\href {\doibase 10.1103/PhysRevD.88.123505} {\bibfield
  {journal} {\bibinfo  {journal} {\prd}\ }\textbf {\bibinfo {volume} {88}},\
  \bibinfo {eid} {123505} (\bibinfo {year} {2013})}\BibitemShut {NoStop}%
\bibitem [{\citenamefont {{Baryakhtar}}\ \emph {et~al.}(2017)\citenamefont
  {{Baryakhtar}}, \citenamefont {{Bramante}}, \citenamefont {{Li}},
  \citenamefont {{Linden}},\ and\ \citenamefont {{Raj}}}]{Baryakhtar17}%
  \BibitemOpen
  \bibfield  {author} {\bibinfo {author} {\bibfnamefont {M.}~\bibnamefont
  {{Baryakhtar}}}, \bibinfo {author} {\bibfnamefont {J.}~\bibnamefont
  {{Bramante}}}, \bibinfo {author} {\bibfnamefont {S.~W.}\ \bibnamefont
  {{Li}}}, \bibinfo {author} {\bibfnamefont {T.}~\bibnamefont {{Linden}}}, \
  and\ \bibinfo {author} {\bibfnamefont {N.}~\bibnamefont {{Raj}}},\ }\href
  {\doibase 10.1103/PhysRevLett.119.131801} {\bibfield  {journal} {\bibinfo
  {journal} {\prl}\ }\textbf {\bibinfo {volume} {119}},\ \bibinfo {eid}
  {131801} (\bibinfo {year} {2017})}\BibitemShut {NoStop}%
\bibitem [{\citenamefont {{Bell}}\ \emph {et~al.}(2021)\citenamefont {{Bell}},
  \citenamefont {{Busoni}}, \citenamefont {{Motta}}, \citenamefont {{Robles}},
  \citenamefont {{Thomas}},\ and\ \citenamefont {{Virgato}}}]{Bell21}%
  \BibitemOpen
  \bibfield  {author} {\bibinfo {author} {\bibfnamefont {N.~F.}\ \bibnamefont
  {{Bell}}}, \bibinfo {author} {\bibfnamefont {G.}~\bibnamefont {{Busoni}}},
  \bibinfo {author} {\bibfnamefont {T.~F.}\ \bibnamefont {{Motta}}}, \bibinfo
  {author} {\bibfnamefont {S.}~\bibnamefont {{Robles}}}, \bibinfo {author}
  {\bibfnamefont {A.~W.}\ \bibnamefont {{Thomas}}}, \ and\ \bibinfo {author}
  {\bibfnamefont {M.}~\bibnamefont {{Virgato}}},\ }\href {\doibase
  10.1103/PhysRevLett.127.111803} {\bibfield  {journal} {\bibinfo  {journal}
  {\prl}\ }\textbf {\bibinfo {volume} {127}},\ \bibinfo {eid} {111803}
  (\bibinfo {year} {2021})}\BibitemShut {NoStop}%
\bibitem [{\citenamefont {{Anzuini}}\ \emph {et~al.}(2021)\citenamefont
  {{Anzuini}}, \citenamefont {{Bell}}, \citenamefont {{Busoni}}, \citenamefont
  {{Motta}}, \citenamefont {{Robles}}, \citenamefont {{Thomas}},\ and\
  \citenamefont {{Virgato}}}]{Anzuini21}%
  \BibitemOpen
  \bibfield  {author} {\bibinfo {author} {\bibfnamefont {F.}~\bibnamefont
  {{Anzuini}}}, \bibinfo {author} {\bibfnamefont {N.~F.}\ \bibnamefont
  {{Bell}}}, \bibinfo {author} {\bibfnamefont {G.}~\bibnamefont {{Busoni}}},
  \bibinfo {author} {\bibfnamefont {T.~F.}\ \bibnamefont {{Motta}}}, \bibinfo
  {author} {\bibfnamefont {S.}~\bibnamefont {{Robles}}}, \bibinfo {author}
  {\bibfnamefont {A.~W.}\ \bibnamefont {{Thomas}}}, \ and\ \bibinfo {author}
  {\bibfnamefont {M.}~\bibnamefont {{Virgato}}},\ }\href {\doibase
  10.1088/1475-7516/2021/11/056} {\bibfield  {journal} {\bibinfo  {journal} {J.
  Cosmol. Astropart. Phys.}\ }\textbf {\bibinfo {volume} {2021}},\ \bibinfo
  {eid} {056} (\bibinfo {year} {2021})}\BibitemShut {NoStop}%
\bibitem [{\citenamefont {{Gonzalez}}\ and\ \citenamefont
  {{Reisenegger}}(2010)}]{Gonzales10}%
  \BibitemOpen
  \bibfield  {author} {\bibinfo {author} {\bibfnamefont {D.}~\bibnamefont
  {{Gonzalez}}}\ and\ \bibinfo {author} {\bibfnamefont {A.}~\bibnamefont
  {{Reisenegger}}},\ }\href {\doibase 10.1051/0004-6361/201015084} {\bibfield
  {journal} {\bibinfo  {journal} {\aap}\ }\textbf {\bibinfo {volume} {522}},\
  \bibinfo {eid} {A16} (\bibinfo {year} {2010})}\BibitemShut {NoStop}%
\bibitem [{\citenamefont {M.$\acute{\rm{A}}$ngeles}\ \emph
  {et~al.}(2012)\citenamefont {M.$\acute{\rm{A}}$ngeles}, \citenamefont
  {P$\acute{\rm{e}}$rez-Garc$\acute{\rm{I}}$a},\ and\ \citenamefont
  {Silk}}]{Angeles12}%
  \BibitemOpen
  \bibfield  {author} {\bibinfo {author} {\bibnamefont
  {M.$\acute{\rm{A}}$ngeles}}, \bibinfo {author} {\bibnamefont
  {P$\acute{\rm{e}}$rez-Garc$\acute{\rm{I}}$a}}, \ and\ \bibinfo {author}
  {\bibfnamefont {J.}~\bibnamefont {Silk}},\ }\href {\doibase
  https://doi.org/10.1016/j.physletb.2012.03.065} {\bibfield  {journal}
  {\bibinfo  {journal} {Phys. Lett. B}\ }\textbf {\bibinfo {volume} {711}},\
  \bibinfo {pages} {6} (\bibinfo {year} {2012})}\BibitemShut {NoStop}%
\bibitem [{\citenamefont {{Herrero}}\ \emph {et~al.}(2019)\citenamefont
  {{Herrero}}, \citenamefont {{P{\'e}rez-Garc{\'\i}a}}, \citenamefont
  {{Silk}},\ and\ \citenamefont {{Albertus}}}]{Herrero19}%
  \BibitemOpen
  \bibfield  {author} {\bibinfo {author} {\bibfnamefont {A.}~\bibnamefont
  {{Herrero}}}, \bibinfo {author} {\bibfnamefont {M.~A.}\ \bibnamefont
  {{P{\'e}rez-Garc{\'\i}a}}}, \bibinfo {author} {\bibfnamefont
  {J.}~\bibnamefont {{Silk}}}, \ and\ \bibinfo {author} {\bibfnamefont
  {C.}~\bibnamefont {{Albertus}}},\ }\href {\doibase
  10.1103/PhysRevD.100.103019} {\bibfield  {journal} {\bibinfo  {journal}
  {\prd}\ }\textbf {\bibinfo {volume} {100}},\ \bibinfo {eid} {103019}
  (\bibinfo {year} {2019})}\BibitemShut {NoStop}%
\bibitem [{\citenamefont {{Garani}}\ \emph {et~al.}(2021)\citenamefont
  {{Garani}}, \citenamefont {{Gupta}},\ and\ \citenamefont {{Raj}}}]{Garani21}%
  \BibitemOpen
  \bibfield  {author} {\bibinfo {author} {\bibfnamefont {R.}~\bibnamefont
  {{Garani}}}, \bibinfo {author} {\bibfnamefont {A.}~\bibnamefont {{Gupta}}}, \
  and\ \bibinfo {author} {\bibfnamefont {N.}~\bibnamefont {{Raj}}},\ }\href
  {\doibase 10.1103/PhysRevD.103.043019} {\bibfield  {journal} {\bibinfo
  {journal} {\prd}\ }\textbf {\bibinfo {volume} {103}},\ \bibinfo {eid}
  {043019} (\bibinfo {year} {2021})}\BibitemShut {NoStop}%
\bibitem [{\citenamefont {{Bose}}\ \emph {et~al.}(2022)\citenamefont {{Bose}},
  \citenamefont {{Maity}},\ and\ \citenamefont {{Ray}}}]{Bose22}%
  \BibitemOpen
  \bibfield  {author} {\bibinfo {author} {\bibfnamefont {D.}~\bibnamefont
  {{Bose}}}, \bibinfo {author} {\bibfnamefont {T.~N.}\ \bibnamefont {{Maity}}},
  \ and\ \bibinfo {author} {\bibfnamefont {T.~S.}\ \bibnamefont {{Ray}}},\
  }\href {\doibase 10.1088/1475-7516/2022/05/001} {\bibfield  {journal}
  {\bibinfo  {journal} {J. Cosmol. Astropart. Phys.}\ }\textbf {\bibinfo
  {volume} {2022}},\ \bibinfo {eid} {001} (\bibinfo {year} {2022})}\BibitemShut
  {NoStop}%
\bibitem [{\citenamefont {{Fujiwara}}\ \emph {et~al.}(2022)\citenamefont
  {{Fujiwara}}, \citenamefont {{Hamaguchi}}, \citenamefont {{Nagata}},\ and\
  \citenamefont {{Zheng}}}]{Fujiwara22}%
  \BibitemOpen
  \bibfield  {author} {\bibinfo {author} {\bibfnamefont {M.}~\bibnamefont
  {{Fujiwara}}}, \bibinfo {author} {\bibfnamefont {K.}~\bibnamefont
  {{Hamaguchi}}}, \bibinfo {author} {\bibfnamefont {N.}~\bibnamefont
  {{Nagata}}}, \ and\ \bibinfo {author} {\bibfnamefont {J.}~\bibnamefont
  {{Zheng}}},\ }\href {\doibase 10.1103/PhysRevD.106.055031} {\bibfield
  {journal} {\bibinfo  {journal} {\prd}\ }\textbf {\bibinfo {volume} {106}},\
  \bibinfo {eid} {055031} (\bibinfo {year} {2022})}\BibitemShut {NoStop}%
\bibitem [{\citenamefont {{Coffey}}\ \emph {et~al.}(2022)\citenamefont
  {{Coffey}}, \citenamefont {{McKeen}}, \citenamefont {{Morrissey}},\ and\
  \citenamefont {{Raj}}}]{Coffey22}%
  \BibitemOpen
  \bibfield  {author} {\bibinfo {author} {\bibfnamefont {J.}~\bibnamefont
  {{Coffey}}}, \bibinfo {author} {\bibfnamefont {D.}~\bibnamefont {{McKeen}}},
  \bibinfo {author} {\bibfnamefont {D.~E.}\ \bibnamefont {{Morrissey}}}, \ and\
  \bibinfo {author} {\bibfnamefont {N.}~\bibnamefont {{Raj}}},\ }\href
  {\doibase 10.1103/PhysRevD.106.115019} {\bibfield  {journal} {\bibinfo
  {journal} {\prd}\ }\textbf {\bibinfo {volume} {106}},\ \bibinfo {eid}
  {115019} (\bibinfo {year} {2022})}\BibitemShut {NoStop}%
\bibitem [{\citenamefont {Ellis}\ \emph {et~al.}(2018)\citenamefont {Ellis}
  \emph {et~al.}}]{Ellis18}%
  \BibitemOpen
  \bibfield  {author} {\bibinfo {author} {\bibfnamefont {J.}~\bibnamefont
  {Ellis}} \emph {et~al.},\ }\href {\doibase 10.1103/PhysRevD.97.123007}
  {\bibfield  {journal} {\bibinfo  {journal} {Phys. Rev. D}\ }\textbf {\bibinfo
  {volume} {97}},\ \bibinfo {pages} {123007} (\bibinfo {year}
  {2018})}\BibitemShut {NoStop}%
\bibitem [{\citenamefont {{Gleason}}\ \emph {et~al.}(2022)\citenamefont
  {{Gleason}}, \citenamefont {{Brown}},\ and\ \citenamefont
  {{Kain}}}]{Gleason22}%
  \BibitemOpen
  \bibfield  {author} {\bibinfo {author} {\bibfnamefont {T.}~\bibnamefont
  {{Gleason}}}, \bibinfo {author} {\bibfnamefont {B.}~\bibnamefont {{Brown}}},
  \ and\ \bibinfo {author} {\bibfnamefont {B.}~\bibnamefont {{Kain}}},\ }\href
  {\doibase 10.1103/PhysRevD.105.023010} {\bibfield  {journal} {\bibinfo
  {journal} {\prd}\ }\textbf {\bibinfo {volume} {105}},\ \bibinfo {eid}
  {023010} (\bibinfo {year} {2022})}\BibitemShut {NoStop}%
\bibitem [{\citenamefont {Arzoumanian}\ \emph {et~al.}(2018)\citenamefont
  {Arzoumanian} \emph {et~al.}}]{Arzoumanian18}%
  \BibitemOpen
  \bibfield  {author} {\bibinfo {author} {\bibfnamefont {Z.}~\bibnamefont
  {Arzoumanian}} \emph {et~al.},\ }\href {\doibase 10.3847/1538-4365/aab5b0}
  {\bibfield  {journal} {\bibinfo  {journal} {Astrophys. J., Suppl. Ser.}\
  }\textbf {\bibinfo {volume} {235}},\ \bibinfo {pages} {37} (\bibinfo {year}
  {2018})}\BibitemShut {NoStop}%
\bibitem [{\citenamefont {Antoniadis}\ \emph {et~al.}(2013)\citenamefont
  {Antoniadis} \emph {et~al.}}]{Antoniades13}%
  \BibitemOpen
  \bibfield  {author} {\bibinfo {author} {\bibfnamefont {J.}~\bibnamefont
  {Antoniadis}} \emph {et~al.},\ }\href {\doibase 10.1126/science.1233232}
  {\bibfield  {journal} {\bibinfo  {journal} {Science}\ }\textbf {\bibinfo
  {volume} {340}},\ \bibinfo {pages} {1233232} (\bibinfo {year}
  {2013})}\BibitemShut {NoStop}%
\bibitem [{\citenamefont {{\it et al.}}(2020)}]{Cromartie20}%
  \BibitemOpen
  \bibfield  {author} {\bibinfo {author} {\bibfnamefont {H.~T.~C.}\
  \bibnamefont {{\it et al.}}},\ }\href {\doibase 10.1038/s41550-019-0880-2}
  {\bibfield  {journal} {\bibinfo  {journal} {Nature Astronomy}\ }\textbf
  {\bibinfo {volume} {4}},\ \bibinfo {pages} {72} (\bibinfo {year}
  {2020})}\BibitemShut {NoStop}%
\bibitem [{\citenamefont {Riley}\ \emph {et~al.}(2019)\citenamefont {Riley}
  \emph {et~al.}}]{Riley19}%
  \BibitemOpen
  \bibfield  {author} {\bibinfo {author} {\bibfnamefont {T.~E.}\ \bibnamefont
  {Riley}} \emph {et~al.},\ }\href {\doibase 10.3847/2041-8213/ab481c}
  {\bibfield  {journal} {\bibinfo  {journal} {Astrophys. J.}\ }\textbf
  {\bibinfo {volume} {887}},\ \bibinfo {pages} {L21} (\bibinfo {year}
  {2019})}\BibitemShut {NoStop}%
\bibitem [{\citenamefont {Miller}\ \emph {et~al.}(2019)\citenamefont {Miller}
  \emph {et~al.}}]{Miller19}%
  \BibitemOpen
  \bibfield  {author} {\bibinfo {author} {\bibfnamefont {M.~C.}\ \bibnamefont
  {Miller}} \emph {et~al.},\ }\href {\doibase 10.3847/2041-8213/ab50c5}
  {\bibfield  {journal} {\bibinfo  {journal} {Astrophys. J.}\ }\textbf
  {\bibinfo {volume} {887}},\ \bibinfo {pages} {L24} (\bibinfo {year}
  {2019})}\BibitemShut {NoStop}%
\bibitem [{\citenamefont {Riley}\ \emph {et~al.}(2021)\citenamefont {Riley}
  \emph {et~al.}}]{Riley21}%
  \BibitemOpen
  \bibfield  {author} {\bibinfo {author} {\bibfnamefont {T.~E.}\ \bibnamefont
  {Riley}} \emph {et~al.},\ }\href {\doibase 10.3847/2041-8213/ac0a81}
  {\bibfield  {journal} {\bibinfo  {journal} {Astrophys. J. Lett.}\ }\textbf
  {\bibinfo {volume} {918}},\ \bibinfo {pages} {L27} (\bibinfo {year}
  {2021})}\BibitemShut {NoStop}%
\bibitem [{\citenamefont {Miller}\ \emph {et~al.}(2021)\citenamefont {Miller}
  \emph {et~al.}}]{Miller21}%
  \BibitemOpen
  \bibfield  {author} {\bibinfo {author} {\bibfnamefont {M.~C.}\ \bibnamefont
  {Miller}} \emph {et~al.},\ }\href {\doibase 10.3847/2041-8213/ac089b}
  {\bibfield  {journal} {\bibinfo  {journal} {Astrophys. J. Lett.}\ }\textbf
  {\bibinfo {volume} {918}},\ \bibinfo {pages} {L28} (\bibinfo {year}
  {2021})}\BibitemShut {NoStop}%
\bibitem [{\citenamefont {{Kodama}}\ and\ \citenamefont
  {{Yamada}}(1972)}]{Kodama72}%
  \BibitemOpen
  \bibfield  {author} {\bibinfo {author} {\bibfnamefont {T.}~\bibnamefont
  {{Kodama}}}\ and\ \bibinfo {author} {\bibfnamefont {M.}~\bibnamefont
  {{Yamada}}},\ }\href {\doibase 10.1143/PTP.47.444} {\bibfield  {journal}
  {\bibinfo  {journal} {Prog. Theor. Phys.}\ }\textbf {\bibinfo {volume}
  {47}},\ \bibinfo {pages} {444} (\bibinfo {year} {1972})}\BibitemShut
  {NoStop}%
\bibitem [{\citenamefont {{Comer}}\ \emph {et~al.}(1999)\citenamefont
  {{Comer}}, \citenamefont {{Langlois}},\ and\ \citenamefont
  {{Lin}}}]{Comer99}%
  \BibitemOpen
  \bibfield  {author} {\bibinfo {author} {\bibfnamefont {G.~L.}\ \bibnamefont
  {{Comer}}}, \bibinfo {author} {\bibfnamefont {D.}~\bibnamefont {{Langlois}}},
  \ and\ \bibinfo {author} {\bibfnamefont {L.~M.}\ \bibnamefont {{Lin}}},\
  }\href {\doibase 10.1103/PhysRevD.60.104025} {\bibfield  {journal} {\bibinfo
  {journal} {\prd}\ }\textbf {\bibinfo {volume} {60}},\ \bibinfo {eid} {104025}
  (\bibinfo {year} {1999})}\BibitemShut {NoStop}%
\bibitem [{\citenamefont {Quddus}\ \emph {et~al.}(2020)\citenamefont {Quddus}
  \emph {et~al.}}]{Quddus20}%
  \BibitemOpen
  \bibfield  {author} {\bibinfo {author} {\bibfnamefont {A.}~\bibnamefont
  {Quddus}} \emph {et~al.},\ }\href {\doibase 10.1088/1361-6471/ab9d36}
  {\bibfield  {journal} {\bibinfo  {journal} {J. Phys. G: Nucl. Part. Phys.}\
  }\textbf {\bibinfo {volume} {47}},\ \bibinfo {pages} {095202} (\bibinfo
  {year} {2020})}\BibitemShut {NoStop}%
\bibitem [{\citenamefont {Wang}\ and\ \citenamefont {Thomas}(2021)}]{Wang21}%
  \BibitemOpen
  \bibfield  {author} {\bibinfo {author} {\bibfnamefont {X.~G.}\ \bibnamefont
  {Wang}}\ and\ \bibinfo {author} {\bibfnamefont {A.~W.}\ \bibnamefont
  {Thomas}},\ }\href {\doibase 10.1103/PhysRevC.103.034606} {\bibfield
  {journal} {\bibinfo  {journal} {Phys. Rev. C}\ }\textbf {\bibinfo {volume}
  {103}},\ \bibinfo {pages} {034606} (\bibinfo {year} {2021})}\BibitemShut
  {NoStop}%
\bibitem [{\citenamefont {{Miao}}\ \emph {et~al.}(2022)\citenamefont {{Miao}},
  \citenamefont {{Zhu}}, \citenamefont {{Li}},\ and\ \citenamefont
  {{Huang}}}]{Miao22}%
  \BibitemOpen
  \bibfield  {author} {\bibinfo {author} {\bibfnamefont {Z.}~\bibnamefont
  {{Miao}}}, \bibinfo {author} {\bibfnamefont {Y.}~\bibnamefont {{Zhu}}},
  \bibinfo {author} {\bibfnamefont {A.}~\bibnamefont {{Li}}}, \ and\ \bibinfo
  {author} {\bibfnamefont {F.}~\bibnamefont {{Huang}}},\ }\href {\doibase
  10.3847/1538-4357/ac8544} {\bibfield  {journal} {\bibinfo  {journal} {\apj}\
  }\textbf {\bibinfo {volume} {936}},\ \bibinfo {eid} {69} (\bibinfo {year}
  {2022})}\BibitemShut {NoStop}%
\bibitem [{\citenamefont {Li}\ \emph {et~al.}(2012)\citenamefont {Li},
  \citenamefont {Huang},\ and\ \citenamefont {Xu}}]{Li12}%
  \BibitemOpen
  \bibfield  {author} {\bibinfo {author} {\bibfnamefont {A.}~\bibnamefont
  {Li}}, \bibinfo {author} {\bibfnamefont {F.}~\bibnamefont {Huang}}, \ and\
  \bibinfo {author} {\bibfnamefont {R.-X.}\ \bibnamefont {Xu}},\ }\href
  {\doibase https://doi.org/10.1016/j.astropartphys.2012.07.006} {\bibfield
  {journal} {\bibinfo  {journal} {Astropart. Phys.}\ }\textbf {\bibinfo
  {volume} {37}},\ \bibinfo {pages} {70} (\bibinfo {year} {2012})}\BibitemShut
  {NoStop}%
\bibitem [{\citenamefont {Leung}\ \emph {et~al.}(2011)\citenamefont {Leung},
  \citenamefont {Chu},\ and\ \citenamefont {Lin}}]{Leung11}%
  \BibitemOpen
  \bibfield  {author} {\bibinfo {author} {\bibfnamefont {S.-C.}\ \bibnamefont
  {Leung}}, \bibinfo {author} {\bibfnamefont {M.-C.}\ \bibnamefont {Chu}}, \
  and\ \bibinfo {author} {\bibfnamefont {L.-M.}\ \bibnamefont {Lin}},\ }\href
  {\doibase 10.1103/PhysRevD.84.107301} {\bibfield  {journal} {\bibinfo
  {journal} {Phys. Rev. D}\ }\textbf {\bibinfo {volume} {84}},\ \bibinfo
  {pages} {107301} (\bibinfo {year} {2011})}\BibitemShut {NoStop}%
\bibitem [{\citenamefont {Maselli}\ \emph {et~al.}(2017)\citenamefont
  {Maselli}, \citenamefont {Pnigouras}, \citenamefont {Nielsen}, \citenamefont
  {Kouvaris},\ and\ \citenamefont {Kokkotas}}]{Maselli17}%
  \BibitemOpen
  \bibfield  {author} {\bibinfo {author} {\bibfnamefont {A.}~\bibnamefont
  {Maselli}}, \bibinfo {author} {\bibfnamefont {P.}~\bibnamefont {Pnigouras}},
  \bibinfo {author} {\bibfnamefont {N.~G.}\ \bibnamefont {Nielsen}}, \bibinfo
  {author} {\bibfnamefont {C.}~\bibnamefont {Kouvaris}}, \ and\ \bibinfo
  {author} {\bibfnamefont {K.~D.}\ \bibnamefont {Kokkotas}},\ }\href {\doibase
  10.1103/PhysRevD.96.023005} {\bibfield  {journal} {\bibinfo  {journal} {Phys.
  Rev. D}\ }\textbf {\bibinfo {volume} {96}},\ \bibinfo {pages} {023005}
  (\bibinfo {year} {2017})}\BibitemShut {NoStop}%
\bibitem [{\citenamefont {Leung}\ \emph {et~al.}(2022)\citenamefont {Leung},
  \citenamefont {Chu},\ and\ \citenamefont {Lin}}]{Leung22}%
  \BibitemOpen
  \bibfield  {author} {\bibinfo {author} {\bibfnamefont {K.-L.}\ \bibnamefont
  {Leung}}, \bibinfo {author} {\bibfnamefont {M.-C.}\ \bibnamefont {Chu}}, \
  and\ \bibinfo {author} {\bibfnamefont {L.-M.}\ \bibnamefont {Lin}},\ }\href
  {\doibase 10.1103/PhysRevD.105.123010} {\bibfield  {journal} {\bibinfo
  {journal} {Phys. Rev. D}\ }\textbf {\bibinfo {volume} {105}},\ \bibinfo
  {pages} {123010} (\bibinfo {year} {2022})}\BibitemShut {NoStop}%
\bibitem [{\citenamefont {Tolos}\ and\ \citenamefont
  {Schaffner-Bielich}(2015)}]{Tolos15}%
  \BibitemOpen
  \bibfield  {author} {\bibinfo {author} {\bibfnamefont {L.}~\bibnamefont
  {Tolos}}\ and\ \bibinfo {author} {\bibfnamefont {J.}~\bibnamefont
  {Schaffner-Bielich}},\ }\href {\doibase 10.1103/PhysRevD.92.123002}
  {\bibfield  {journal} {\bibinfo  {journal} {Phys. Rev. D}\ }\textbf {\bibinfo
  {volume} {92}},\ \bibinfo {pages} {123002} (\bibinfo {year}
  {2015})}\BibitemShut {NoStop}%
\bibitem [{\citenamefont {Dengler}\ \emph {et~al.}(2022)\citenamefont {Dengler}
  \emph {et~al.}}]{Dengler22}%
  \BibitemOpen
  \bibfield  {author} {\bibinfo {author} {\bibfnamefont {Y.}~\bibnamefont
  {Dengler}} \emph {et~al.},\ }\href {\doibase 10.1103/PhysRevD.105.043013}
  {\bibfield  {journal} {\bibinfo  {journal} {Phys. Rev. D}\ }\textbf {\bibinfo
  {volume} {105}},\ \bibinfo {pages} {043013} (\bibinfo {year}
  {2022})}\BibitemShut {NoStop}%
\bibitem [{\citenamefont {Kain}(2021)}]{Kain21}%
  \BibitemOpen
  \bibfield  {author} {\bibinfo {author} {\bibfnamefont {B.}~\bibnamefont
  {Kain}},\ }\href {\doibase 10.1103/PhysRevD.103.043009} {\bibfield  {journal}
  {\bibinfo  {journal} {Phys. Rev. D}\ }\textbf {\bibinfo {volume} {103}},\
  \bibinfo {pages} {043009} (\bibinfo {year} {2021})}\BibitemShut {NoStop}%
\bibitem [{\citenamefont {Colpi}\ \emph {et~al.}(1986)\citenamefont {Colpi},
  \citenamefont {Shapiro},\ and\ \citenamefont {Wasserman}}]{Colpi86}%
  \BibitemOpen
  \bibfield  {author} {\bibinfo {author} {\bibfnamefont {M.}~\bibnamefont
  {Colpi}}, \bibinfo {author} {\bibfnamefont {S.~L.}\ \bibnamefont {Shapiro}},
  \ and\ \bibinfo {author} {\bibfnamefont {I.}~\bibnamefont {Wasserman}},\
  }\href {\doibase 10.1103/PhysRevLett.57.2485} {\bibfield  {journal} {\bibinfo
   {journal} {Phys. Rev. Lett.}\ }\textbf {\bibinfo {volume} {57}},\ \bibinfo
  {pages} {2485} (\bibinfo {year} {1986})}\BibitemShut {NoStop}%
\bibitem [{\citenamefont {{Chavanis}}\ and\ \citenamefont
  {{Harko}}(2012)}]{Chavanis12}%
  \BibitemOpen
  \bibfield  {author} {\bibinfo {author} {\bibfnamefont {P.-H.}\ \bibnamefont
  {{Chavanis}}}\ and\ \bibinfo {author} {\bibfnamefont {T.}~\bibnamefont
  {{Harko}}},\ }\href {\doibase 10.1103/PhysRevD.86.064011} {\bibfield
  {journal} {\bibinfo  {journal} {\prd}\ }\textbf {\bibinfo {volume} {86}},\
  \bibinfo {eid} {064011} (\bibinfo {year} {2012})}\BibitemShut {NoStop}%
\bibitem [{\citenamefont {{Rafiei Karkevandi}}\ \emph
  {et~al.}(2022)\citenamefont {{Rafiei Karkevandi}}, \citenamefont {{Shakeri}},
  \citenamefont {{Sagun}},\ and\ \citenamefont {{Ivanytskyi}}}]{Karkevandi22}%
  \BibitemOpen
  \bibfield  {author} {\bibinfo {author} {\bibfnamefont {D.}~\bibnamefont
  {{Rafiei Karkevandi}}}, \bibinfo {author} {\bibfnamefont {S.}~\bibnamefont
  {{Shakeri}}}, \bibinfo {author} {\bibfnamefont {V.}~\bibnamefont {{Sagun}}},
  \ and\ \bibinfo {author} {\bibfnamefont {O.}~\bibnamefont {{Ivanytskyi}}},\
  }\href {\doibase 10.1103/PhysRevD.105.023001} {\bibfield  {journal} {\bibinfo
   {journal} {\prd}\ }\textbf {\bibinfo {volume} {105}},\ \bibinfo {eid}
  {023001} (\bibinfo {year} {2022})}\BibitemShut {NoStop}%
\bibitem [{\citenamefont {{Kouvaris}}\ and\ \citenamefont
  {{Tinyakov}}(2010)}]{Kouvaris10}%
  \BibitemOpen
  \bibfield  {author} {\bibinfo {author} {\bibfnamefont {C.}~\bibnamefont
  {{Kouvaris}}}\ and\ \bibinfo {author} {\bibfnamefont {P.}~\bibnamefont
  {{Tinyakov}}},\ }\href {\doibase 10.1103/PhysRevD.82.063531} {\bibfield
  {journal} {\bibinfo  {journal} {\prd}\ }\textbf {\bibinfo {volume} {82}},\
  \bibinfo {eid} {063531} (\bibinfo {year} {2010})}\BibitemShut {NoStop}%
\bibitem [{\citenamefont {{Bramante}}\ and\ \citenamefont
  {{Elahi}}(2015)}]{Bramante15}%
  \BibitemOpen
  \bibfield  {author} {\bibinfo {author} {\bibfnamefont {J.}~\bibnamefont
  {{Bramante}}}\ and\ \bibinfo {author} {\bibfnamefont {F.}~\bibnamefont
  {{Elahi}}},\ }\href {\doibase 10.1103/PhysRevD.91.115001} {\bibfield
  {journal} {\bibinfo  {journal} {\prd}\ }\textbf {\bibinfo {volume} {91}},\
  \bibinfo {eid} {115001} (\bibinfo {year} {2015})}\BibitemShut {NoStop}%
\bibitem [{\citenamefont {{Deliyergiyev}}\ \emph {et~al.}(2019)\citenamefont
  {{Deliyergiyev}}, \citenamefont {{Del Popolo}}, \citenamefont {{Tolos}},
  \citenamefont {{Le Delliou}}, \citenamefont {{Lee}},\ and\ \citenamefont
  {{Burgio}}}]{Deliyergiyev19}%
  \BibitemOpen
  \bibfield  {author} {\bibinfo {author} {\bibfnamefont {M.}~\bibnamefont
  {{Deliyergiyev}}}, \bibinfo {author} {\bibfnamefont {A.}~\bibnamefont {{Del
  Popolo}}}, \bibinfo {author} {\bibfnamefont {L.}~\bibnamefont {{Tolos}}},
  \bibinfo {author} {\bibfnamefont {M.}~\bibnamefont {{Le Delliou}}}, \bibinfo
  {author} {\bibfnamefont {X.}~\bibnamefont {{Lee}}}, \ and\ \bibinfo {author}
  {\bibfnamefont {F.}~\bibnamefont {{Burgio}}},\ }\href {\doibase
  10.1103/PhysRevD.99.063015} {\bibfield  {journal} {\bibinfo  {journal}
  {\prd}\ }\textbf {\bibinfo {volume} {99}},\ \bibinfo {eid} {063015} (\bibinfo
  {year} {2019})}\BibitemShut {NoStop}%
\bibitem [{\citenamefont {{Goldman}}\ \emph {et~al.}(2013)\citenamefont
  {{Goldman}}, \citenamefont {{Mohapatra}}, \citenamefont {{Nussinov}},
  \citenamefont {{Rosenbaum}},\ and\ \citenamefont {{Teplitz}}}]{Goldman13}%
  \BibitemOpen
  \bibfield  {author} {\bibinfo {author} {\bibfnamefont {I.}~\bibnamefont
  {{Goldman}}}, \bibinfo {author} {\bibfnamefont {R.~N.}\ \bibnamefont
  {{Mohapatra}}}, \bibinfo {author} {\bibfnamefont {S.}~\bibnamefont
  {{Nussinov}}}, \bibinfo {author} {\bibfnamefont {D.}~\bibnamefont
  {{Rosenbaum}}}, \ and\ \bibinfo {author} {\bibfnamefont {V.}~\bibnamefont
  {{Teplitz}}},\ }\href {\doibase 10.1016/j.physletb.2013.07.017} {\bibfield
  {journal} {\bibinfo  {journal} {\plb}\ }\textbf {\bibinfo {volume} {725}},\
  \bibinfo {pages} {200} (\bibinfo {year} {2013})}\BibitemShut {NoStop}%
\bibitem [{\citenamefont {{Kouvaris}}\ and\ \citenamefont
  {{Nielsen}}(2015)}]{Kouvaris15}%
  \BibitemOpen
  \bibfield  {author} {\bibinfo {author} {\bibfnamefont {C.}~\bibnamefont
  {{Kouvaris}}}\ and\ \bibinfo {author} {\bibfnamefont {N.~G.}\ \bibnamefont
  {{Nielsen}}},\ }\href {\doibase 10.1103/PhysRevD.92.063526} {\bibfield
  {journal} {\bibinfo  {journal} {\prd}\ }\textbf {\bibinfo {volume} {92}},\
  \bibinfo {eid} {063526} (\bibinfo {year} {2015})}\BibitemShut {NoStop}%
\bibitem [{\citenamefont {{Eby}}\ \emph {et~al.}(2016)\citenamefont {{Eby}},
  \citenamefont {{Kouvaris}}, \citenamefont {{Nielsen}},\ and\ \citenamefont
  {{Wijewardhana}}}]{Eby16}%
  \BibitemOpen
  \bibfield  {author} {\bibinfo {author} {\bibfnamefont {J.}~\bibnamefont
  {{Eby}}}, \bibinfo {author} {\bibfnamefont {C.}~\bibnamefont {{Kouvaris}}},
  \bibinfo {author} {\bibfnamefont {N.~G.}\ \bibnamefont {{Nielsen}}}, \ and\
  \bibinfo {author} {\bibfnamefont {L.~C.~R.}\ \bibnamefont {{Wijewardhana}}},\
  }\href {\doibase 10.1007/JHEP02(2016)028} {\bibfield  {journal} {\bibinfo
  {journal} {J. High Energy Phys.}\ }\textbf {\bibinfo {volume} {2016}},\
  \bibinfo {eid} {28} (\bibinfo {year} {2016})}\BibitemShut {NoStop}%
\bibitem [{\citenamefont {{Nelson}}\ \emph {et~al.}(2019)\citenamefont
  {{Nelson}}, \citenamefont {{Reddy}},\ and\ \citenamefont
  {{Zhou}}}]{Nelson19}%
  \BibitemOpen
  \bibfield  {author} {\bibinfo {author} {\bibfnamefont {A.~E.}\ \bibnamefont
  {{Nelson}}}, \bibinfo {author} {\bibfnamefont {S.}~\bibnamefont {{Reddy}}}, \
  and\ \bibinfo {author} {\bibfnamefont {D.}~\bibnamefont {{Zhou}}},\ }\href
  {\doibase 10.1088/1475-7516/2019/07/012} {\bibfield  {journal} {\bibinfo
  {journal} {J. Cosmol. Astropart. Phys.}\ }\textbf {\bibinfo {volume}
  {2019}},\ \bibinfo {eid} {012} (\bibinfo {year} {2019})}\BibitemShut
  {NoStop}%
\bibitem [{\citenamefont {{Di Giovanni}}\ \emph {et~al.}(2020)\citenamefont
  {{Di Giovanni}}, \citenamefont {{Fakhry}}, \citenamefont {{Sanchis-Gual}},
  \citenamefont {{Degollado}},\ and\ \citenamefont {{Font}}}]{DiGiovanni20}%
  \BibitemOpen
  \bibfield  {author} {\bibinfo {author} {\bibfnamefont {F.}~\bibnamefont {{Di
  Giovanni}}}, \bibinfo {author} {\bibfnamefont {S.}~\bibnamefont {{Fakhry}}},
  \bibinfo {author} {\bibfnamefont {N.}~\bibnamefont {{Sanchis-Gual}}},
  \bibinfo {author} {\bibfnamefont {J.~C.}\ \bibnamefont {{Degollado}}}, \ and\
  \bibinfo {author} {\bibfnamefont {J.~A.}\ \bibnamefont {{Font}}},\ }\href
  {\doibase 10.1103/PhysRevD.102.084063} {\bibfield  {journal} {\bibinfo
  {journal} {\prd}\ }\textbf {\bibinfo {volume} {102}},\ \bibinfo {eid}
  {084063} (\bibinfo {year} {2020})}\BibitemShut {NoStop}%
\bibitem [{\citenamefont {Schunck}\ and\ \citenamefont
  {Mielke}(2003)}]{Schunck03}%
  \BibitemOpen
  \bibfield  {author} {\bibinfo {author} {\bibfnamefont {F.~E.}\ \bibnamefont
  {Schunck}}\ and\ \bibinfo {author} {\bibfnamefont {E.~W.}\ \bibnamefont
  {Mielke}},\ }\href {\doibase 10.1088/0264-9381/20/20/201} {\bibfield
  {journal} {\bibinfo  {journal} {Class. Quantum Grav.}\ }\textbf {\bibinfo
  {volume} {20}},\ \bibinfo {pages} {R301} (\bibinfo {year}
  {2003})}\BibitemShut {NoStop}%
\bibitem [{\citenamefont {{Liebling}}\ and\ \citenamefont
  {{Palenzuela}}(2012)}]{Liebling12}%
  \BibitemOpen
  \bibfield  {author} {\bibinfo {author} {\bibfnamefont {S.~L.}\ \bibnamefont
  {{Liebling}}}\ and\ \bibinfo {author} {\bibfnamefont {C.}~\bibnamefont
  {{Palenzuela}}},\ }\href {\doibase 10.12942/lrr-2012-6} {\bibfield  {journal}
  {\bibinfo  {journal} {Living Rev. Rel.}\ }\textbf {\bibinfo {volume} {15}},\
  \bibinfo {eid} {6} (\bibinfo {year} {2012})}\BibitemShut {NoStop}%
\bibitem [{\citenamefont {Del~Popolo}\ \emph {et~al.}(2020)\citenamefont
  {Del~Popolo}, \citenamefont {Le~Delliou},\ and\ \citenamefont
  {Deliyergiyev}}]{Delpopolo20}%
  \BibitemOpen
  \bibfield  {author} {\bibinfo {author} {\bibfnamefont {A.}~\bibnamefont
  {Del~Popolo}}, \bibinfo {author} {\bibfnamefont {M.}~\bibnamefont
  {Le~Delliou}}, \ and\ \bibinfo {author} {\bibfnamefont {M.}~\bibnamefont
  {Deliyergiyev}},\ }\href {https://www.mdpi.com/2218-1997/6/12/222} {\bibfield
   {journal} {\bibinfo  {journal} {Universe}\ }\textbf {\bibinfo {volume} {6}}
  (\bibinfo {year} {2020})}\BibitemShut {NoStop}%
\bibitem [{\citenamefont {{Del Popolo}}\ \emph {et~al.}(2020)\citenamefont
  {{Del Popolo}}, \citenamefont {{Deliyergiyev}},\ and\ \citenamefont {{Le
  Delliou}}}]{Delpopolo20b}%
  \BibitemOpen
  \bibfield  {author} {\bibinfo {author} {\bibfnamefont {A.}~\bibnamefont {{Del
  Popolo}}}, \bibinfo {author} {\bibfnamefont {M.}~\bibnamefont
  {{Deliyergiyev}}}, \ and\ \bibinfo {author} {\bibfnamefont {M.}~\bibnamefont
  {{Le Delliou}}},\ }\href {\doibase 10.1016/j.dark.2020.100622} {\bibfield
  {journal} {\bibinfo  {journal} {Phys. Dark Universe}\ }\textbf {\bibinfo
  {volume} {30}},\ \bibinfo {eid} {100622} (\bibinfo {year}
  {2020})}\BibitemShut {NoStop}%
\bibitem [{\citenamefont {Das}\ \emph {et~al.}(2021)\citenamefont {Das},
  \citenamefont {Kumar}, \citenamefont {Kumar}, \citenamefont {Biswal},\ and\
  \citenamefont {Patra}}]{Das21}%
  \BibitemOpen
  \bibfield  {author} {\bibinfo {author} {\bibfnamefont {H.}~\bibnamefont
  {Das}}, \bibinfo {author} {\bibfnamefont {A.}~\bibnamefont {Kumar}}, \bibinfo
  {author} {\bibfnamefont {B.}~\bibnamefont {Kumar}}, \bibinfo {author}
  {\bibfnamefont {S.}~\bibnamefont {Biswal}}, \ and\ \bibinfo {author}
  {\bibfnamefont {S.}~\bibnamefont {Patra}},\ }\href {\doibase
  10.1088/1475-7516/2021/01/007} {\bibfield  {journal} {\bibinfo  {journal} {J.
  Cosmol. Astropart. Phys.}\ }\textbf {\bibinfo {volume} {2021}},\ \bibinfo
  {pages} {007} (\bibinfo {year} {2021})}\BibitemShut {NoStop}%
\bibitem [{\citenamefont {Yang}\ \emph {et~al.}(2021)\citenamefont {Yang},
  \citenamefont {Pi},\ and\ \citenamefont {Zheng}}]{Yang21}%
  \BibitemOpen
  \bibfield  {author} {\bibinfo {author} {\bibfnamefont {S.-H.}\ \bibnamefont
  {Yang}}, \bibinfo {author} {\bibfnamefont {C.-M.}\ \bibnamefont {Pi}}, \ and\
  \bibinfo {author} {\bibfnamefont {X.-P.}\ \bibnamefont {Zheng}},\ }\href
  {\doibase 10.1103/PhysRevD.104.083016} {\bibfield  {journal} {\bibinfo
  {journal} {Phys. Rev. D}\ }\textbf {\bibinfo {volume} {104}},\ \bibinfo
  {pages} {083016} (\bibinfo {year} {2021})}\BibitemShut {NoStop}%
\bibitem [{\citenamefont {{Maity}}\ and\ \citenamefont
  {{Queiroz}}(2021)}]{Maity21}%
  \BibitemOpen
  \bibfield  {author} {\bibinfo {author} {\bibfnamefont {T.~N.}\ \bibnamefont
  {{Maity}}}\ and\ \bibinfo {author} {\bibfnamefont {F.~S.}\ \bibnamefont
  {{Queiroz}}},\ }\href {\doibase 10.1103/PhysRevD.104.083019} {\bibfield
  {journal} {\bibinfo  {journal} {\prd}\ }\textbf {\bibinfo {volume} {104}},\
  \bibinfo {eid} {083019} (\bibinfo {year} {2021})}\BibitemShut {NoStop}%
\bibitem [{\citenamefont {{Bramante}}\ and\ \citenamefont
  {{Linden}}(2014)}]{Bramante14}%
  \BibitemOpen
  \bibfield  {author} {\bibinfo {author} {\bibfnamefont {J.}~\bibnamefont
  {{Bramante}}}\ and\ \bibinfo {author} {\bibfnamefont {T.}~\bibnamefont
  {{Linden}}},\ }\href {\doibase 10.1103/PhysRevLett.113.191301} {\bibfield
  {journal} {\bibinfo  {journal} {\prl}\ }\textbf {\bibinfo {volume} {113}},\
  \bibinfo {eid} {191301} (\bibinfo {year} {2014})}\BibitemShut {NoStop}%
\bibitem [{\citenamefont {Yagi}\ and\ \citenamefont
  {Yunes}(2013{\natexlab{a}})}]{Yagi13}%
  \BibitemOpen
  \bibfield  {author} {\bibinfo {author} {\bibfnamefont {K.}~\bibnamefont
  {Yagi}}\ and\ \bibinfo {author} {\bibfnamefont {N.}~\bibnamefont {Yunes}},\
  }\href {\doibase 10.1103/PhysRevD.88.023009} {\bibfield  {journal} {\bibinfo
  {journal} {Phys. Rev. D}\ }\textbf {\bibinfo {volume} {88}},\ \bibinfo
  {pages} {023009} (\bibinfo {year} {2013}{\natexlab{a}})}\BibitemShut
  {NoStop}%
\bibitem [{\citenamefont {Yagi}\ and\ \citenamefont
  {Yunes}(2013{\natexlab{b}})}]{Yagi13b}%
  \BibitemOpen
  \bibfield  {author} {\bibinfo {author} {\bibfnamefont {K.}~\bibnamefont
  {Yagi}}\ and\ \bibinfo {author} {\bibfnamefont {N.}~\bibnamefont {Yunes}},\
  }\href {\doibase 10.1126/science.1236462} {\bibfield  {journal} {\bibinfo
  {journal} {Science}\ }\textbf {\bibinfo {volume} {341}},\ \bibinfo {pages}
  {365} (\bibinfo {year} {2013}{\natexlab{b}})}\BibitemShut {NoStop}%
\bibitem [{\citenamefont {Yagi}\ and\ \citenamefont {Yunes}(2017)}]{Yagi17}%
  \BibitemOpen
  \bibfield  {author} {\bibinfo {author} {\bibfnamefont {K.}~\bibnamefont
  {Yagi}}\ and\ \bibinfo {author} {\bibfnamefont {N.}~\bibnamefont {Yunes}},\
  }\href {\doibase https://doi.org/10.1016/j.physrep.2017.03.002} {\bibfield
  {journal} {\bibinfo  {journal} {Phys. Rep.}\ }\textbf {\bibinfo {volume}
  {681}},\ \bibinfo {pages} {1} (\bibinfo {year} {2017})}\BibitemShut {NoStop}%
\bibitem [{\citenamefont {Zhang}\ \emph {et~al.}(2020)\citenamefont {Zhang},
  \citenamefont {Huang}, \citenamefont {Tsao},\ and\ \citenamefont
  {Lin}}]{Zhang20}%
  \BibitemOpen
  \bibfield  {author} {\bibinfo {author} {\bibfnamefont {K.}~\bibnamefont
  {Zhang}}, \bibinfo {author} {\bibfnamefont {G.-Z.}\ \bibnamefont {Huang}},
  \bibinfo {author} {\bibfnamefont {J.-S.}\ \bibnamefont {Tsao}}, \ and\
  \bibinfo {author} {\bibfnamefont {F.-L.}\ \bibnamefont {Lin}},\ }\href
  {\doibase 10.1140/epjc/s10052-022-10335-8} {\bibfield  {journal} {\bibinfo
  {journal} {Eur. Phys. J. C}\ }\textbf {\bibinfo {volume} {82}},\ \bibinfo
  {pages} {366} (\bibinfo {year} {2020})}\BibitemShut {NoStop}%
\bibitem [{\citenamefont {Abbott}\ \emph {et~al.}(2017)\citenamefont {Abbott}
  \emph {et~al.}}]{Abbott17}%
  \BibitemOpen
  \bibfield  {author} {\bibinfo {author} {\bibfnamefont {B.~P.}\ \bibnamefont
  {Abbott}} \emph {et~al.} (\bibinfo {collaboration} {LIGO Scientific
  Collaboration and Virgo Collaboration}),\ }\href {\doibase
  10.1103/PhysRevLett.119.161101} {\bibfield  {journal} {\bibinfo  {journal}
  {Phys. Rev. Lett.}\ }\textbf {\bibinfo {volume} {119}},\ \bibinfo {pages}
  {161101} (\bibinfo {year} {2017})}\BibitemShut {NoStop}%
\bibitem [{\citenamefont {Abbott}\ \emph {et~al.}(2020)\citenamefont {Abbott}
  \emph {et~al.}}]{Abbott20}%
  \BibitemOpen
  \bibfield  {author} {\bibinfo {author} {\bibfnamefont {B.~P.}\ \bibnamefont
  {Abbott}} \emph {et~al.},\ }\href {\doibase 10.3847/2041-8213/ab75f5}
  {\bibfield  {journal} {\bibinfo  {journal} {Astrophys. J. Lett.}\ }\textbf
  {\bibinfo {volume} {892}},\ \bibinfo {pages} {L3} (\bibinfo {year}
  {2020})}\BibitemShut {NoStop}%
\bibitem [{\citenamefont {{Husain}}\ and\ \citenamefont
  {{Thomas}}(2021)}]{Husain21}%
  \BibitemOpen
  \bibfield  {author} {\bibinfo {author} {\bibfnamefont {W.}~\bibnamefont
  {{Husain}}}\ and\ \bibinfo {author} {\bibfnamefont {A.~W.}\ \bibnamefont
  {{Thomas}}},\ }\href {\doibase 10.1088/1475-7516/2021/10/086} {\bibfield
  {journal} {\bibinfo  {journal} {J. Cosmol. Astropart. Phys.}\ }\textbf
  {\bibinfo {volume} {2021}},\ \bibinfo {eid} {086} (\bibinfo {year}
  {2021})}\BibitemShut {NoStop}%
\bibitem [{\citenamefont {{Das}}\ \emph {et~al.}(2021)\citenamefont {{Das}},
  \citenamefont {{Kumar}}, \citenamefont {{Biswal}},\ and\ \citenamefont
  {{Patra}}}]{Das21d}%
  \BibitemOpen
  \bibfield  {author} {\bibinfo {author} {\bibfnamefont {H.~C.}\ \bibnamefont
  {{Das}}}, \bibinfo {author} {\bibfnamefont {A.}~\bibnamefont {{Kumar}}},
  \bibinfo {author} {\bibfnamefont {S.~K.}\ \bibnamefont {{Biswal}}}, \ and\
  \bibinfo {author} {\bibfnamefont {S.~K.}\ \bibnamefont {{Patra}}},\ }\href
  {\doibase 10.1103/PhysRevD.104.123006} {\bibfield  {journal} {\bibinfo
  {journal} {\prd}\ }\textbf {\bibinfo {volume} {104}},\ \bibinfo {eid}
  {123006} (\bibinfo {year} {2021})}\BibitemShut {NoStop}%
\bibitem [{\citenamefont {{Jeukenne}}\ \emph {et~al.}(1976)\citenamefont
  {{Jeukenne}}, \citenamefont {{Lejeune}},\ and\ \citenamefont
  {{Mahaux}}}]{Jeukenne76}%
  \BibitemOpen
  \bibfield  {author} {\bibinfo {author} {\bibfnamefont {J.~P.}\ \bibnamefont
  {{Jeukenne}}}, \bibinfo {author} {\bibfnamefont {A.}~\bibnamefont
  {{Lejeune}}}, \ and\ \bibinfo {author} {\bibfnamefont {C.}~\bibnamefont
  {{Mahaux}}},\ }\href {\doibase 10.1016/0370-1573(76)90017-X} {\bibfield
  {journal} {\bibinfo  {journal} {\physrep}\ }\textbf {\bibinfo {volume}
  {25}},\ \bibinfo {pages} {83} (\bibinfo {year} {1976})}\BibitemShut {NoStop}%
\bibitem [{\citenamefont {Baldo}(1999)}]{Baldo99}%
  \BibitemOpen
  \bibinfo {editor} {\bibfnamefont {M.}~\bibnamefont {Baldo}},\ ed.,\
  \href@noop {} {\emph {\bibinfo {title} {Nuclear Methods and The Nuclear
  Equation of State}}}\ (\bibinfo  {publisher} {World Scientific},\ \bibinfo
  {address} {Singapore},\ \bibinfo {year} {1999})\BibitemShut {NoStop}%
\bibitem [{\citenamefont {{Baldo}}\ and\ \citenamefont
  {{Burgio}}(2012)}]{Baldo12}%
  \BibitemOpen
  \bibfield  {author} {\bibinfo {author} {\bibfnamefont {M.}~\bibnamefont
  {{Baldo}}}\ and\ \bibinfo {author} {\bibfnamefont {G.~F.}\ \bibnamefont
  {{Burgio}}},\ }\href {\doibase 10.1088/0034-4885/75/2/026301} {\bibfield
  {journal} {\bibinfo  {journal} {\rpp}\ }\textbf {\bibinfo {volume} {75}},\
  \bibinfo {eid} {026301} (\bibinfo {year} {2012})}\BibitemShut {NoStop}%
\bibitem [{\citenamefont {Song}\ \emph {et~al.}(1998)\citenamefont {Song},
  \citenamefont {Baldo}, \citenamefont {Giansiracusa},\ and\ \citenamefont
  {Lombardo}}]{Song98}%
  \BibitemOpen
  \bibfield  {author} {\bibinfo {author} {\bibfnamefont {H.}~\bibnamefont
  {Song}}, \bibinfo {author} {\bibfnamefont {M.}~\bibnamefont {Baldo}},
  \bibinfo {author} {\bibfnamefont {G.}~\bibnamefont {Giansiracusa}}, \ and\
  \bibinfo {author} {\bibfnamefont {U.}~\bibnamefont {Lombardo}},\ }\href
  {\doibase 10.1103/PhysRevLett.81.1584} {\bibfield  {journal} {\bibinfo
  {journal} {{Phys. Rev. Lett.}}\ }\textbf {\bibinfo {volume} {81}},\ \bibinfo
  {pages} {1584} (\bibinfo {year} {1998})}\BibitemShut {NoStop}%
\bibitem [{\citenamefont {{Baldo}}\ \emph {et~al.}(2000)\citenamefont
  {{Baldo}}, \citenamefont {{Giansiracusa}}, \citenamefont {{Lombardo}},\ and\
  \citenamefont {{Song}}}]{Song00}%
  \BibitemOpen
  \bibfield  {author} {\bibinfo {author} {\bibfnamefont {M.}~\bibnamefont
  {{Baldo}}}, \bibinfo {author} {\bibfnamefont {G.}~\bibnamefont
  {{Giansiracusa}}}, \bibinfo {author} {\bibfnamefont {U.}~\bibnamefont
  {{Lombardo}}}, \ and\ \bibinfo {author} {\bibfnamefont {H.~Q.}\ \bibnamefont
  {{Song}}},\ }\href {\doibase 10.1016/S0370-2693(99)01463-X} {\bibfield
  {journal} {\bibinfo  {journal} {\plb}\ }\textbf {\bibinfo {volume} {473}},\
  \bibinfo {pages} {1} (\bibinfo {year} {2000})}\BibitemShut {NoStop}%
\bibitem [{\citenamefont {{Lu}}\ \emph {et~al.}(2017)\citenamefont {{Lu}},
  \citenamefont {{Li}}, \citenamefont {{Chen}}, \citenamefont {{Baldo}},\ and\
  \citenamefont {{Schulze}}}]{Lu17}%
  \BibitemOpen
  \bibfield  {author} {\bibinfo {author} {\bibfnamefont {J.-J.}\ \bibnamefont
  {{Lu}}}, \bibinfo {author} {\bibfnamefont {Z.-H.}\ \bibnamefont {{Li}}},
  \bibinfo {author} {\bibfnamefont {C.-Y.}\ \bibnamefont {{Chen}}}, \bibinfo
  {author} {\bibfnamefont {M.}~\bibnamefont {{Baldo}}}, \ and\ \bibinfo
  {author} {\bibfnamefont {H.-J.}\ \bibnamefont {{Schulze}}},\ }\href {\doibase
  10.1103/PhysRevC.96.044309} {\bibfield  {journal} {\bibinfo  {journal}
  {\prc}\ }\textbf {\bibinfo {volume} {96}},\ \bibinfo {eid} {044309} (\bibinfo
  {year} {2017})}\BibitemShut {NoStop}%
\bibitem [{\citenamefont {{Lu}}\ \emph {et~al.}(2018)\citenamefont {{Lu}},
  \citenamefont {{Li}}, \citenamefont {{Chen}}, \citenamefont {{Baldo}},\ and\
  \citenamefont {{Schulze}}}]{Lu18}%
  \BibitemOpen
  \bibfield  {author} {\bibinfo {author} {\bibfnamefont {J.-J.}\ \bibnamefont
  {{Lu}}}, \bibinfo {author} {\bibfnamefont {Z.-H.}\ \bibnamefont {{Li}}},
  \bibinfo {author} {\bibfnamefont {C.-Y.}\ \bibnamefont {{Chen}}}, \bibinfo
  {author} {\bibfnamefont {M.}~\bibnamefont {{Baldo}}}, \ and\ \bibinfo
  {author} {\bibfnamefont {H.-J.}\ \bibnamefont {{Schulze}}},\ }\href {\doibase
  10.1103/PhysRevC.98.064322} {\bibfield  {journal} {\bibinfo  {journal}
  {\prc}\ }\textbf {\bibinfo {volume} {98}},\ \bibinfo {eid} {064322} (\bibinfo
  {year} {2018})}\BibitemShut {NoStop}%
\bibitem [{\citenamefont {Wiringa}\ \emph {et~al.}(1995)\citenamefont
  {Wiringa}, \citenamefont {Stoks},\ and\ \citenamefont
  {Schiavilla}}]{Wiringa95}%
  \BibitemOpen
  \bibfield  {author} {\bibinfo {author} {\bibfnamefont {R.~B.}\ \bibnamefont
  {Wiringa}}, \bibinfo {author} {\bibfnamefont {V.}~\bibnamefont {Stoks}}, \
  and\ \bibinfo {author} {\bibfnamefont {R.}~\bibnamefont {Schiavilla}},\
  }\href {\doibase 10.1103/PhysRevC.51.38} {\bibfield  {journal} {\bibinfo
  {journal} {\prc}\ }\textbf {\bibinfo {volume} {51}},\ \bibinfo {pages} {38}
  (\bibinfo {year} {1995})}\BibitemShut {NoStop}%
\bibitem [{\citenamefont {Grang\'e}\ \emph {et~al.}(1989)\citenamefont
  {Grang\'e}, \citenamefont {Lejeune}, \citenamefont {Martzolff},\ and\
  \citenamefont {Mathiot}}]{Grange89}%
  \BibitemOpen
  \bibfield  {author} {\bibinfo {author} {\bibfnamefont {P.}~\bibnamefont
  {Grang\'e}}, \bibinfo {author} {\bibfnamefont {A.}~\bibnamefont {Lejeune}},
  \bibinfo {author} {\bibfnamefont {M.}~\bibnamefont {Martzolff}}, \ and\
  \bibinfo {author} {\bibfnamefont {J.-F.}\ \bibnamefont {Mathiot}},\ }\href
  {\doibase 10.1103/PhysRevC.40.1040} {\bibfield  {journal} {\bibinfo
  {journal} {Phys. Rev. C}\ }\textbf {\bibinfo {volume} {40}},\ \bibinfo
  {pages} {1040} (\bibinfo {year} {1989})}\BibitemShut {NoStop}%
\bibitem [{\citenamefont {{Zuo}}\ \emph {et~al.}(2002)\citenamefont {{Zuo}},
  \citenamefont {{Lejeune}}, \citenamefont {{Lombardo}},\ and\ \citenamefont
  {{Mathiot}}}]{Zuo02}%
  \BibitemOpen
  \bibfield  {author} {\bibinfo {author} {\bibfnamefont {W.}~\bibnamefont
  {{Zuo}}}, \bibinfo {author} {\bibfnamefont {A.}~\bibnamefont {{Lejeune}}},
  \bibinfo {author} {\bibfnamefont {U.}~\bibnamefont {{Lombardo}}}, \ and\
  \bibinfo {author} {\bibfnamefont {J.~F.}\ \bibnamefont {{Mathiot}}},\ }\href
  {https://doi.org/10.1140/epja/i2002-10031-y} {\bibfield  {journal} {\bibinfo
  {journal} {Eur. Phys. J. A}\ }\textbf {\bibinfo {volume} {14}},\ \bibinfo
  {pages} {469} (\bibinfo {year} {2002})}\BibitemShut {NoStop}%
\bibitem [{\citenamefont {{Li}}\ \emph {et~al.}(2008)\citenamefont {{Li}},
  \citenamefont {{Lombardo}}, \citenamefont {{Schulze}},\ and\ \citenamefont
  {{Zuo}}}]{Li08a}%
  \BibitemOpen
  \bibfield  {author} {\bibinfo {author} {\bibfnamefont {Z.~H.}\ \bibnamefont
  {{Li}}}, \bibinfo {author} {\bibfnamefont {U.}~\bibnamefont {{Lombardo}}},
  \bibinfo {author} {\bibfnamefont {H.-J.}\ \bibnamefont {{Schulze}}}, \ and\
  \bibinfo {author} {\bibfnamefont {W.}~\bibnamefont {{Zuo}}},\ }\href
  {\doibase 10.1103/PhysRevC.77.034316} {\bibfield  {journal} {\bibinfo
  {journal} {\prc}\ }\textbf {\bibinfo {volume} {77}},\ \bibinfo {eid} {034316}
  (\bibinfo {year} {2008})}\BibitemShut {NoStop}%
\bibitem [{\citenamefont {{Li}}\ and\ \citenamefont {{Schulze}}(2008)}]{Li08b}%
  \BibitemOpen
  \bibfield  {author} {\bibinfo {author} {\bibfnamefont {Z.~H.}\ \bibnamefont
  {{Li}}}\ and\ \bibinfo {author} {\bibfnamefont {H.-J.}\ \bibnamefont
  {{Schulze}}},\ }\href {\doibase 10.1103/PhysRevC.78.028801} {\bibfield
  {journal} {\bibinfo  {journal} {\prc}\ }\textbf {\bibinfo {volume} {78}},\
  \bibinfo {eid} {028801} (\bibinfo {year} {2008})}\BibitemShut {NoStop}%
\bibitem [{\citenamefont {{Wei}}\ \emph {et~al.}(2020)\citenamefont {{Wei}},
  \citenamefont {{Lu}}, \citenamefont {{Burgio}}, \citenamefont {{Li}},\ and\
  \citenamefont {{Schulze}}}]{Wei20}%
  \BibitemOpen
  \bibfield  {author} {\bibinfo {author} {\bibfnamefont {J.-B.}\ \bibnamefont
  {{Wei}}}, \bibinfo {author} {\bibfnamefont {J.-J.}\ \bibnamefont {{Lu}}},
  \bibinfo {author} {\bibfnamefont {G.~F.}\ \bibnamefont {{Burgio}}}, \bibinfo
  {author} {\bibfnamefont {Z.-H.}\ \bibnamefont {{Li}}}, \ and\ \bibinfo
  {author} {\bibfnamefont {H.-J.}\ \bibnamefont {{Schulze}}},\ }\href {\doibase
  10.1140/epja/s10050-020-00058-3} {\bibfield  {journal} {\bibinfo  {journal}
  {Eur. Phys. J. A}\ }\textbf {\bibinfo {volume} {56}},\ \bibinfo {eid} {63}
  (\bibinfo {year} {2020})}\BibitemShut {NoStop}%
\bibitem [{\citenamefont {{Burgio}}\ \emph {et~al.}(2021)\citenamefont
  {{Burgio}}, \citenamefont {{Schulze}}, \citenamefont {{Vida{\~n}a}},\ and\
  \citenamefont {{Wei}}}]{Burgio21b}%
  \BibitemOpen
  \bibfield  {author} {\bibinfo {author} {\bibfnamefont {G.~F.}\ \bibnamefont
  {{Burgio}}}, \bibinfo {author} {\bibfnamefont {H.-J.}\ \bibnamefont
  {{Schulze}}}, \bibinfo {author} {\bibfnamefont {I.}~\bibnamefont
  {{Vida{\~n}a}}}, \ and\ \bibinfo {author} {\bibfnamefont {J.-B.}\
  \bibnamefont {{Wei}}},\ }\href {\doibase 10.3390/sym13030400} {\bibfield
  {journal} {\bibinfo  {journal} {Symmetry}\ }\textbf {\bibinfo {volume}
  {13}},\ \bibinfo {pages} {400} (\bibinfo {year} {2021})}\BibitemShut
  {NoStop}%
\bibitem [{\citenamefont {{Liu}}\ \emph {et~al.}(2022)\citenamefont {{Liu}},
  \citenamefont {{Zhang}}, \citenamefont {{Li}}, \citenamefont {{Wei}},
  \citenamefont {{Burgio}},\ and\ \citenamefont {{Schulze}}}]{Liu22}%
  \BibitemOpen
  \bibfield  {author} {\bibinfo {author} {\bibfnamefont {H.-M.}\ \bibnamefont
  {{Liu}}}, \bibinfo {author} {\bibfnamefont {J.}~\bibnamefont {{Zhang}}},
  \bibinfo {author} {\bibfnamefont {Z.-H.}\ \bibnamefont {{Li}}}, \bibinfo
  {author} {\bibfnamefont {J.-B.}\ \bibnamefont {{Wei}}}, \bibinfo {author}
  {\bibfnamefont {G.~F.}\ \bibnamefont {{Burgio}}}, \ and\ \bibinfo {author}
  {\bibfnamefont {H.~J.}\ \bibnamefont {{Schulze}}},\ }\href {\doibase
  10.1103/PhysRevC.106.025801} {\bibfield  {journal} {\bibinfo  {journal}
  {\prc}\ }\textbf {\bibinfo {volume} {106}},\ \bibinfo {eid} {025801}
  (\bibinfo {year} {2022})}\BibitemShut {NoStop}%
\bibitem [{\citenamefont {Negele}\ and\ \citenamefont
  {Vautherin}(1973)}]{Negele71}%
  \BibitemOpen
  \bibfield  {author} {\bibinfo {author} {\bibfnamefont {J.~W.}\ \bibnamefont
  {Negele}}\ and\ \bibinfo {author} {\bibfnamefont {D.}~\bibnamefont
  {Vautherin}},\ }\href {\doibase https://doi.org/10.1016/0375-9474(73)90349-7}
  {\bibfield  {journal} {\bibinfo  {journal} {Nucl. Phys. A}\ }\textbf
  {\bibinfo {volume} {207}},\ \bibinfo {pages} {298} (\bibinfo {year}
  {1973})}\BibitemShut {NoStop}%
\bibitem [{\citenamefont {Baym}\ \emph {et~al.}(1971)\citenamefont {Baym},
  \citenamefont {Pethick},\ and\ \citenamefont {Sutherland}}]{Baym71}%
  \BibitemOpen
  \bibfield  {author} {\bibinfo {author} {\bibfnamefont {G.}~\bibnamefont
  {Baym}}, \bibinfo {author} {\bibfnamefont {C.}~\bibnamefont {Pethick}}, \
  and\ \bibinfo {author} {\bibfnamefont {P.}~\bibnamefont {Sutherland}},\
  }\href {\doibase 10.1086/151216} {\bibfield  {journal} {\bibinfo  {journal}
  {\apj}\ }\textbf {\bibinfo {volume} {170}},\ \bibinfo {pages} {299} (\bibinfo
  {year} {1971})}\BibitemShut {NoStop}%
\bibitem [{\citenamefont {Feynman}\ \emph {et~al.}(1949)\citenamefont
  {Feynman}, \citenamefont {Metropolis},\ and\ \citenamefont
  {Teller}}]{Feynman49}%
  \BibitemOpen
  \bibfield  {author} {\bibinfo {author} {\bibfnamefont {R.~P.}\ \bibnamefont
  {Feynman}}, \bibinfo {author} {\bibfnamefont {N.}~\bibnamefont {Metropolis}},
  \ and\ \bibinfo {author} {\bibfnamefont {E.}~\bibnamefont {Teller}},\ }\href
  {\doibase 10.1103/PhysRev.75.1561} {\bibfield  {journal} {\bibinfo  {journal}
  {Phys. Rev.}\ }\textbf {\bibinfo {volume} {75}},\ \bibinfo {pages} {1561}
  (\bibinfo {year} {1949})}\BibitemShut {NoStop}%
\bibitem [{\citenamefont {{P. Demorest {\it et al.}}}(2010)}]{Demorest10}%
  \BibitemOpen
  \bibfield  {author} {\bibinfo {author} {\bibnamefont {{P. Demorest {\it et
  al.}}}},\ }\href {\doibase 10.1038/nature09466} {\bibfield  {journal}
  {\bibinfo  {journal} {Nature}\ }\textbf {\bibinfo {volume} {467}},\ \bibinfo
  {pages} {1081} (\bibinfo {year} {2010})}\BibitemShut {NoStop}%
\bibitem [{\citenamefont {{\it et al.}}(2016)}]{Fonseca16}%
  \BibitemOpen
  \bibfield  {author} {\bibinfo {author} {\bibfnamefont {E.~F.}\ \bibnamefont
  {{\it et al.}}},\ }\href {\doibase 10.3847/0004-637X/832/2/167} {\bibfield
  {journal} {\bibinfo  {journal} {Astrophys. J}\ }\textbf {\bibinfo {volume}
  {832}},\ \bibinfo {pages} {167} (\bibinfo {year} {2016})}\BibitemShut
  {NoStop}%
\bibitem [{\citenamefont {{Shibata}}\ \emph {et~al.}(2017)\citenamefont
  {{Shibata}}, \citenamefont {{Fujibayashi}}, \citenamefont {{Hotokezaka}},
  \citenamefont {{Kiuchi}}, \citenamefont {{Kyutoku}}, \citenamefont
  {{Sekiguchi}},\ and\ \citenamefont {{Tanaka}}}]{Shibata17}%
  \BibitemOpen
  \bibfield  {author} {\bibinfo {author} {\bibfnamefont {M.}~\bibnamefont
  {{Shibata}}}, \bibinfo {author} {\bibfnamefont {S.}~\bibnamefont
  {{Fujibayashi}}}, \bibinfo {author} {\bibfnamefont {K.}~\bibnamefont
  {{Hotokezaka}}}, \bibinfo {author} {\bibfnamefont {K.}~\bibnamefont
  {{Kiuchi}}}, \bibinfo {author} {\bibfnamefont {K.}~\bibnamefont {{Kyutoku}}},
  \bibinfo {author} {\bibfnamefont {Y.}~\bibnamefont {{Sekiguchi}}}, \ and\
  \bibinfo {author} {\bibfnamefont {M.}~\bibnamefont {{Tanaka}}},\ }\href
  {\doibase 10.1103/PhysRevD.96.123012} {\bibfield  {journal} {\bibinfo
  {journal} {\prd}\ }\textbf {\bibinfo {volume} {96}},\ \bibinfo {eid} {123012}
  (\bibinfo {year} {2017})}\BibitemShut {NoStop}%
\bibitem [{\citenamefont {Margalit}\ and\ \citenamefont
  {Metzger}(2017)}]{Margalit17}%
  \BibitemOpen
  \bibfield  {author} {\bibinfo {author} {\bibfnamefont {B.}~\bibnamefont
  {Margalit}}\ and\ \bibinfo {author} {\bibfnamefont {B.~D.}\ \bibnamefont
  {Metzger}},\ }\href {\doibase 10.3847/2041-8213/aa991c} {\bibfield  {journal}
  {\bibinfo  {journal} {Astrophys. J. Lett.}\ }\textbf {\bibinfo {volume}
  {850}},\ \bibinfo {pages} {L19} (\bibinfo {year} {2017})}\BibitemShut
  {NoStop}%
\bibitem [{\citenamefont {{L. Rezzolla, E. R. Most, and L. R.
  Weih}}(2018)}]{Rezzolla18}%
  \BibitemOpen
  \bibfield  {author} {\bibinfo {author} {\bibnamefont {{L. Rezzolla, E. R.
  Most, and L. R. Weih}}},\ }\href {\doibase 10.3847/2041-8213/aaa401}
  {\bibfield  {journal} {\bibinfo  {journal} {Astrophys. J}\ }\textbf {\bibinfo
  {volume} {852}},\ \bibinfo {pages} {L25} (\bibinfo {year}
  {2018})}\BibitemShut {NoStop}%
\bibitem [{\citenamefont {{Shibata}}\ \emph {et~al.}(2019)\citenamefont
  {{Shibata}}, \citenamefont {{Zhou}}, \citenamefont {{Kiuchi}},\ and\
  \citenamefont {{Fujibayashi}}}]{Shibata19}%
  \BibitemOpen
  \bibfield  {author} {\bibinfo {author} {\bibfnamefont {M.}~\bibnamefont
  {{Shibata}}}, \bibinfo {author} {\bibfnamefont {E.}~\bibnamefont {{Zhou}}},
  \bibinfo {author} {\bibfnamefont {K.}~\bibnamefont {{Kiuchi}}}, \ and\
  \bibinfo {author} {\bibfnamefont {S.}~\bibnamefont {{Fujibayashi}}},\ }\href
  {\doibase 10.1103/PhysRevD.100.023015} {\bibfield  {journal} {\bibinfo
  {journal} {\prd}\ }\textbf {\bibinfo {volume} {100}},\ \bibinfo {eid}
  {023015} (\bibinfo {year} {2019})}\BibitemShut {NoStop}%
\bibitem [{\citenamefont {Khadkikar}\ \emph {et~al.}(2021)\citenamefont
  {Khadkikar}, \citenamefont {Raduta}, \citenamefont {Oertel},\ and\
  \citenamefont {Sedrakian}}]{Khadkikar21}%
  \BibitemOpen
  \bibfield  {author} {\bibinfo {author} {\bibfnamefont {S.}~\bibnamefont
  {Khadkikar}}, \bibinfo {author} {\bibfnamefont {A.~R.}\ \bibnamefont
  {Raduta}}, \bibinfo {author} {\bibfnamefont {M.}~\bibnamefont {Oertel}}, \
  and\ \bibinfo {author} {\bibfnamefont {A.}~\bibnamefont {Sedrakian}},\ }\href
  {\doibase 10.1103/PhysRevC.103.055811} {\bibfield  {journal} {\bibinfo
  {journal} {Phys. Rev. C}\ }\textbf {\bibinfo {volume} {103}},\ \bibinfo
  {pages} {055811} (\bibinfo {year} {2021})}\BibitemShut {NoStop}%
\bibitem [{\citenamefont {{Bauswein}}\ \emph {et~al.}(2021)\citenamefont
  {{Bauswein}}, \citenamefont {{Blacker}}, \citenamefont {{Lioutas}},
  \citenamefont {{Soultanis}}, \citenamefont {{Vijayan}},\ and\ \citenamefont
  {{Stergioulas}}}]{Bauswein21}%
  \BibitemOpen
  \bibfield  {author} {\bibinfo {author} {\bibfnamefont {A.}~\bibnamefont
  {{Bauswein}}}, \bibinfo {author} {\bibfnamefont {S.}~\bibnamefont
  {{Blacker}}}, \bibinfo {author} {\bibfnamefont {G.}~\bibnamefont
  {{Lioutas}}}, \bibinfo {author} {\bibfnamefont {T.}~\bibnamefont
  {{Soultanis}}}, \bibinfo {author} {\bibfnamefont {V.}~\bibnamefont
  {{Vijayan}}}, \ and\ \bibinfo {author} {\bibfnamefont {N.}~\bibnamefont
  {{Stergioulas}}},\ }\href {\doibase 10.1103/PhysRevD.103.123004} {\bibfield
  {journal} {\bibinfo  {journal} {\prd}\ }\textbf {\bibinfo {volume} {103}},\
  \bibinfo {eid} {123004} (\bibinfo {year} {2021})}\BibitemShut {NoStop}%
\bibitem [{\citenamefont {{Figura}}\ \emph {et~al.}(2021)\citenamefont
  {{Figura}}, \citenamefont {{Li}}, \citenamefont {{Lu}}, \citenamefont
  {{Burgio}}, \citenamefont {{Li}},\ and\ \citenamefont
  {{Schulze}}}]{Figura21}%
  \BibitemOpen
  \bibfield  {author} {\bibinfo {author} {\bibfnamefont {A.}~\bibnamefont
  {{Figura}}}, \bibinfo {author} {\bibfnamefont {F.}~\bibnamefont {{Li}}},
  \bibinfo {author} {\bibfnamefont {J.-J.}\ \bibnamefont {{Lu}}}, \bibinfo
  {author} {\bibfnamefont {G.~F.}\ \bibnamefont {{Burgio}}}, \bibinfo {author}
  {\bibfnamefont {Z.-H.}\ \bibnamefont {{Li}}}, \ and\ \bibinfo {author}
  {\bibfnamefont {H.-J.}\ \bibnamefont {{Schulze}}},\ }\href {\doibase
  10.1103/PhysRevD.103.083012} {\bibfield  {journal} {\bibinfo  {journal}
  {\prd}\ }\textbf {\bibinfo {volume} {103}},\ \bibinfo {eid} {083012}
  (\bibinfo {year} {2021})}\BibitemShut {NoStop}%
\bibitem [{\citenamefont {{Burgio}}\ \emph {et~al.}(2018)\citenamefont
  {{Burgio}}, \citenamefont {{Drago}}, \citenamefont {{Pagliara}},
  \citenamefont {{Schulze}},\ and\ \citenamefont {{Wei}}}]{Burgio18}%
  \BibitemOpen
  \bibfield  {author} {\bibinfo {author} {\bibfnamefont {G.~F.}\ \bibnamefont
  {{Burgio}}}, \bibinfo {author} {\bibfnamefont {A.}~\bibnamefont {{Drago}}},
  \bibinfo {author} {\bibfnamefont {G.}~\bibnamefont {{Pagliara}}}, \bibinfo
  {author} {\bibfnamefont {H.-J.}\ \bibnamefont {{Schulze}}}, \ and\ \bibinfo
  {author} {\bibfnamefont {J.-B.}\ \bibnamefont {{Wei}}},\ }\href {\doibase
  10.3847/1538-4357/aac6ee} {\bibfield  {journal} {\bibinfo  {journal}
  {Astrophys. J.}\ }\textbf {\bibinfo {volume} {860}},\ \bibinfo {eid} {139}
  (\bibinfo {year} {2018})}\BibitemShut {NoStop}%
\bibitem [{\citenamefont {Wei}\ \emph {et~al.}(2019)\citenamefont {Wei},
  \citenamefont {Figura}, \citenamefont {Burgio}, \citenamefont {Chen},\ and\
  \citenamefont {Schulze}}]{Wei19}%
  \BibitemOpen
  \bibfield  {author} {\bibinfo {author} {\bibfnamefont {J.-B.}\ \bibnamefont
  {Wei}}, \bibinfo {author} {\bibfnamefont {A.}~\bibnamefont {Figura}},
  \bibinfo {author} {\bibfnamefont {G.~F.}\ \bibnamefont {Burgio}}, \bibinfo
  {author} {\bibfnamefont {H.}~\bibnamefont {Chen}}, \ and\ \bibinfo {author}
  {\bibfnamefont {H.-J.}\ \bibnamefont {Schulze}},\ }\href {\doibase
  10.1088/1361-6471/aaf95c} {\bibfield  {journal} {\bibinfo  {journal} {J.
  Phys. G: Nucl. Part. Phys.}\ }\textbf {\bibinfo {volume} {46}},\ \bibinfo
  {pages} {034001} (\bibinfo {year} {2019})}\BibitemShut {NoStop}%
\bibitem [{\citenamefont {{Pang}}\ \emph {et~al.}(2021)\citenamefont {{Pang}},
  \citenamefont {{Tews}}, \citenamefont {{Coughlin}}, \citenamefont {{Bulla}},
  \citenamefont {{Van Den Broeck}},\ and\ \citenamefont {{Dietrich}}}]{Pang21}%
  \BibitemOpen
  \bibfield  {author} {\bibinfo {author} {\bibfnamefont {P.~T.~H.}\
  \bibnamefont {{Pang}}}, \bibinfo {author} {\bibfnamefont {I.}~\bibnamefont
  {{Tews}}}, \bibinfo {author} {\bibfnamefont {M.~W.}\ \bibnamefont
  {{Coughlin}}}, \bibinfo {author} {\bibfnamefont {M.}~\bibnamefont {{Bulla}}},
  \bibinfo {author} {\bibfnamefont {C.}~\bibnamefont {{Van Den Broeck}}}, \
  and\ \bibinfo {author} {\bibfnamefont {T.}~\bibnamefont {{Dietrich}}},\
  }\href {\doibase 10.3847/1538-4357/ac19ab} {\bibfield  {journal} {\bibinfo
  {journal} {\apj}\ }\textbf {\bibinfo {volume} {922}},\ \bibinfo {eid} {14}
  (\bibinfo {year} {2021})}\BibitemShut {NoStop}%
\bibitem [{\citenamefont {{Raaijmakers}}\ \emph {et~al.}(2021)\citenamefont
  {{Raaijmakers}} \emph {et~al.}}]{Raaijmakers21}%
  \BibitemOpen
  \bibfield  {author} {\bibinfo {author} {\bibfnamefont {G.}~\bibnamefont
  {{Raaijmakers}}} \emph {et~al.},\ }\href {\doibase 10.3847/2041-8213/ac089a}
  {\bibfield  {journal} {\bibinfo  {journal} {\apjl}\ }\textbf {\bibinfo
  {volume} {918}},\ \bibinfo {eid} {L29} (\bibinfo {year} {2021})}\BibitemShut
  {NoStop}%
\bibitem [{\citenamefont {Abbott}\ \emph {et~al.}(2018)\citenamefont {Abbott}
  \emph {et~al.}}]{Abbott18}%
  \BibitemOpen
  \bibfield  {author} {\bibinfo {author} {\bibfnamefont {B.~P.}\ \bibnamefont
  {Abbott}} \emph {et~al.},\ }\href {\doibase 10.1103/PhysRevLett.121.161101}
  {\bibfield  {journal} {\bibinfo  {journal} {\prl}\ }\textbf {\bibinfo
  {volume} {121}},\ \bibinfo {eid} {161101} (\bibinfo {year}
  {2018})}\BibitemShut {NoStop}%
\bibitem [{\citenamefont {{Davoudiasl}}\ \emph {et~al.}(2011)\citenamefont
  {{Davoudiasl}}, \citenamefont {{Morrissey}}, \citenamefont {{Sigurdson}},\
  and\ \citenamefont {{Tulin}}}]{Davoudiasl11}%
  \BibitemOpen
  \bibfield  {author} {\bibinfo {author} {\bibfnamefont {H.}~\bibnamefont
  {{Davoudiasl}}}, \bibinfo {author} {\bibfnamefont {D.~E.}\ \bibnamefont
  {{Morrissey}}}, \bibinfo {author} {\bibfnamefont {K.}~\bibnamefont
  {{Sigurdson}}}, \ and\ \bibinfo {author} {\bibfnamefont {S.}~\bibnamefont
  {{Tulin}}},\ }\href {\doibase 10.1103/PhysRevD.84.096008} {\bibfield
  {journal} {\bibinfo  {journal} {\prd}\ }\textbf {\bibinfo {volume} {84}},\
  \bibinfo {eid} {096008} (\bibinfo {year} {2011})}\BibitemShut {NoStop}%
\bibitem [{\citenamefont {{Baym}}\ \emph {et~al.}(2018)\citenamefont {{Baym}},
  \citenamefont {{Beck}}, \citenamefont {{Geltenbort}},\ and\ \citenamefont
  {{Shelton}}}]{Baym18}%
  \BibitemOpen
  \bibfield  {author} {\bibinfo {author} {\bibfnamefont {G.}~\bibnamefont
  {{Baym}}}, \bibinfo {author} {\bibfnamefont {D.~H.}\ \bibnamefont {{Beck}}},
  \bibinfo {author} {\bibfnamefont {P.}~\bibnamefont {{Geltenbort}}}, \ and\
  \bibinfo {author} {\bibfnamefont {J.}~\bibnamefont {{Shelton}}},\ }\href
  {\doibase 10.1103/PhysRevLett.121.061801} {\bibfield  {journal} {\bibinfo
  {journal} {\prl}\ }\textbf {\bibinfo {volume} {121}},\ \bibinfo {eid}
  {061801} (\bibinfo {year} {2018})}\BibitemShut {NoStop}%
\bibitem [{\citenamefont {Hartle}(1967)}]{Hartle67}%
  \BibitemOpen
  \bibfield  {author} {\bibinfo {author} {\bibfnamefont {J.~B.}\ \bibnamefont
  {Hartle}},\ }\href {\doibase 10.1086/149400} {\bibfield  {journal} {\bibinfo
  {journal} {Astrophys. J.}\ }\textbf {\bibinfo {volume} {150}},\ \bibinfo
  {pages} {1005} (\bibinfo {year} {1967})}\BibitemShut {NoStop}%
\bibitem [{\citenamefont {Hinderer}\ \emph {et~al.}(2010)\citenamefont
  {Hinderer}, \citenamefont {Lackey}, \citenamefont {Lang},\ and\ \citenamefont
  {Read}}]{Hinderer10}%
  \BibitemOpen
  \bibfield  {author} {\bibinfo {author} {\bibfnamefont {T.}~\bibnamefont
  {Hinderer}}, \bibinfo {author} {\bibfnamefont {B.~D.}\ \bibnamefont
  {Lackey}}, \bibinfo {author} {\bibfnamefont {R.~N.}\ \bibnamefont {Lang}}, \
  and\ \bibinfo {author} {\bibfnamefont {J.~S.}\ \bibnamefont {Read}},\ }\href
  {\doibase 10.1103/PhysRevD.81.123016} {\bibfield  {journal} {\bibinfo
  {journal} {Phys. Rev. D}\ }\textbf {\bibinfo {volume} {81}},\ \bibinfo
  {pages} {123016} (\bibinfo {year} {2010})}\BibitemShut {NoStop}%
\bibitem [{\citenamefont {Binnington}\ and\ \citenamefont
  {Poisson}(2009)}]{Binnington09}%
  \BibitemOpen
  \bibfield  {author} {\bibinfo {author} {\bibfnamefont {T.}~\bibnamefont
  {Binnington}}\ and\ \bibinfo {author} {\bibfnamefont {E.}~\bibnamefont
  {Poisson}},\ }\href {\doibase 10.1103/PhysRevD.80.084018} {\bibfield
  {journal} {\bibinfo  {journal} {Phys. Rev. D}\ }\textbf {\bibinfo {volume}
  {80}},\ \bibinfo {pages} {084018} (\bibinfo {year} {2009})}\BibitemShut
  {NoStop}%
\bibitem [{\citenamefont {Damour}\ and\ \citenamefont
  {Nagar}(2010)}]{Damour10}%
  \BibitemOpen
  \bibfield  {author} {\bibinfo {author} {\bibfnamefont {T.}~\bibnamefont
  {Damour}}\ and\ \bibinfo {author} {\bibfnamefont {A.}~\bibnamefont {Nagar}},\
  }\href {\doibase 10.1103/PhysRevD.81.084016} {\bibfield  {journal} {\bibinfo
  {journal} {Phys. Rev. D}\ }\textbf {\bibinfo {volume} {81}},\ \bibinfo
  {pages} {084016} (\bibinfo {year} {2010})}\BibitemShut {NoStop}%
\bibitem [{\citenamefont {Postnikov}\ \emph {et~al.}(2010)\citenamefont
  {Postnikov}, \citenamefont {Prakash},\ and\ \citenamefont
  {Lattimer}}]{Postnikov10}%
  \BibitemOpen
  \bibfield  {author} {\bibinfo {author} {\bibfnamefont {S.}~\bibnamefont
  {Postnikov}}, \bibinfo {author} {\bibfnamefont {M.}~\bibnamefont {Prakash}},
  \ and\ \bibinfo {author} {\bibfnamefont {J.~M.}\ \bibnamefont {Lattimer}},\
  }\href {\doibase 10.1103/PhysRevD.82.024016} {\bibfield  {journal} {\bibinfo
  {journal} {Phys. Rev. D}\ }\textbf {\bibinfo {volume} {82}},\ \bibinfo
  {pages} {024016} (\bibinfo {year} {2010})}\BibitemShut {NoStop}%
\end{thebibliography}%

\end{document}